\documentclass[useAMS,usegraphicx,usenatbib]{mn2e} 
\usepackage{natbib}
\usepackage{amsmath}
\usepackage{color}
\usepackage{url}
\usepackage{ulem}
\usepackage{multirow}
\usepackage{amsmath}
\usepackage{wasysym}
\usepackage{amssymb}% http://ctan.org/pkg/amssymb
\usepackage{pifont}% http://ctan.org/pkg/pifont

\bibpunct{(}{)}{;}{a}{}{,}

\title[AGN in late mergers are heavily obscured]{Growing supermassive black holes in the late stages of galaxy mergers are heavily obscured}
\author[C. Ricci et al.]{C. Ricci$^{1,2,3}$\thanks{E-mail:
cricci@astro.puc.cl},  F. E. Bauer$^{1,2,4,5}$, E. Treister$^{1,2}$, K. Schawinski$^{6}$, G. C. Privon$^{1,2}$,\newauthor
L. Blecha$^{7}$, P. Arevalo$^{8}$, L. Armus$^{9}$, F. Harrison$^{10}$, L. C. Ho$^{3,11}$, K. Iwasawa$^{12,13}$, \newauthor
D. B. Sanders$^{14}$, D. Stern$^{15}$\\
$^{1}$Instituto de Astrof\'{\i}sica, Facultad de F\'{i}sica, Pontificia Universidad Cat\'{o}lica de Chile, Casilla 306, Santiago 22, Chile\\
$^{2}$EMBIGGEN Anillo, Concepcion, Chile\\
$^{3}$Kavli Institute for Astronomy and Astrophysics, Peking University, Beijing 100871, China\\
$^{4}$Space Science Institute, 4750 Walnut Street, Suite 205, Boulder, Colorado 80301, USA\\
$^{5}$Millenium Institute of Astrophysics, Santiago, Chile\\
$^{6}$Institute for Astronomy, Department of Physics, ETH Zurich, Wolfgang-Pauli-Strasse 27, CH-8093 Zurich, Switzerland\\
$^{7}$University of Maryland Dept. of Astronomy, 1113 PSC, Bldg. 415, College Park, MD 20742, USA\\
$^{8}$Instituto de F\'isica y Astronom\'ia, Facultad de Ciencias, Universidad de Valpara\'iso, Gran Bretana N¼ 1111, Playa Ancha, Valpara\'iso, Chile\\
$^{9}$Infrared Processing and Analysis Center, California Institute of Technology, 1200 E. California Boulevard, Pasadena, CA 91125, USA\\
$^{10}$Cahill Center for Astronomy and Astrophysics, California Institute of Technology, Pasadena, CA 91125, USA\\
$^{11}$Department of Astronomy, School of Physics, Peking University, Beijing 100871, China \\
$^{12}$Institut de Ci\`encies del Cosmos, Universitat de Barceloa, IEEC-UB, Mart\'i i Franqu\`es, 1, 08028 Barcelona, Spain\\
$^{13}$ICREA, Pg. Llu\'is Companys, 23, 08010 Barcelona, Spain\\
$^{14}$Institute for Astronomy, 2680 Woodlawn Drive, University of Hawaii, Honolulu, HI 96822, USA\\
$^{15}$Jet Propulsion Laboratory, California Institute of Technology, Pasadena, CA 91109, USA
 }
\begin{document}
\date{Received; accepted}

\pagerange{\pageref{firstpage}--\pageref{lastpage}} \pubyear{2017}

\maketitle

\label{firstpage}

\begin{abstract}
Mergers of galaxies are thought to cause significant gas inflows to the inner parsecs, which can activate rapid accretion onto supermassive black holes (SMBHs), giving rise to Active Galactic Nuclei (AGN). During a significant fraction of this process, SMBHs are predicted to be enshrouded by gas and dust.
Studying 52 galactic nuclei in infrared-selected local Luminous and Ultra-luminous infrared galaxies in different merger stages in the hard X-ray band, where radiation is less affected by absorption, we find that the amount of material around SMBHs increases during the last phases of the merger. We find that the fraction of Compton-thick (CT, $N_{\rm\,H}\geq 10^{24}\rm\,cm^{-2}$) AGN in late merger galaxies is higher ($f_{\rm\,CT}=65^{+12}_{-13}\%$) than in local hard X-ray selected AGN ($f_{\rm\,CT}=27\pm 4\%$), and that obscuration reaches its maximum when the nuclei of the two merging galaxies are at a projected distance of $D_{12}\simeq0.4-10.8$\, kiloparsecs ($f_{\rm\,CT}=77_{-17}^{+13}\%$).
We also find that all AGN of our sample in late merger galaxies have $N_{\rm\,H}> 10^{23}\rm\,cm^{-2}$, which implies that the obscuring material covers $95^{+4}_{-8}\%$ of the X-ray source.
These observations show that the material is most effectively funnelled from the galactic scale to the inner tens of parsecs during the late stages of galaxy mergers, and that the close environment of SMBHs in advanced mergers is richer in gas and dust with respect to that of SMBHs in isolated galaxies, and cannot be explained by the classical AGN unification model in which the torus is responsible for the obscuration.

\end{abstract}	
               
  \begin{keywords}
        galaxies: active --- X-rays: general --- galaxies: nuclei --- galaxies: Seyfert --- quasars: general --- infrared: galaxies

\end{keywords}

   %______________________________________________________________
   
\section{Introduction}

\begin{figure*}
\centering
\centering
\textbf{\Large Early merger galaxies}\par\smallskip
\begin{minipage}{.49\textwidth}
\centering
\includegraphics[width=8.2cm]{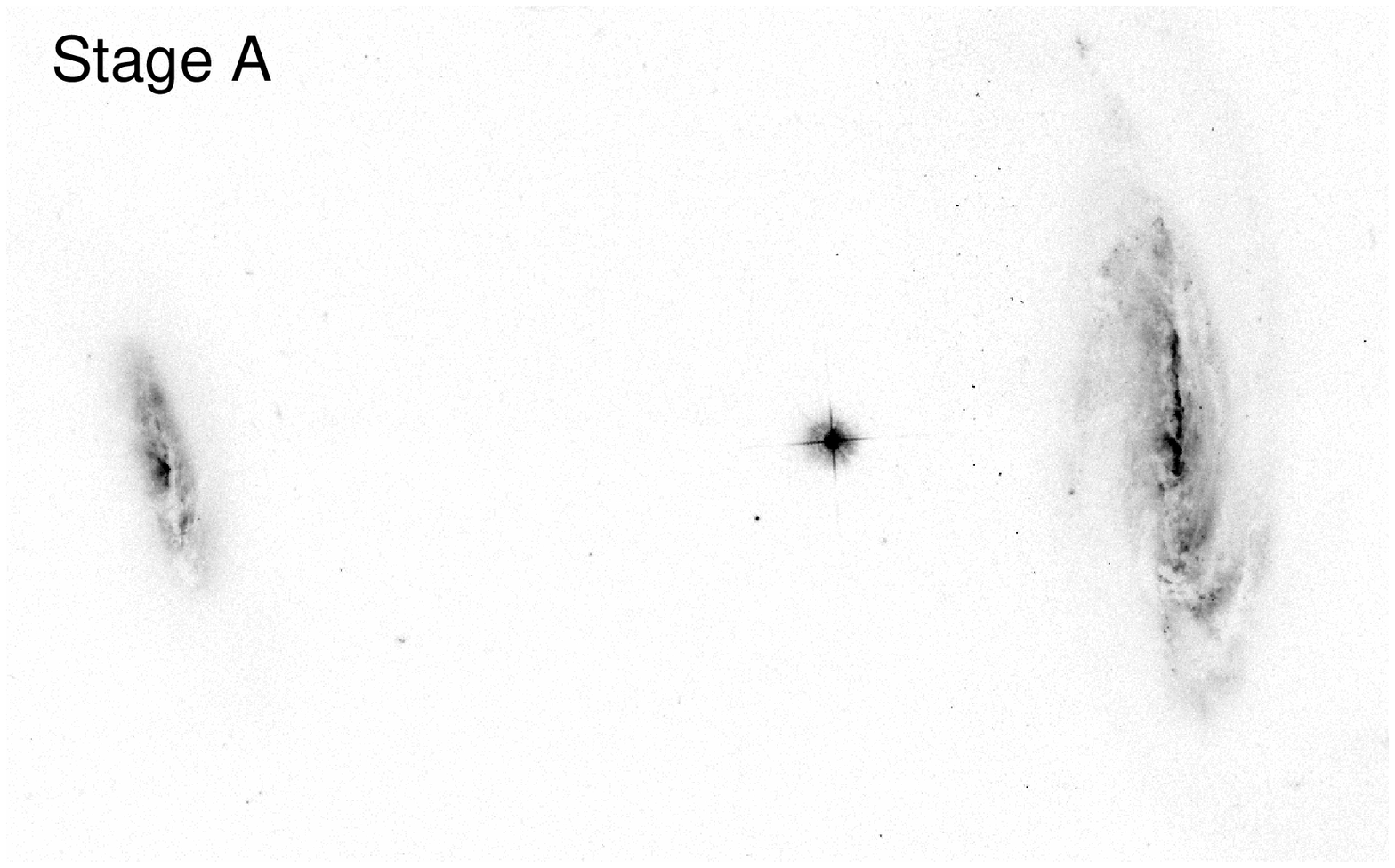}\end{minipage}
\begin{minipage}{.49\textwidth}
\centering
\includegraphics[width=8.2cm]{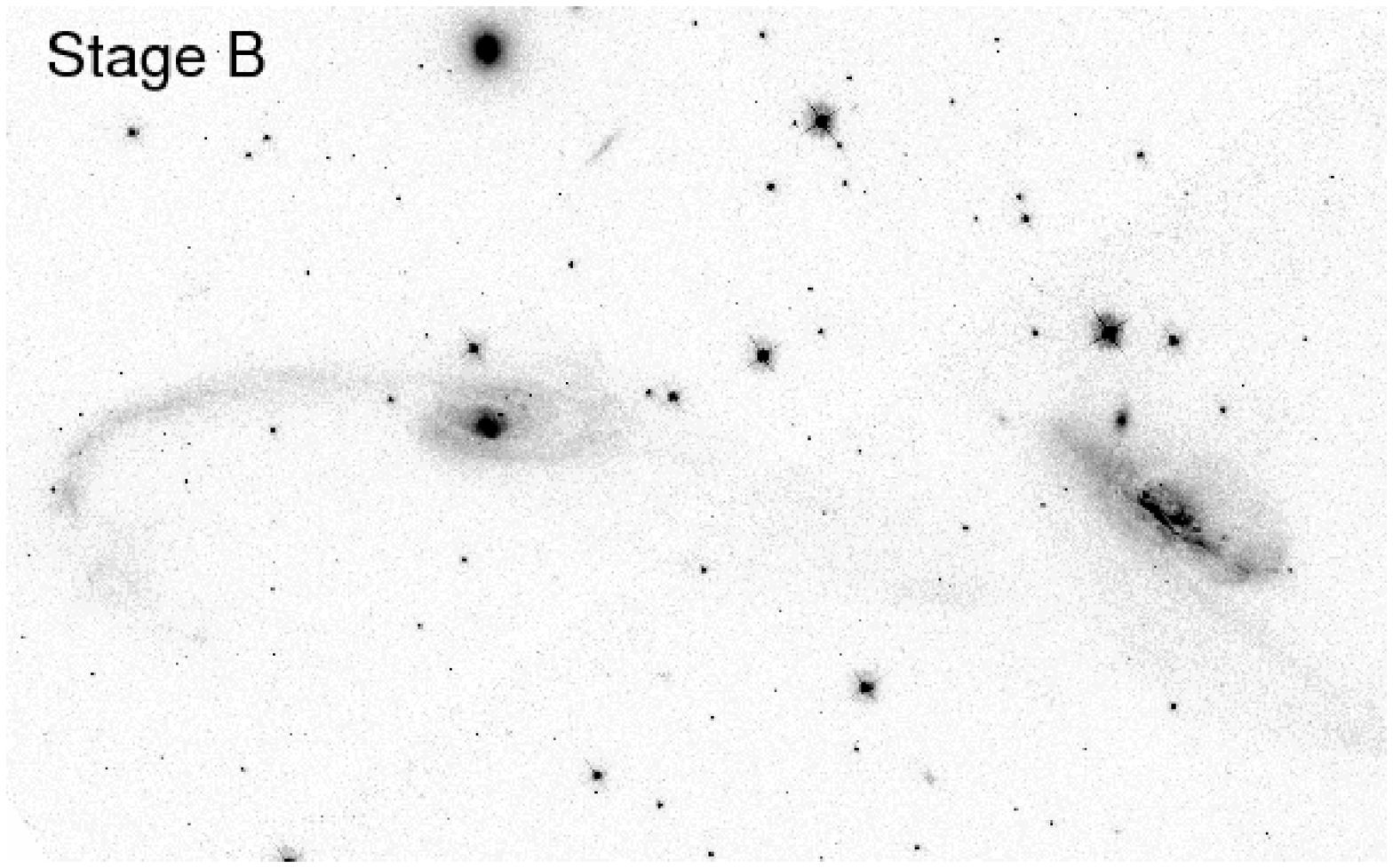}\end{minipage}
\par\smallskip
\centering
\par\smallskip
\par\smallskip
\par\smallskip
\textbf{\Large Late merger galaxies}\par\smallskip
\begin{minipage}{.49\textwidth}
\centering
\includegraphics[width=8.2cm]{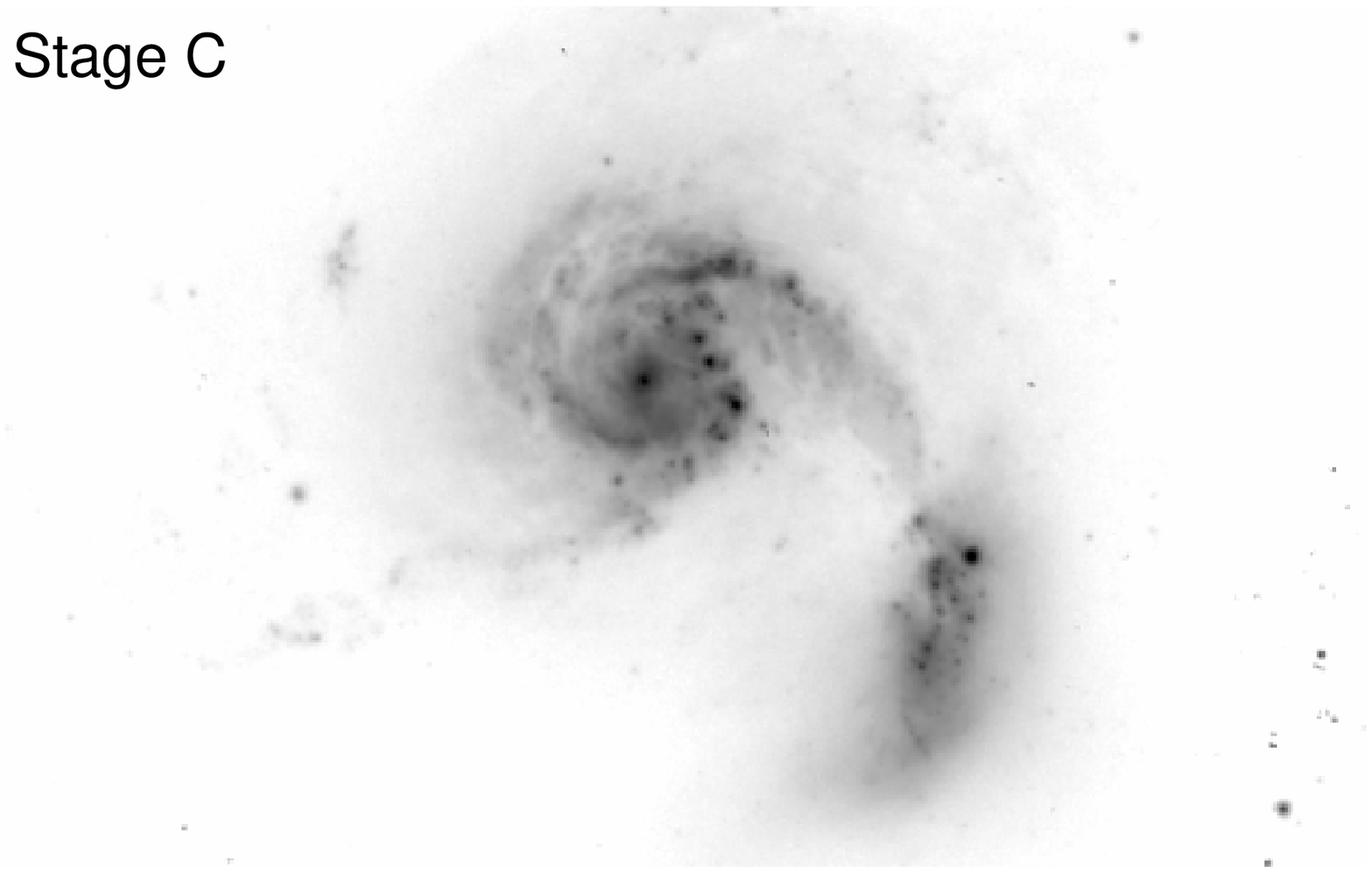}\end{minipage}
\begin{minipage}{.49\textwidth}
\centering
\includegraphics[width=8.2cm]{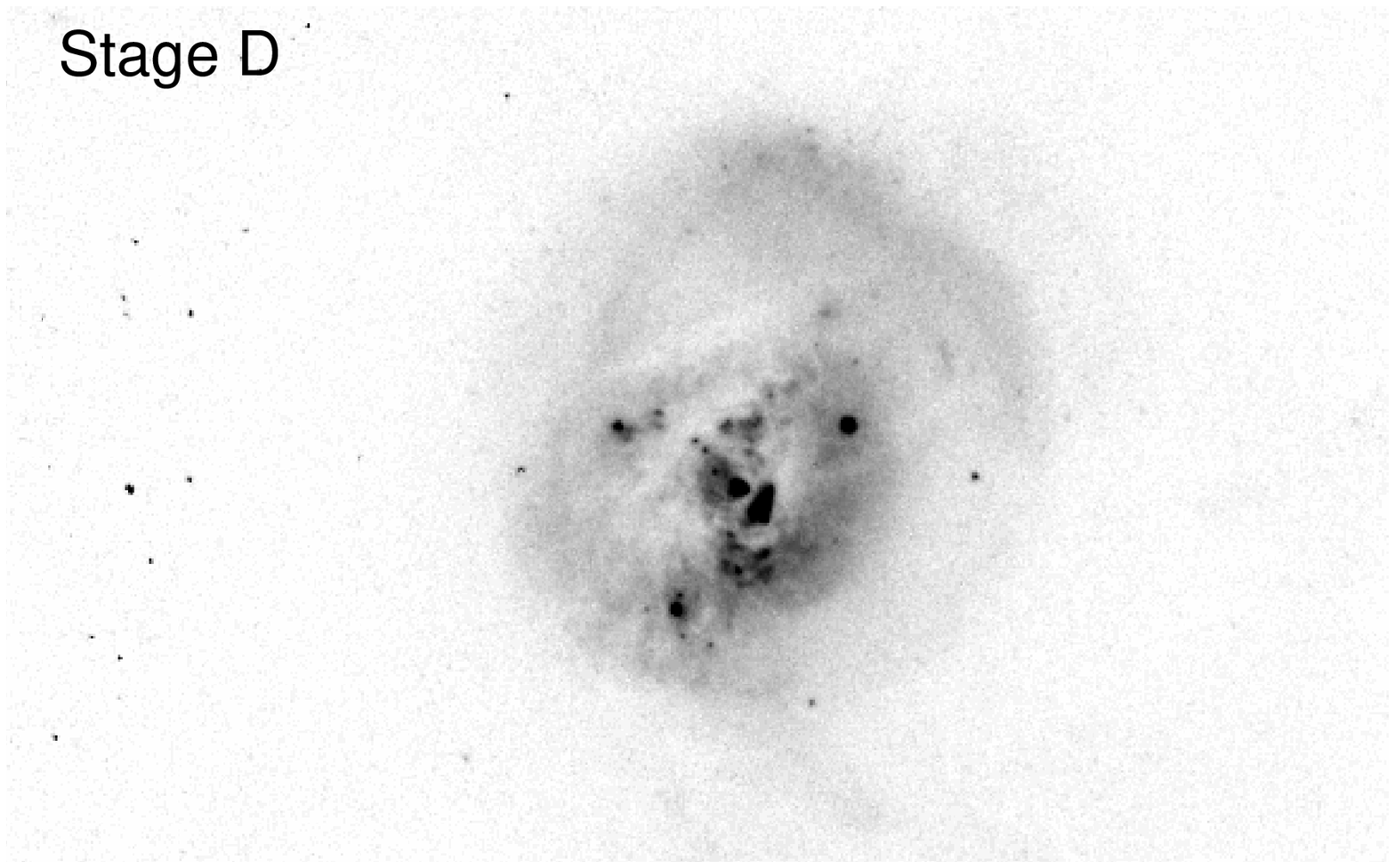}\end{minipage}
%
% %% caption
 \begin{minipage}{1\textwidth}
  \caption{Merger stages of U/LIRGs from the GOALS sample. \textit{Hubble Space Telescope} images of four systems in different merger stages: IRAS\,F11231+1456 (stage\,A, F435W), ESO\,069$-$IG006 (stage\,B, F435W), NGC\,6090 (stage\,C, F814N) and IRAS\,F23365+3604 (stage\,D, F435W). Objects in stage\,\,A and B are are \textit{early mergers} while those in stage\,\,C and D are \textit{late mergers} (see $\S$\ref{SM_sample}).}
\label{fig:mstages}
 \end{minipage}
\end{figure*}

Over the past twenty years, accumulated evidence shows that galaxies and the supermassive black holes (SMBHs) at their centres grow together \citep{Ferrarese:2000kq,Gebhardt:2000fj,Kormendy:2013uf} during a phase in which the SMBH is accreting matter and is observed as an Active Galactic Nucleus (AGN). 
It is however still uncertain how accretion onto SMBHs is triggered, since the gas and dust need to lose $\sim 99\%$ of their angular momentum before reaching the vicinity of the SMBH (e.g., \citealp{Shlosman:1990cs}). The merger of two or more galaxies has been long thought to be one of the main mechanisms capable of funnelling material from the kpc- to the pc-scale environment of SMBHs, while simultaneously triggering vigorous star formation \citep{Sanders:1988ys}. This has been predicted by analytical studies as well as by numerical simulations (e.g., \citealp{Di-Matteo:2005qv}), which indicate that tidal torques lead to the inflow of gas into the innermost regions \citep{Hernquist:1989rm}. In agreement with this idea, \cite{Treister:2012kk} found that the fraction of AGN in mergers increases with the AGN luminosity (see also \citealp{Bessiere:2012jk,Schawinski:2012yg,Hickox:2014vn,Glikman:2015lk}), and observations of optically-selected \citep{Ellison:2011yf}, IR-selected \citep{Satyapal:2014th} and X-ray selected \citep{Silverman:2011ek,Koss:2012dz} samples of pairs have shown that the fraction of dual AGN is larger in systems with close companions. Studies of radio-loud AGN have also shown that galaxy interactions might play a role in triggering AGN activity \citep{Ramos-Almeida:2011vl}, as well as in the formation of strong jets \citep{Chiaberge:2015sf}.

One of the main predictions of merger-triggered accretion is that during the final phases of a merger most of the accretion would happen when the SMBH is completely enshrouded by gas and dust (e.g., \citealp{Hopkins:2006gf}). Recent studies have also shown that there seems to be an increase in the amount of obscuring material in mergers of galaxies. Studying pairs of galaxies, \cite{Satyapal:2014th} found that there is an enhanced fraction of IR-selected AGN with respect to optically-selected AGN in advanced mergers, which suggests that a large number of objects are missed by optical classification because they are heavily obscured. 
\begin{table*}
\centering
\caption{Sample used for our work. The table shows (1) the IRAS name of the source, (2) the counterparts, (3) the redshift, (4) the merger stage and (5) the separation between the two nuclei in arcsec and (6) in kpc. In (5) and (6) we report ``S" for objects for which a single nucleus is observed.}\label{tab:sample}
\begin{center}
\begin{tabular}{llcccc}
\noalign{\smallskip}
\hline \hline \noalign{\smallskip}
\multicolumn{1}{c}{(1)}  & \multicolumn{1}{c}{(2)} & (3) & (4) & (5) & (6)  \\
\noalign{\smallskip}
{\it IRAS} name & \multicolumn{1}{c}{Source} & z  & M &  $D_{12}$ (\arcsec) & $D_{12}$ (kpc) \\
\noalign{\smallskip}
\hline \noalign{\smallskip}
\noalign{\smallskip}
F00085$-$1223  &   NGC\,34                               &  0.019617   & D     &  S     & S     \\      
\noalign{\smallskip}	
F00163$-$1039  &   Arp\,256      \& MCG$-$02$-$01$-$052                     & 0.027152       & B    &  56.1     & 33.1    \\    
\noalign{\smallskip}	
 F00506+7248 &   MCG+12$-$02$-$001           &  0.015698  & C    & 0.9      & 0.3  	  \\         
\noalign{\smallskip}	
F02069$-$1022  &   NGC\,833	\&  NGC\,835			& 0.012889  & A    &   55.9   & 15.7      \\  
\noalign{\smallskip}	
F05054+1718  &   CGCG 468$-$002E \& CGCG 468$-$002W				&  0.018193  & B     &  29.6    & 11.3     \\
\noalign{\smallskip}	
F05189$-$2524 &  IRAS\,05189$-$2524			& 0.042563   & D    &    S    & S   	   \\  
\noalign{\smallskip}	
F08572+3915  &   IRAS\,08572+3915 (NW \& SE)						&  0.058350  &  D    &    4.4    & 5.6  	  \\  
\noalign{\smallskip}	
F09320+6134  &   UGC\,05101							&  0.039367  & D     &   S     &  S  	  \\  
\noalign{\smallskip}	
F09333+4841  &   MCG+08$-$18$-$013   \& MCG+08$-$18$-$012 	&  0.025941 & A &   66.5    & 35.4   \\     
\noalign{\smallskip}	
 F10015$-$0614 &   NGC\,3110     \& MCG$-$01$-$26$-$013              	& 0.016858	 & A   &  108.9    &  37.7      \\             
\noalign{\smallskip}	
 F10257$-$4339  &   NGC\,3256					& 0.009354   & D    &   5.1   & 1.0   	 \\  
\noalign{\smallskip}	
F10565+2448  &   IRAS\,10565+2448						& 0.043100   &  D      &   7.4   &  6.7  	  \\  
\noalign{\smallskip}	
F11257+5850 &   Arp\,299	(NGC3690W \& NGC3690E)							& 0.010220   & C    &    21.3  & 4.6  	 \\
\noalign{\smallskip}	
 F12043$-$3140 &   ESO\,440$-$IG058N   \&  ESO\,440$-$IG058S                &  0.023413      & B   &   12.0    & 5.9     \\          
\noalign{\smallskip}	
  F12540+5708 &   Mrk\,231									&0.042170    & D      &   S    & S      \\   
\noalign{\smallskip}	
F12590+2934 &   NGC\,4922N   \&  NGC\,4922S                          &  0.023169   & C    &  21.2    & 10.8    	 \\      
\noalign{\smallskip}	
 13120$-$5453  &   IRAS\,13120$-$5453						&  0.030761  & D     &  S     & S     \\  
\noalign{\smallskip}	
F13197$-$1627	&	MCG$-$03$-$34$-$064 \& MCG$-$03$-$34$-$063										&	0.021328		& A	&   106.2   &  37.7	\\
\noalign{\smallskip}	
F13428+5608  &   Mrk\,273								& 0.037780    &  D    &    0.9  & 0.7       \\   
\noalign{\smallskip}	
F14378$-$3651  &   IRAS\,14378$-$3651						&  0.067637  & D       &    S   &  S      \\  
\noalign{\smallskip}	
F14544$-$4255  &   IC\,4518A	\& IC\,4518B					& 0.016261   & B    &   36.0   & 12.0     \\  
\noalign{\smallskip}	
F15327+2340  &   Arp\,220W \& Arp\,220E								& 0.018126   & D     &   1.0   & 0.4   	  \\   
\noalign{\smallskip}	
F16504+0228  &   NGC\,6240N \& NGC\,6240S											& 0.024480   &  D     &   1.4   & 0.7     	  \\	
\noalign{\smallskip}	
F16577+5900 &   NGC\,6286    \& NGC\,6285                          &  0.018349     &  B   &    91.1   & 35.8  	   \\    
\noalign{\smallskip}	
F17138$-$1017  &   IRAS\,F17138$-$1017          &  0.017335   & D      &    S   & S       \\     
\noalign{\smallskip}	
20264+2533  &   MCG $+$04$-$48$-$002 \& NGC\,6921					& 0.013900   & A   &   91.4   & 27.1       \\  
\noalign{\smallskip}	
F21453$-$3511  &   NGC\,7130								&0.016151   & D     &   S    & S    \\
\noalign{\smallskip}	
F23007+0836  &   NGC\,7469	\& IC\,5283				& 0.016317   &  A    & 79.7     & 26.8   	  \\
\noalign{\smallskip}	
F23254+0830  &   NGC\,7674	\& NGC\,7674A								& 0.028924  &  A	   &  34.1    & 20.7  	  \\
\noalign{\smallskip}	
 23262+0314 &   NGC\,7679	\& NGC\,7682				& 0.017139  & A    &   269.7   & 97.3       \\
\noalign{\smallskip}
\hline
\noalign{\smallskip}
\end{tabular}
\end{center}
\end{table*} 
Studying galaxies at $z\sim 1$, \cite{Kocevski:2015vh} found an increase in the fraction of galaxies undergoing mergers, or with interactions signatures, in AGN with $N_{\rm\,H}\geq 3\times 10^{23}\rm\,cm^{-2}$ with respect to unobscured AGN. Tentative evidence of an excess of merging systems in AGN with $N_{\rm\,H}\geq 2\times 10^{23}\rm\,cm^{-2}$ was found by \cite{Del-Moro:2016kq} studying mid-IR quasars at $z\sim2$. Analysing the X-ray spectra of heavily obscured AGN in the COSMOS field, \cite{Lanzuisi:2015uo} found evidence of an increase in the fraction of mergers in CT sources with respect to X-ray selected samples. \cite{Koss:2016kq} also discussed a possible increase in the fraction of advanced mergers in Compton-thick (CT, $N_{\rm\,H}\geq 10^{24}\rm\,cm^{-2}$) AGN, finding that two out of the nine CT AGN selected by using the spectral curvature approach are in close mergers ($D_{12}<10$\,kpc). This fraction ($26^{+14}_{-12}\%$)\footnote{The value was calculated following \cite{Cameron:2011cl}, and the uncertainties represent the 16th and 84th quantiles of a binomial distribution.} is higher than that typically found for {\it Swift}/BAT AGN ($8\pm2\%$; \citealp{Koss:2010rc}).

\begin{table*}
\begin{center}
%\footnotesize
\caption[]{ The table reports (1) the IRAS name, (2) the counterparts, (3) the values of the column density in the direction of the AGN reported in the literature, and (4) the reference for $N_{\rm\,H}$.}
\label{tab:results_other}
\begin{tabular}{llcc}
\hline \hline \noalign{\smallskip}
 \noalign{\smallskip}
\multicolumn{1}{c}{(1)} & (2) & (3) & (4)  \\
\noalign{\smallskip}
{\it IRAS} name & Source &  $N_{\rm\,H}$ ($10^{22}\rm\,cm^{-2}$)& Reference  \\
\noalign{\smallskip}
\hline \noalign{\smallskip}
\noalign{\smallskip}
 F02069$-$1022 & 	NGC\,833					&		$28\pm3$			&	\cite{Oda:2017aa}		\\
\noalign{\smallskip}
			  & 	NGC\,835$^a$				&		$55^{+15}_{-10}$			&		\cite{Oda:2017aa}	\\
\noalign{\smallskip}
			 & 	NGC\,835$^b$				&		$30\pm2$			&	\cite{Oda:2017aa}		\\
\noalign{\smallskip}
\noalign{\smallskip}
F05189$-$2524  &  IRAS\,05189$-$2524		&  $12.7^{+1.0}_{-0.7}$     & \cite{Teng:2015vn}  \\ 
\noalign{\smallskip}
\noalign{\smallskip}
 F08572+3915 &  IRAS\,08572+3915	&  -- 	    & \cite{Teng:2015vn}  \\ 
\noalign{\smallskip}
\noalign{\smallskip}
F09320+6134  &  UGC\,5101	&  $132^{+32}_{-37}$     & \cite{Oda:2016aa}  \\  
\noalign{\smallskip}
\noalign{\smallskip}
F10257$-$4339  &  NGC\,3256	& --	 	& \cite{Lehmer:2015ys} \\
\noalign{\smallskip}
\noalign{\smallskip}
F10565+2448  &  IRAS\,10565+2448	&  -- 	   & \cite{Teng:2015vn}  \\  
\noalign{\smallskip}
\noalign{\smallskip}
 F11257+5850 &  NGC\,3690W		 &  $350_{-20}^{NC}$     & \cite{Ptak:2015nx}  \\ 
\noalign{\smallskip}
\noalign{\smallskip}
F12540+5708  &  Mrk\,231	 & $14.5^{+3.1}_{-2.4}$   &         \cite{Teng:2014oq} \\ 
\noalign{\smallskip}
\noalign{\smallskip}
13120$-$5453  &  IRAS\,13120$-$5453		&  $316^{+233}_{-129}$       & \cite{Teng:2015vn}	\\ 
\noalign{\smallskip}
\noalign{\smallskip}
F13428+5608  &  Mrk\,273				 &  $44^{+10}_{-6}$     & \cite{Teng:2015vn}  \\
\noalign{\smallskip}
\noalign{\smallskip}
F14378$-$3651  &  IRAS\,14378$-$3651	 &  -- 	& 	\cite{Teng:2015vn}  \\ 
\noalign{\smallskip}
\noalign{\smallskip}
F15327+2340  &  Arp\,220W	 	 &  $\geq 530$       & \cite{Teng:2015vn} 	\\ 
\noalign{\smallskip}
\noalign{\smallskip}
F16504+0228  &  NGC\,6240N	& $155^{+72}_{-23}$	&    \cite{Puccetti:2016cj}	 \\
\noalign{\smallskip}
			  &  NGC\,6240S	& $147^{+21}_{-17}$	& 	\cite{Puccetti:2016cj}  \\
\noalign{\smallskip}
\noalign{\smallskip}
F16577+5900  &  NGC\,6286	&  $111^{+109}_{-41}$ 	   &  \cite{Ricci:2016zr} \\  
\noalign{\smallskip}
\noalign{\smallskip}
F23254+0830  &  NGC\,7674	&  $\gtrsim 300$ 	   & \cite{Gandhi:2016bq}  \\  
%
%6
%
\hline
\noalign{\smallskip}
\multicolumn{4}{l}{{\bf Notes.} NC: value not constrained; $^a$ {\it XMM-Newton} observation; $^b$ {\it NuSTAR} and {\it Chandra} observation.}
\end{tabular}
\end{center}
\end{table*}

 With the aim of understanding how galaxy mergers affect the environment of AGN, we have studied a sample of 30 luminous\footnote{$L_{\rm\,IR}(8-1000\,\mu\rm m)=10^{11}-10^{12}$ $L_{\odot}$} and ultra-luminous\footnote{$L_{\rm\,IR}(8-1000\,\mu\rm m)>10^{12}$ $L_{\odot}$} IR galaxies (U/LIRGs) in different merger stages (for a total of 52 individual galactic nuclei) at hard X-ray energies ($\geq 10$\,keV). Hard X-rays are one of the best energy bands to detect and characterise the most obscured and possibly CT AGN, due to the lower opacity of the obscuring material with respect to softer X-ray energies.  The recent launch of the NASA mission {\it NuSTAR}  \citep{Harrison:2013lq}, the first focussing hard X-ray (3--79\,keV) telescope in orbit, has allowed the study of even some of the most obscured and elusive growth episodes of AGN (e.g., \citealp{Annuar:2015ve,Balokovic:2014dq,Boorman:2016fj,Brightman:2015cr,Gandhi:2014ys,Lansbury:2014cl,Lansbury:2015vn,Ricci:2016kq,Ricci:2016zr}). The objects in our sample were selected from the Great Observatories All-sky LIRG Survey (GOALS, \citealp{Armus:2009db}). GOALS is a local ($z<0.088$) galaxy sample selected from the {\it Infrared Astronomical Satellite} ({\it IRAS}) revised bright Galaxy Survey \citep{Sanders:2003rw}. The sample has a rich collection of ancillary data, from the radio to the X-rays (e.g., \citealp{Inami:2010pf,Iwasawa:2011fk,Petric:2011zt,Romero-Canizales:2012fv,Diaz-Santos:2013lg,Stierwalt:2013eu,Privon:2015qp,Privon:2016mz}), and the {\it IRAS} infrared selection guarantees that the sample is not biased by obscuration.

In this work we report the results of broad-band X-ray spectroscopy for all the U/LIRGs of the sample by combining {\it NuSTAR} and {\it Swift}/BAT data with archival {\it Chandra}, {\it XMM-Newton} and {\it Swift}/XRT data. From X-ray spectroscopy and multi-wavelength properties, clear evidence of AGN emission is found in 25 nuclei, of which 13 are early mergers and 12 are late mergers. The X-ray spectra of these AGN were self-consistently modelled to take into account both absorption (including Compton scattering and photoelectric absorption) and reprocessing of the X-ray radiation from the circumnuclear material. This approach provides strong constraints on the line-of-sight column density. The paper is structured as follows. In $\S$\ref{SM_sample} we describe the sample of galaxies used and the different merger stages, in $\S$\ref{SM_datared} we report details about the X-ray data used and the data reduction, and in $\S$\ref{SM_spectral} we discuss the spectral fitting procedure adopted. The observed increase in the fraction of CT AGN along the merger sequence (with advanced mergers being typically more obscured than isolated galaxies) is discussed in $\S$\ref{sect:discussion}, while in $\S$\ref{sect:summary} we summarise our results and present our conclusions. In a forthcoming paper (Ricci et al. in prep.) we will discuss the relation between the multi-wavelength properties of the U/LIRGs of our sample and the bolometric luminosity of the AGN. Throughout the paper we adopt standard cosmological parameters ($H_{0}=70\rm\,km\,s^{-1}\,Mpc^{-1}$, $\Omega_{\mathrm{m}}=0.3$, $\Omega_{\Lambda}=0.7$). Unless otherwise stated, all uncertainties are quoted at the 90\% confidence level.

\begin{figure}
 \centering
 \begin{minipage}{.48\textwidth}
 \centering
\includegraphics[width=9cm]{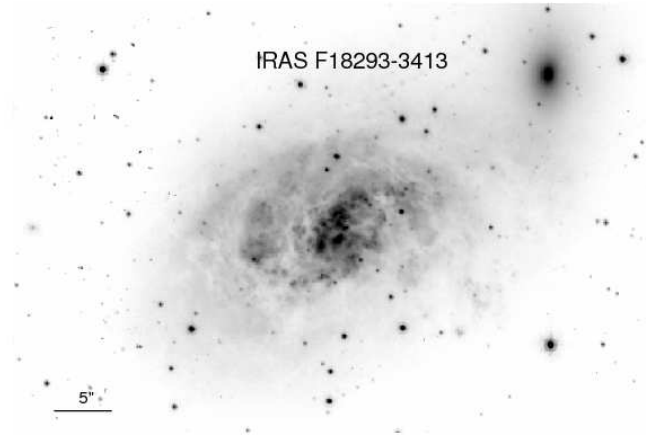}\end{minipage}
  \begin{minipage}{.48\textwidth}
   \caption{{\it HST} image of IRAS\,F18293$-$3413 ({\it HST} F814W; $5\arcsec\simeq 1.8$\,kpc).}
 \label{fig:image_18293}
  \end{minipage}
 \end{figure}

\section{Sample and merger stages}
\label{SM_sample}

The GOALS sample contains 180 LIRGs and 22 ULIRGs selected from the {\it IRAS} Revised Bright Galaxy Sample \citep{Sanders:2003rw}, which is a complete sample of extragalactic objects with 60$\mu$m flux $>5.24$\,Jy. The galaxies in GOALS represent a large, statistically complete sample of U/LIRGs. Objects in GOALS have been observed by {\it Spitzer} and {\it Herschel} \citep{Inami:2010pf,Petric:2011zt,Diaz-Santos:2011tg,Inami:2013il,Diaz-Santos:2013lg,Stierwalt:2013eu,Diaz-Santos:2014kl,Stierwalt:2014dg}. 
Our sample contains all GOALS sources observed by {\it NuSTAR} and for which data were available as of March 2016. This includes:  i) sources located at $<120$\,Mpc, and with $\log (L_{\rm\,IR}/L_{\odot}) > 11.3$, amongst which ten LIRGs in different merger stages that were awarded to our group during AO--1 (for a total of 200 ks, PI: F. E. Bauer); ii) the nearest ($z<0.078$) ULIRGs from \citet{Sanders:2003rw} (see \citealp{Teng:2015vn} for details); iii) sources detected by {\it Swift}/BAT and observed as a part of the campaign aimed at characterising local hard X-ray selected sources with {\it NuSTAR}.
We excluded the four U/LIRGs classified as non-mergers ({\it N} stage) by \citet{Stierwalt:2013eu} since we are interested in studying the evolution of obscuration along the merger sequence. We also included three systems (IRAS\,F23007+0836, IRAS\,23262+0314, IRAS\,F13197$-$1627) that were detected at hard X-rays by {\it Swift}/BAT \citep{Baumgartner:2013ek} but have not been observed by {\it NuSTAR}, to avoid any possible bias in the {\it NuSTAR} selection of BAT sources.

We divided GOALS galaxies into four subsamples, based on the merger stage determined by \citet{Stierwalt:2013eu} and \citet{Haan:2011fk} from visual inspection of {\it Spitzer} IRAC 3.6$\mu$m and {\it Hubble Space Telescope} images (see Figure\,\ref{fig:mstages} and Table\,\ref{tab:sample}). The merger sequence we considered is the following:
\smallskip

\noindent {\bf Stage\,\,A} are systems in pre-merger, i.e. galaxy pairs prior to a first encounter.
\smallskip

\noindent {\bf Stage\,\,B} are objects in the initial phases of merger, i.e. galaxies after the first-encounter, with galaxy disks still symmetric and showing signs of tidal tails.
\smallskip

\noindent {\bf Stage\,\,C} are systems in the mid-stage of merger, showing amorphous disks, tidal tails, and other signs of merger activity.
\smallskip

\noindent {\bf Stage\,\,D} are sources in the final merger stages, showing the two nuclei in a common envelope or only a single nucleus.
\smallskip

 \begin{figure}
 \centering
 \begin{minipage}{.48\textwidth}
 \centering
\includegraphics[width=9cm]{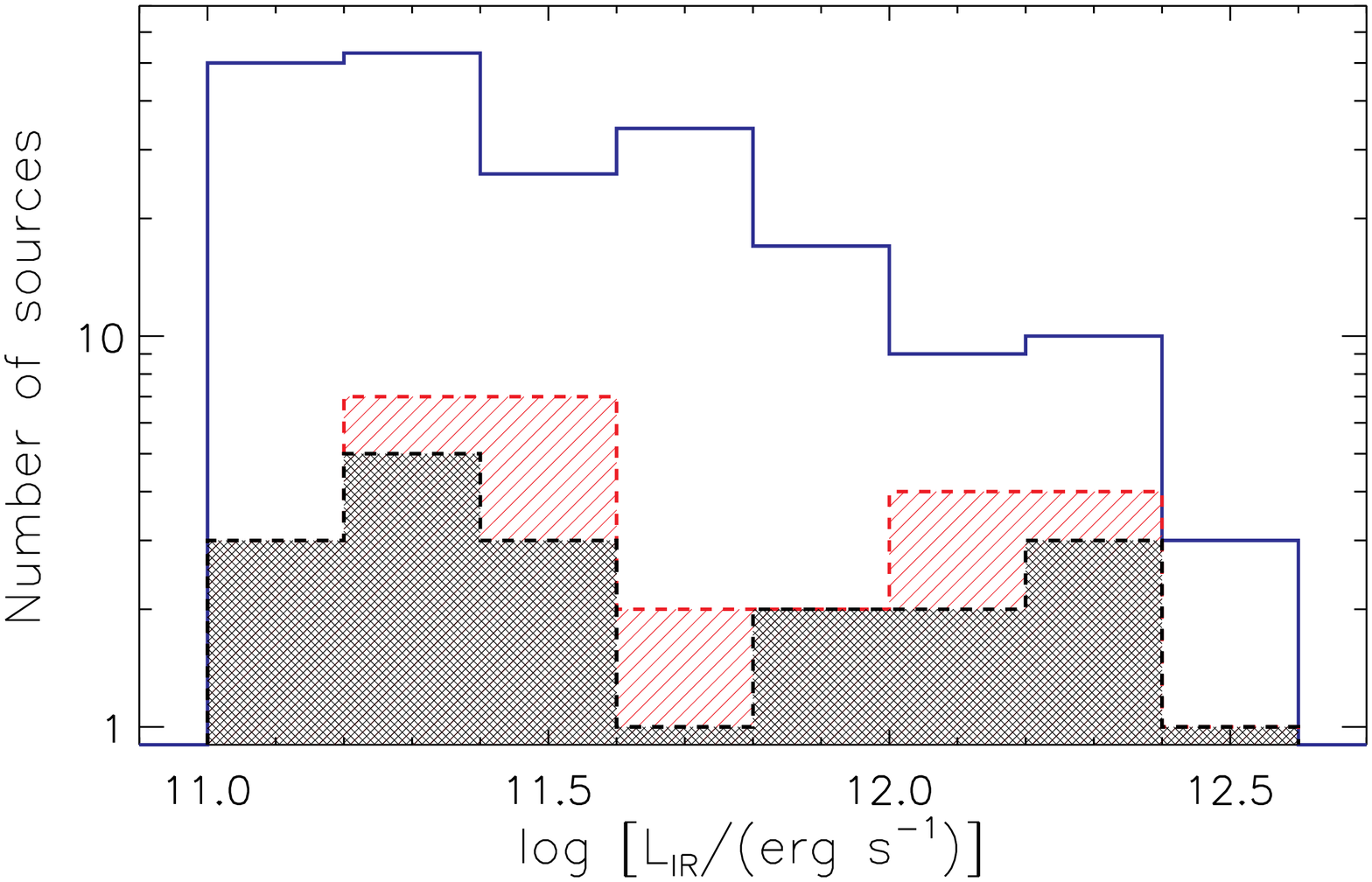}\end{minipage}
\par\smallskip
 \begin{minipage}{.48\textwidth}
 \centering
\includegraphics[width=9cm]{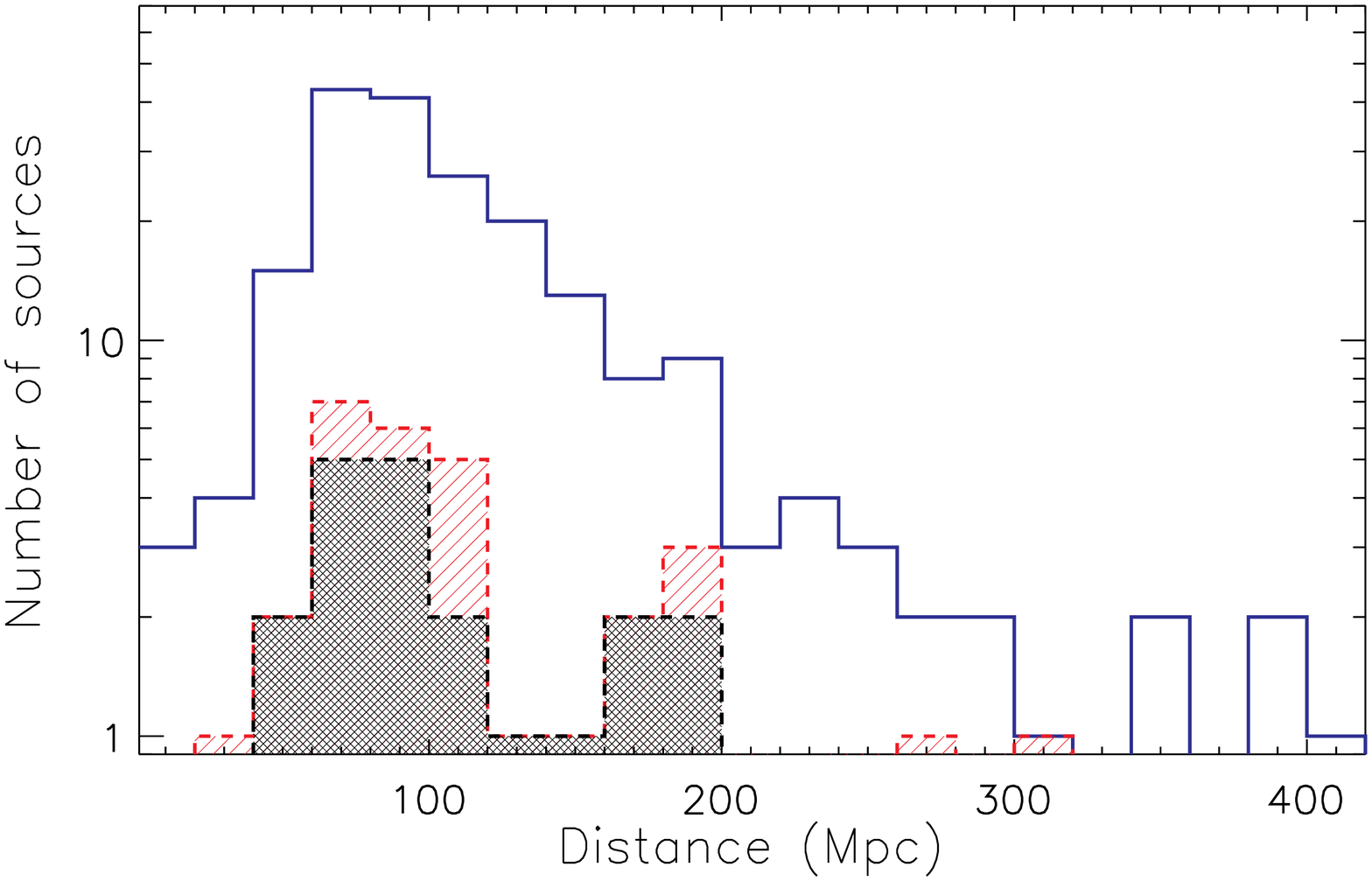}\end{minipage}

 %% caption
  \begin{minipage}{.48\textwidth}
   \caption{{\it Top panel:} Histogram of the 8--1000\,$\mu$m luminosity ($L_{\rm\,IR}$) of all the systems in the GOALS sample (blue empty histogram), of the GOALS objects in our sample (red histogram), and of the AGN in our sample (black filled histogram). {\it Bottom panel:} same as top panel for the distances.}
 \label{fig:HistogramLIR}
  \end{minipage}
 \end{figure}

\begin{table*}
\centering
\caption{X-ray observations log. The columns report (1) the IRAS name of the source, (2) the counterparts, (3) the X-ray facility, (4) the observation ID, and (5) the exposure time. For {\it XMM-Newton} the exposures listed are those of EPIC/PN, MOS1 and MOS2, while for {\it NuSTAR} those of FPMA and FPMB.}\label{tab:obslog}
\begin{center}
\begin{tabular}{llccc}
\noalign{\smallskip}
\hline \hline \noalign{\smallskip}
\multicolumn{1}{c}{(1)} & \multicolumn{1}{l}{(2)} & (3) & (4) & (5) \\
\noalign{\smallskip}
{\it IRAS} name & Source  & Facility & Observation ID &Exposure [ks] \\
\noalign{\smallskip}
\hline \noalign{\smallskip}
F00085$-$1223 &NGC\,34                               & {\it NuSTAR}    & 60101068002   & 21.4/21.4 \\
 &					   & {\it Chandra}     &  15061  & 14.8 \\
 &					   & {\it XMM-Newton}     &  0150480501  & 12.0/16.7/17.2 \\
\noalign{\smallskip}
\noalign{\smallskip}
F00163$-$1039 & Arp\,256   \& MCG$-$02$-$01$-$052                           & {\it NuSTAR}      & 60101069002   & 20.5/20.5 \\
 &					  & {\it Chandra}  	   &  13823  & 29.6  \\
\noalign{\smallskip}
 \noalign{\smallskip}
F00506+7248 & MCG+12$-$02$-$001                          & {\it NuSTAR}     & 60101070002   & 25.5/25.5  \\
 & 						          & {\it Chandra}   	  & 15062   & 14.3  \\
\noalign{\smallskip}
\noalign{\smallskip}
F05054+1718 & CGCG 468$-$002E \& CGCG 468$-$002W		& {\it NuSTAR}  	  &  60006011002  & 15.5/15.5 \\
 &				& {\it Swift}/XRT   & 49706   & 72.1  \\
\noalign{\smallskip}
\noalign{\smallskip}
F09333+4841 & MCG+08$-$18$-$013   \& MCG+08$-$18$-$012     & {\it NuSTAR}       &   60101071002 & 19.1/19.1  \\  
  &      					   & {\it Chandra}      & 15067   & 13.8 \\
\noalign{\smallskip}
\noalign{\smallskip}
F10015$-$0614 & NGC\,3110    \& MCG$-$01$-$26$-$013                          & {\it NuSTAR}      & 60101072002   & 17.9/17.8 \\ 
 &					     & {\it Chandra}  	   & 15069   &  14.9 \\
 &					    & {\it XMM-Newton}    &  0550460201  & 11.0/17.8/18.0 \\
\noalign{\smallskip}
\noalign{\smallskip}
F12043$-$3140 & ESO\,440$-$IG058N \& ESO\,440$-$IG058S                          &  {\it NuSTAR}     & 60101073002   & 28.6/28.6 \\
 &						   & {\it Chandra}   &  15064   & 14.8  \\
\noalign{\smallskip}
\noalign{\smallskip}
F12590+2934 & NGC\,4922N \&   NGC\,4922S                             & {\it NuSTAR}     &  60101074002 &  20.7/20.6 \\
 &					      & {\it Chandra}     & 15065    &  14.9 \\
\noalign{\smallskip}
\noalign{\smallskip}
F13197$-$1627 & MCG$-$03$-$34$-$064 \& MCG$-$03$-$34$-$063	 & {\it XMM-Newton}    &  0506340101  & 66.5/85.4/85.9 \\
\noalign{\smallskip}
\noalign{\smallskip}
F14544$-$4255 & IC 4518A \& IC 4518B 		& {\it NuSTAR}    &  60061260002  & 7.8/7.8  \\
 &					   	& {\it XMM-Newton}     &  0406410101  & 21.2/24.4  \\
\noalign{\smallskip}
\noalign{\smallskip}
F16577+5900 & NGC\,6286  \& NGC\,6285                            & {\it NuSTAR}  	 & 60101075002   & 17.5/17.5 \\
 &					 & {\it Chandra}   &  10566      & 14.0 \\
\noalign{\smallskip}
\noalign{\smallskip}
F17138$-$1017 & IRAS\,F17138$-$1017                      & {\it NuSTAR}    &  60101076002 & 25.9/25.9 \\
 & 					   & {\it Chandra}     &  15063  & 14.8 \\
\noalign{\smallskip}
\noalign{\smallskip}
20264+2533 & MCG +04$-$48$-$002 \&  NGC\,6921						   & {\it NuSTAR}   &   60061300002 & 19.5/19.5  \\
 &					   								& {\it XMM-Newton}  	    &  0312192301   & 6.5/12.4/12.7 \\
\noalign{\smallskip}
\noalign{\smallskip}
F21453$-$3511 & NGC 7130							   & {\it NuSTAR}   &  60061347002  & 21.2/21.2  \\
 &					   				& {\it Chandra}     &   2188 &  38.6 \\
\noalign{\smallskip}
\noalign{\smallskip}
F23007+0836 & NGC\,7469	\& IC\,5283				 	  &  {\it XMM-Newton}   &  0207090101   &  59.2\\
\noalign{\smallskip}
\noalign{\smallskip}
23262+0314 & NGC\,7679 \& NGC\,7682			   & {\it XMM-Newton} &    0301150501     & 13.1/19.3/19.1 \\
\noalign{\smallskip}
\hline
\end{tabular}
\end{center}
\end{table*}

\noindent Objects in stage\,\,A and B can be considered {\it early mergers}, and those in stage\,\,C and D {\it late mergers}. We used the merger classifications provided by \citet{Stierwalt:2013eu}, with the exception of the following three systems.
\begin{itemize}
\item IRAS\,F14544$-$4255 (IC\,4518A and IC\,4518B) is in the GOALS sample but, while {\it Spitzer} imaging is available, this system was not classified by \citet{Stierwalt:2013eu}. Optical and IR images show that the two galaxies are clearly separated (36\,arcsec or 12.0\,kpc) and show signs of tidal interactions (see Appendix\,\ref{sect:F14544-4255}). They were therefore classified as being in stage\,\,B.
\item IRAS\,F18293$-$3413 is classified as being in stage\,\,C by \citet{Stierwalt:2013eu}. {\it Hubble Space Telescope} ({\it HST}) imaging (Fig.\,\ref{fig:image_18293}) shows, however, that the two systems do not share a common envelope, and the source was classified as being in a pre-merger stage by \citet{Haan:2011fk}. Given the much smaller size of the companion object, we reclassify this object as stage\,\,N.
\item IRAS\,F21453$-$3511 (NGC\,7130), which was initially classified as a non-merger has been recently shown to present post-merger features by \citet{Davies:2014dz}. We therefore reclassify this source as stage\,\,D with a single nucleus.
\end{itemize}

Overall our sample consists of 30 systems, of which eight show a single nucleus and the remaining 22 show two nuclei, for a total of 52 galactic nuclei. Our sample is a representative subsample of GOALS galaxies. The histogram of the 8-1000\,$\mu$m luminosities ($L_{\rm\,IR}$) of all objects in the GOALS sample and of the GOALS galaxies in our sample are showed in the top panel of Fig.\,\ref{fig:HistogramLIR} (blue and red lines, respectively). The two samples have similar luminosities, and performing a Kolmogorov-Smirnov (KS) test we find a p-value of 0.13, which implies that our sample does not have a distribution of $L_{\rm\,IR}$ significantly different from that of the GOALS sample. Similarly, the distribution of distances of our sample does not differ from that of GOALS galaxies (bottom panel of Fig.\,\ref{fig:HistogramLIR}), and the KS test results in a p-value of 0.91. The two panels of Fig.\,\ref{fig:HistogramLIR} also show the distributions of the AGN in our sample (black lines). A KS test between the luminosity (distance) distributions of AGN and of all the galaxies in the GOALS sample shows that the two samples are not significantly different, resulting in a p-value of 0.16 (0.60).

\begin{figure*}
\centering
\begin{minipage}{.48\textwidth}
\centering
\textbf{IRAS\,F00085$-$1223}\par\smallskip
\includegraphics[width=9cm]{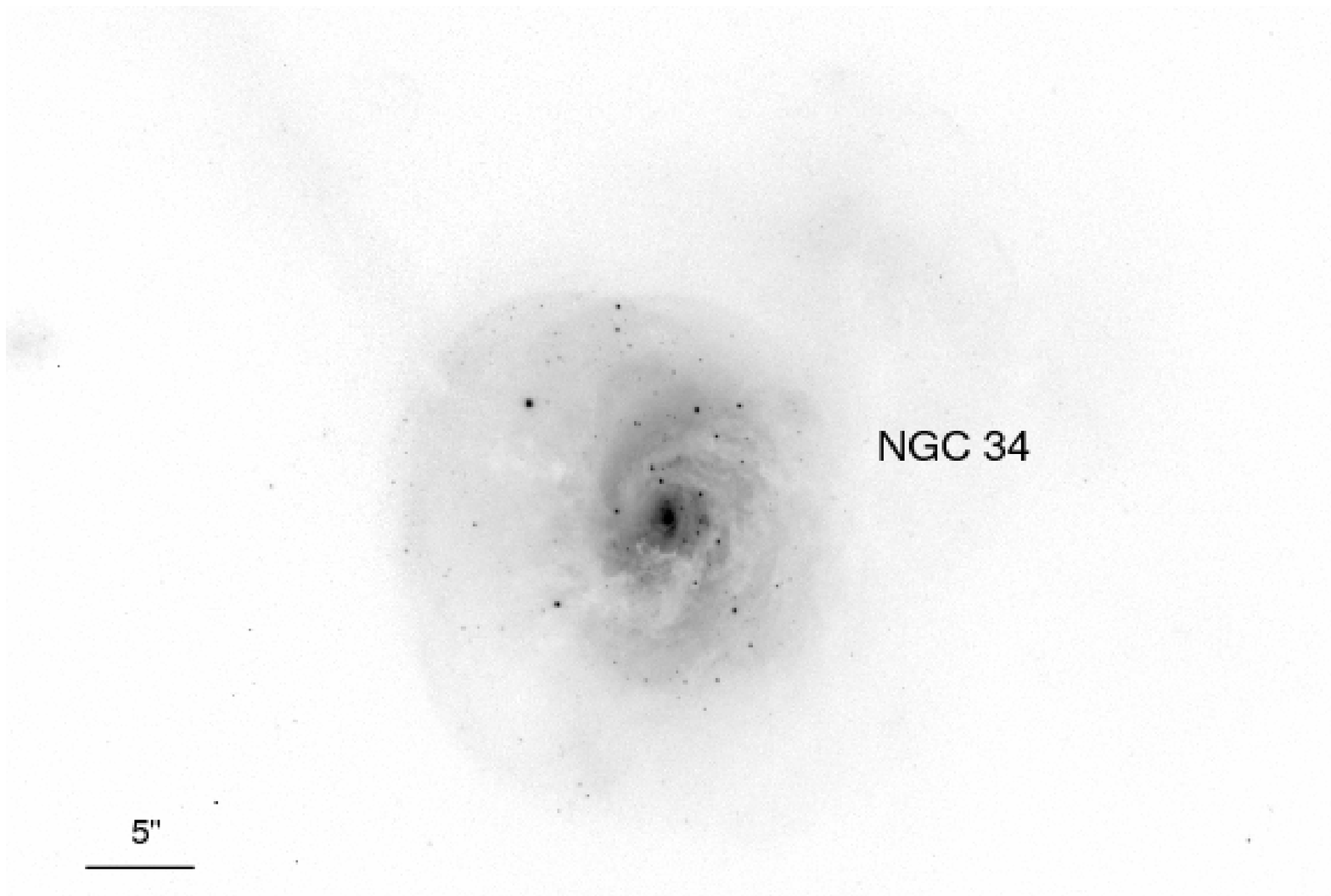}\end{minipage}
\begin{minipage}{.48\textwidth}
\centering
\textbf{IRAS\,F00163$-$1039}\par\smallskip
\includegraphics[width=9cm]{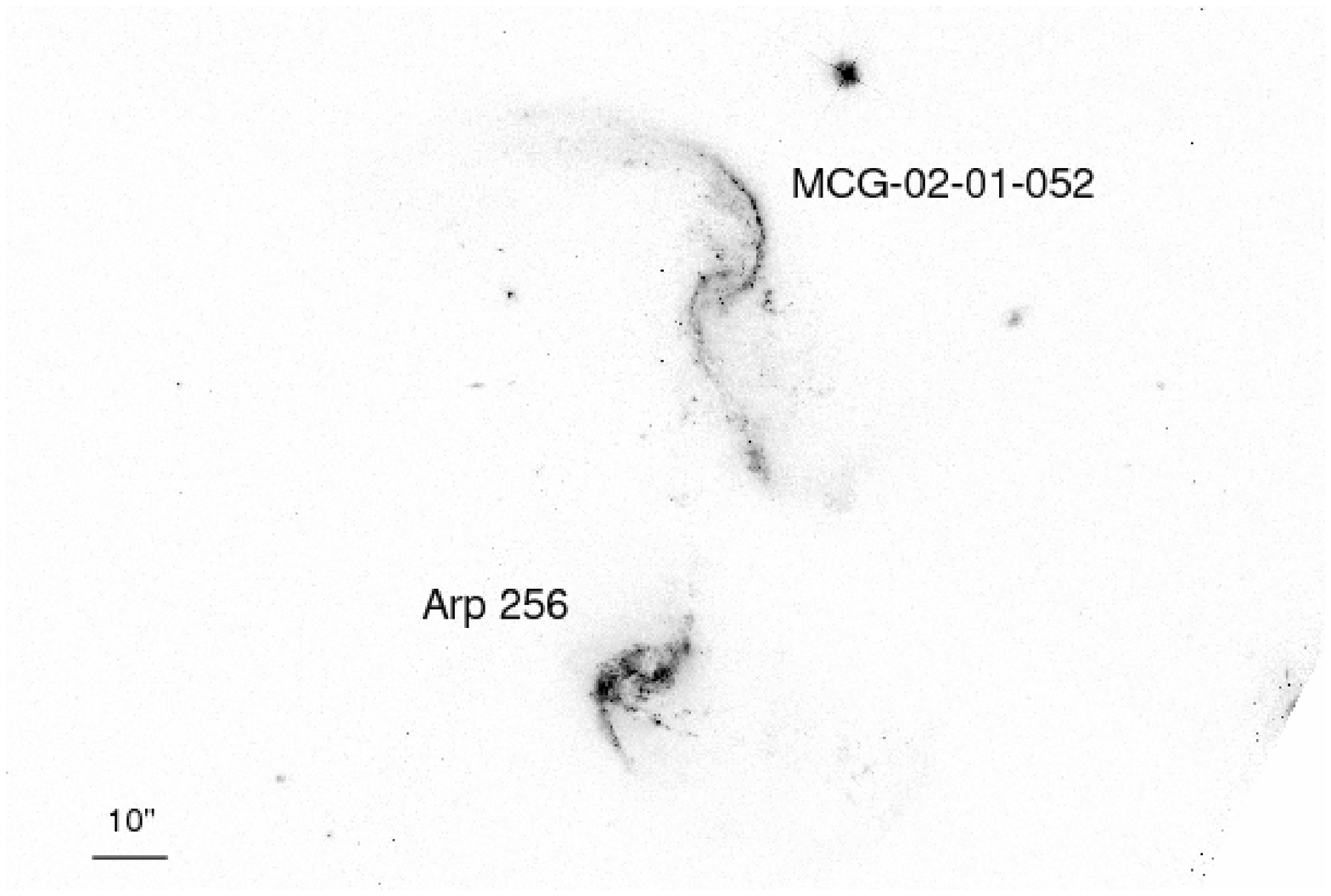}\end{minipage}
\par\smallskip
\par\smallskip
\par\smallskip
\par\smallskip
\par\smallskip
\begin{minipage}{.48\textwidth}
\centering
\textbf{IRAS\,F00506+7248}\par\smallskip
\includegraphics[width=9cm]{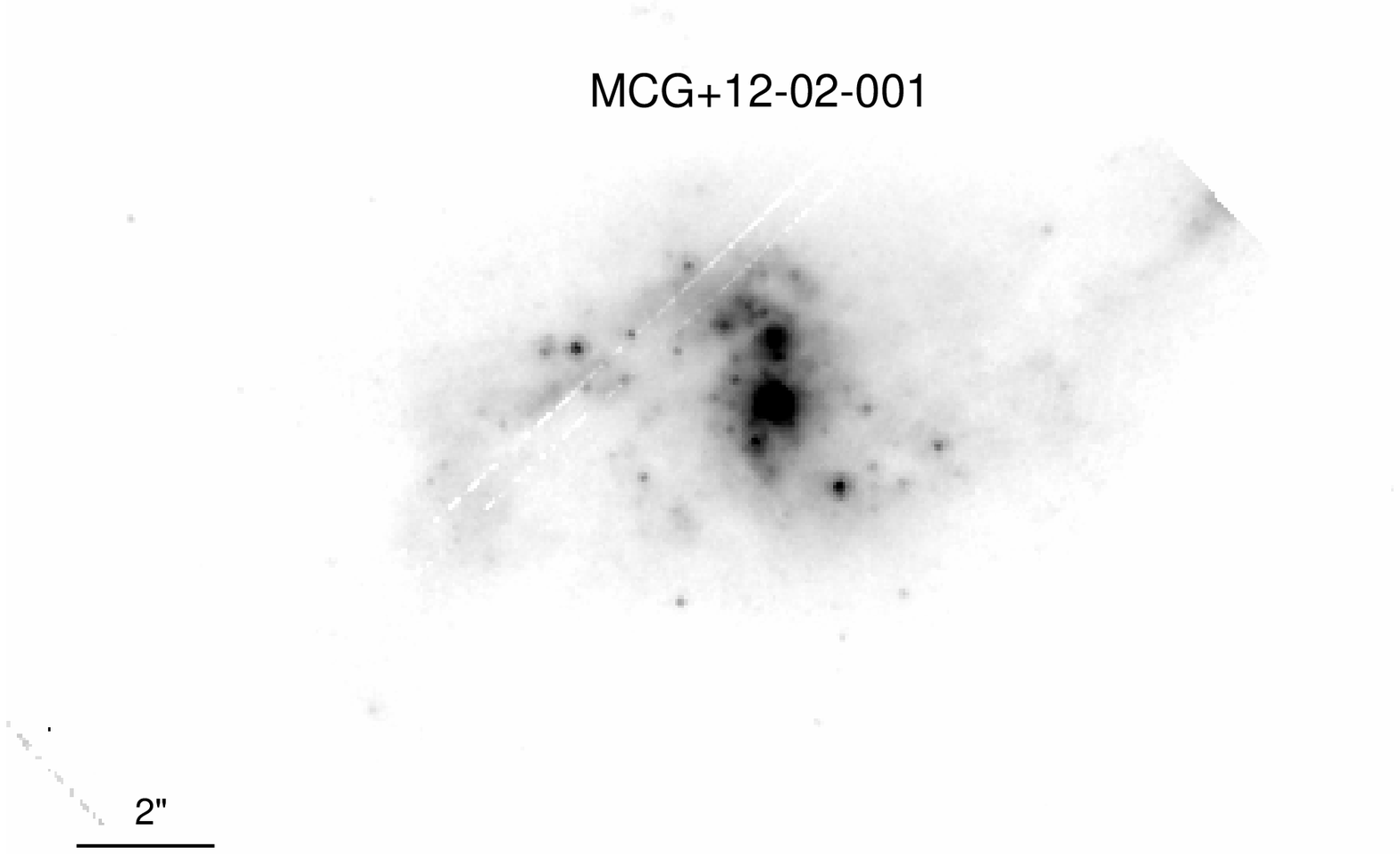}\end{minipage}
\begin{minipage}{.48\textwidth}
\centering
\textbf{IRAS\,F02069$-$1022}\par\smallskip
\includegraphics[width=9cm]{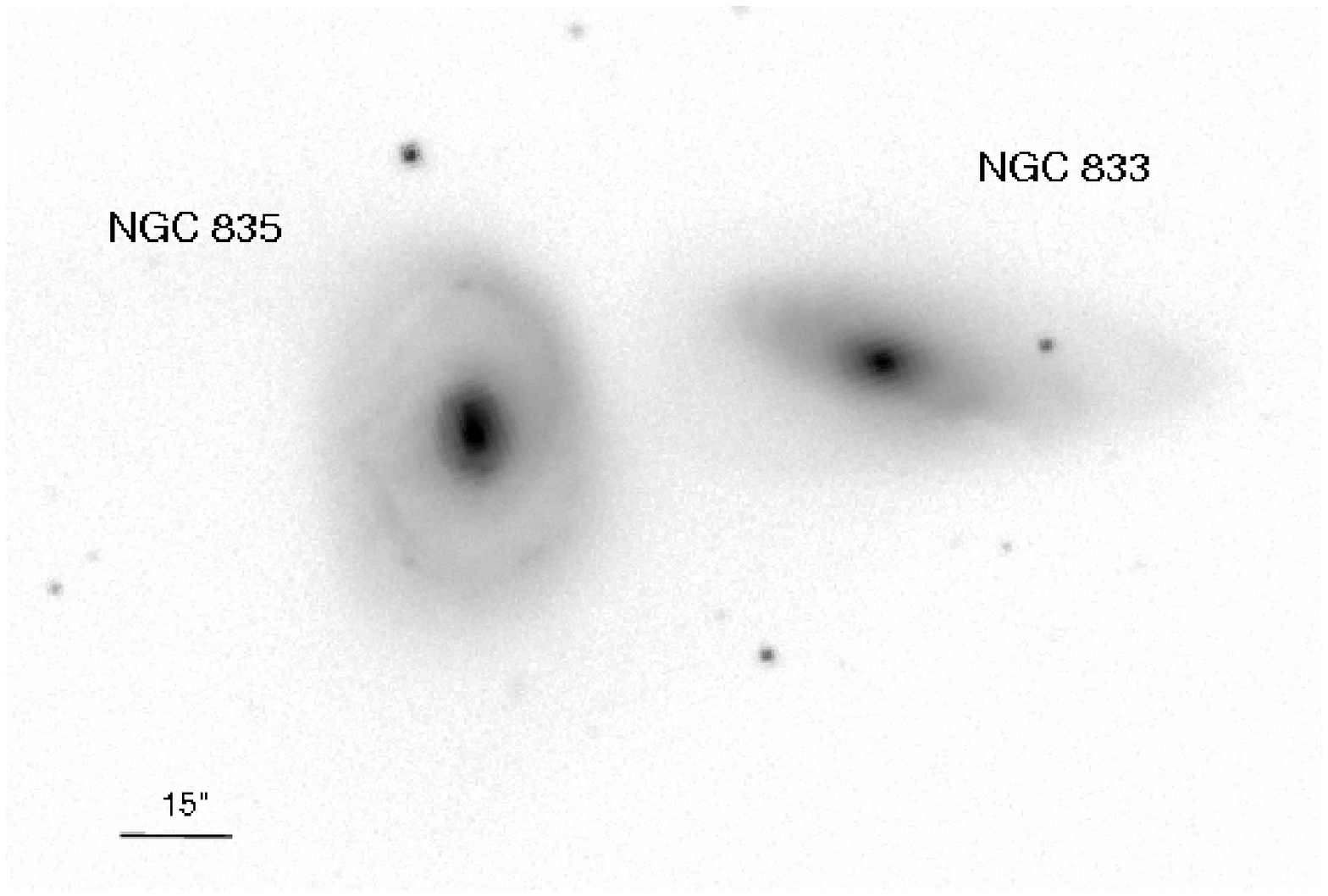}\end{minipage}
\par\smallskip
\par\smallskip
\par\smallskip
\par\smallskip
\par\smallskip
\begin{minipage}{.48\textwidth}
\centering
\textbf{IRAS\,F05054+1718}\par\smallskip
\includegraphics[width=9cm]{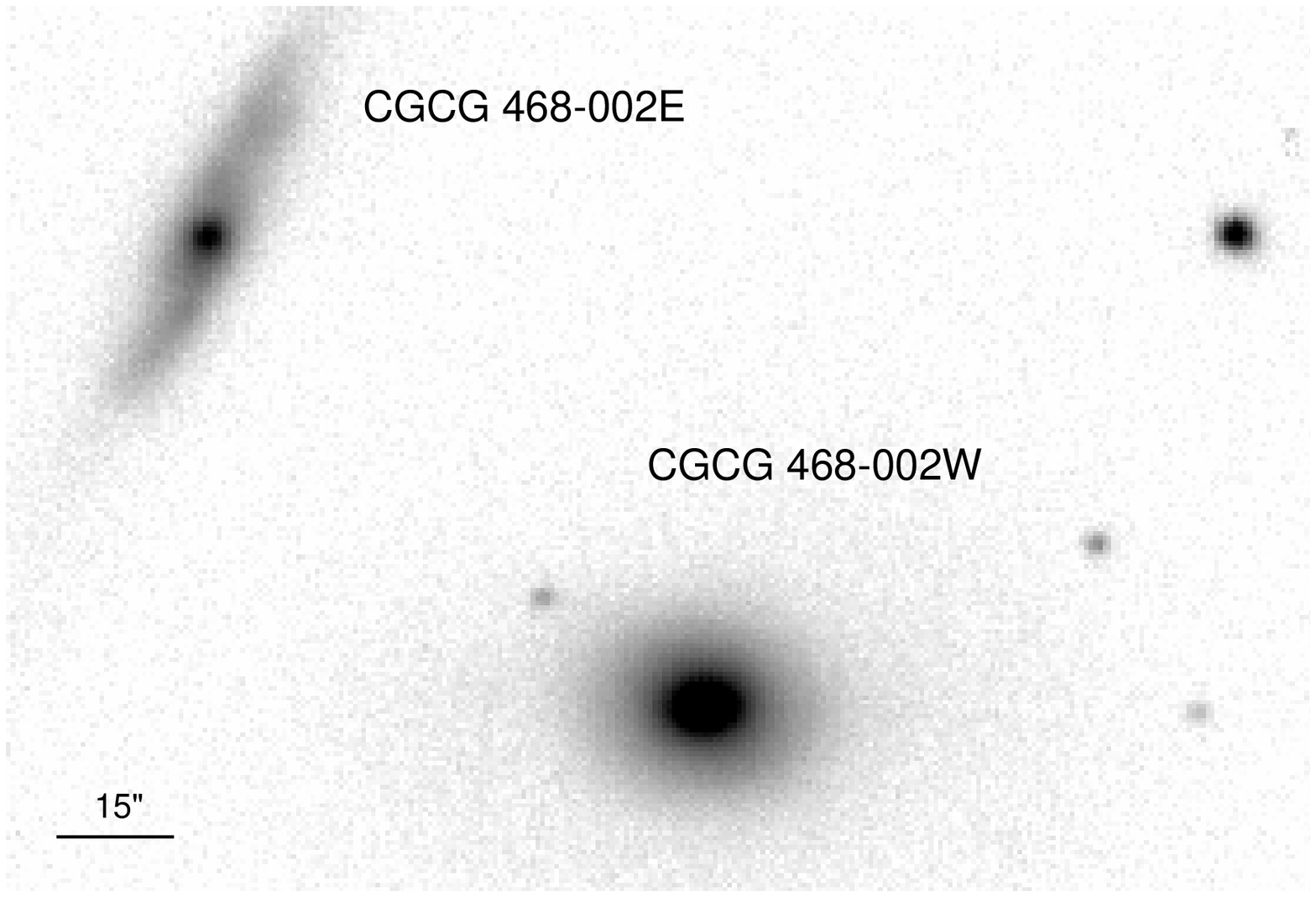}\end{minipage}
\begin{minipage}{.48\textwidth}
\centering
\textbf{IRAS\,F05189$-$2524}\par\smallskip
\includegraphics[width=9cm]{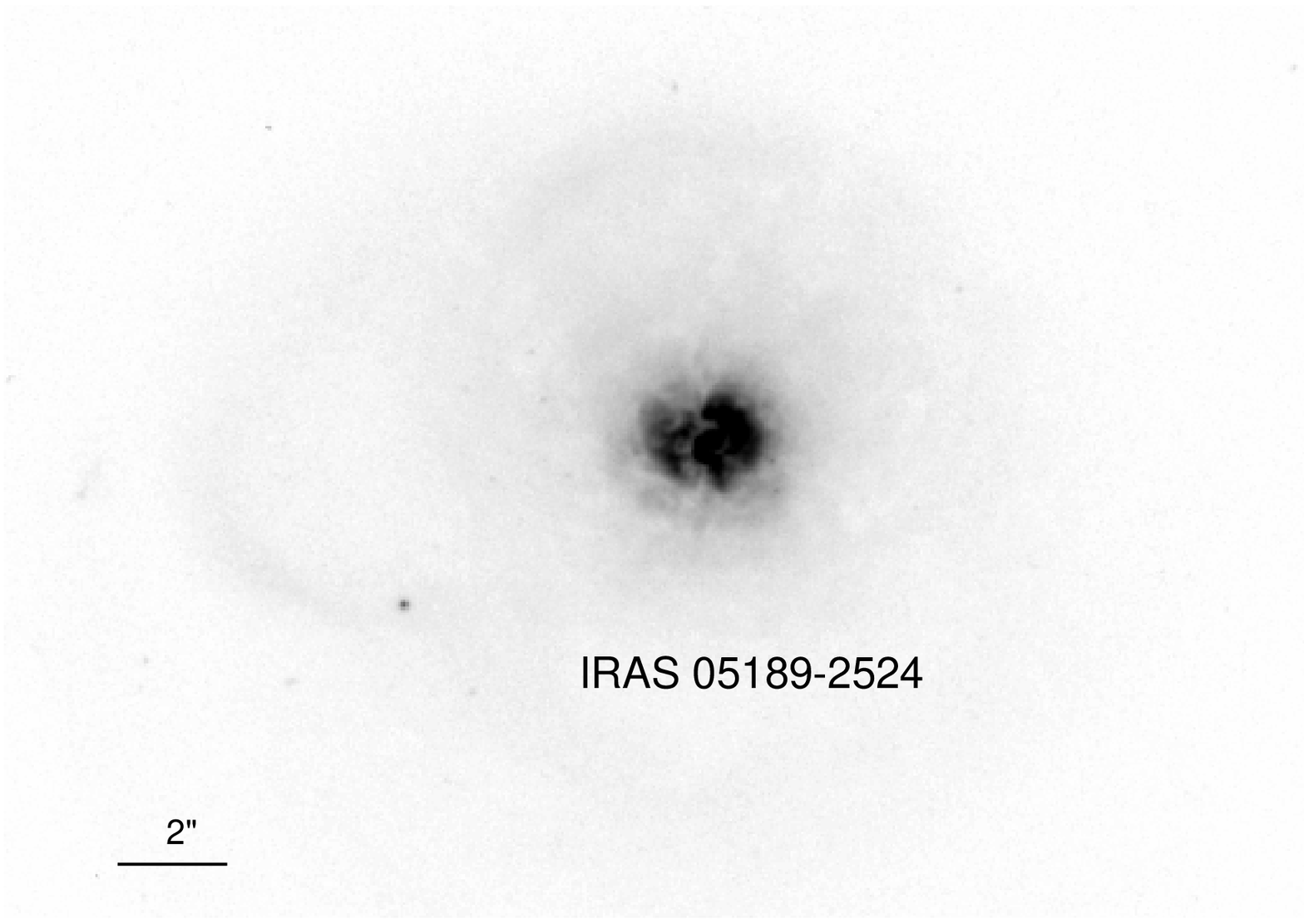}\end{minipage}
 \begin{minipage}{1\textwidth}
  \caption{Fields around IRAS\,F00085$-$1223 ({\it HST} F435W; $5\arcsec\simeq 1.9$\,kpc), IRAS\,F00163$-$1039 ({\it HST} F435W; $10\arcsec\simeq5.9$\,kpc), IRAS\,F00506+7248 ({\it HST} F110W; $2\arcsec\simeq0.7$\,kpc), IRAS\,F02069$-$1022 (SDSS $r$ band; $15\arcsec\simeq4.2$\,kpc), IRAS\,F05054+1718 (UKIDSS K band; $15\arcsec\simeq5.7$\,kpc), IRAS\,F05189$-$2524 ({\it HST} F435W; $2\arcsec\simeq 1.8$\,kpc). In all images North is to the top and East is to the left.}
\label{fig:images}
 \end{minipage}
\end{figure*}

\begin{figure*}
\centering
\begin{minipage}{.48\textwidth}
\centering
\textbf{IRAS\,08572$+$3915}\par\smallskip
\includegraphics[width=9cm]{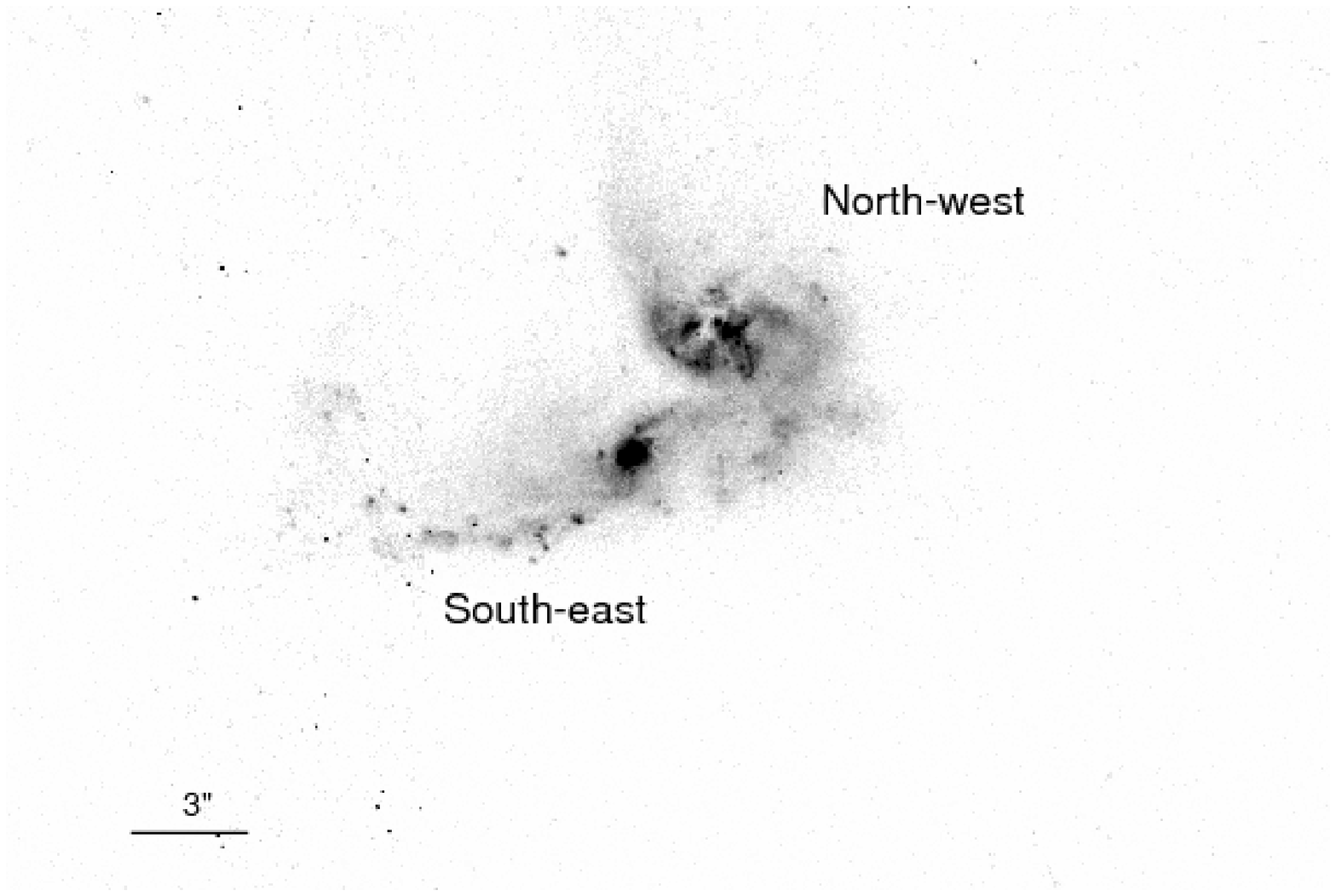}\end{minipage}
\begin{minipage}{.48\textwidth}
\centering
\textbf{IRAS\,F09320+6134}\par\smallskip
\includegraphics[width=9cm]{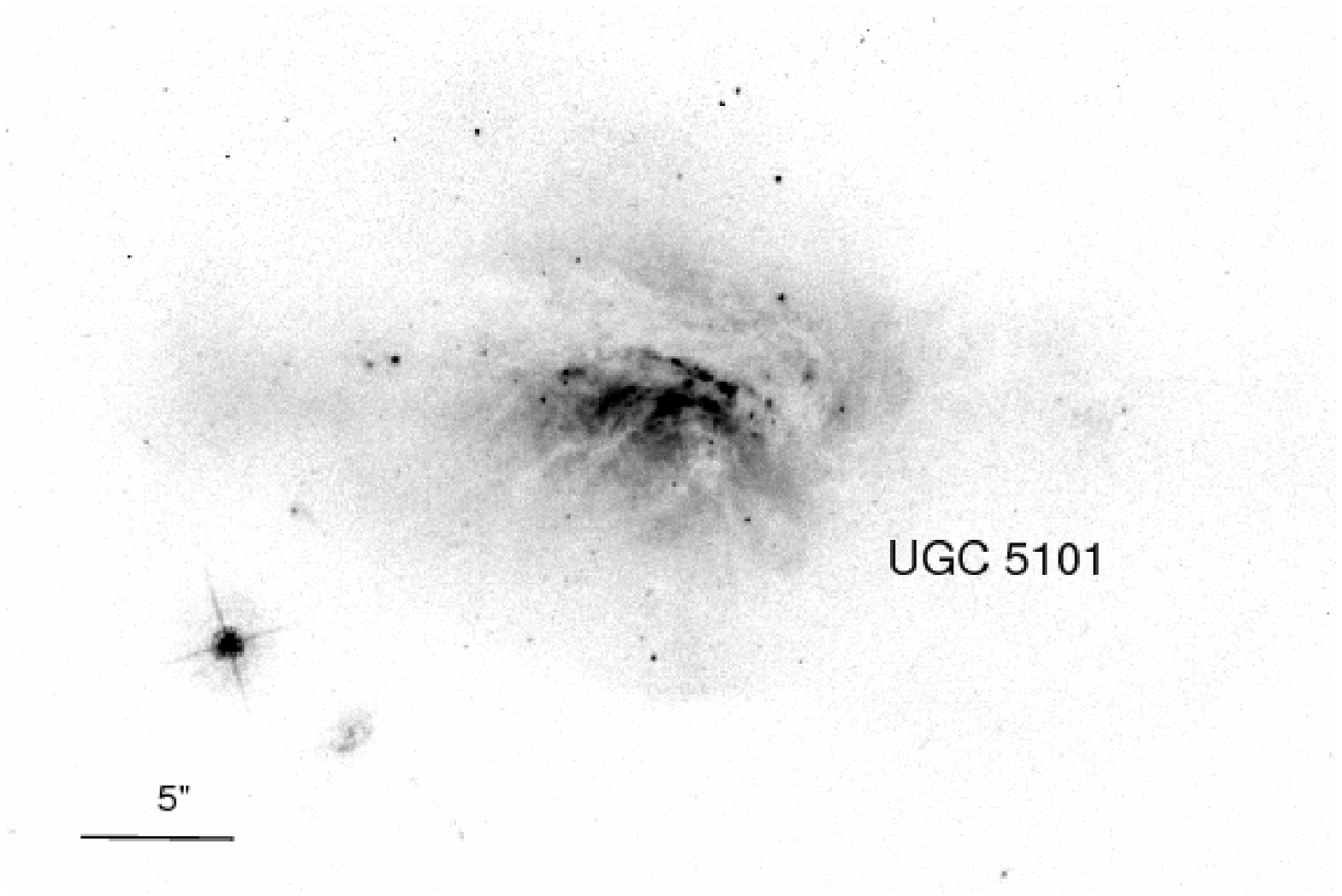}\end{minipage}
\par\smallskip
\par\smallskip
\par\smallskip
\par\smallskip
\par\smallskip
\begin{minipage}{.48\textwidth}
\centering
\textbf{IRAS\,F09333+4841}\par\smallskip
\includegraphics[width=9cm]{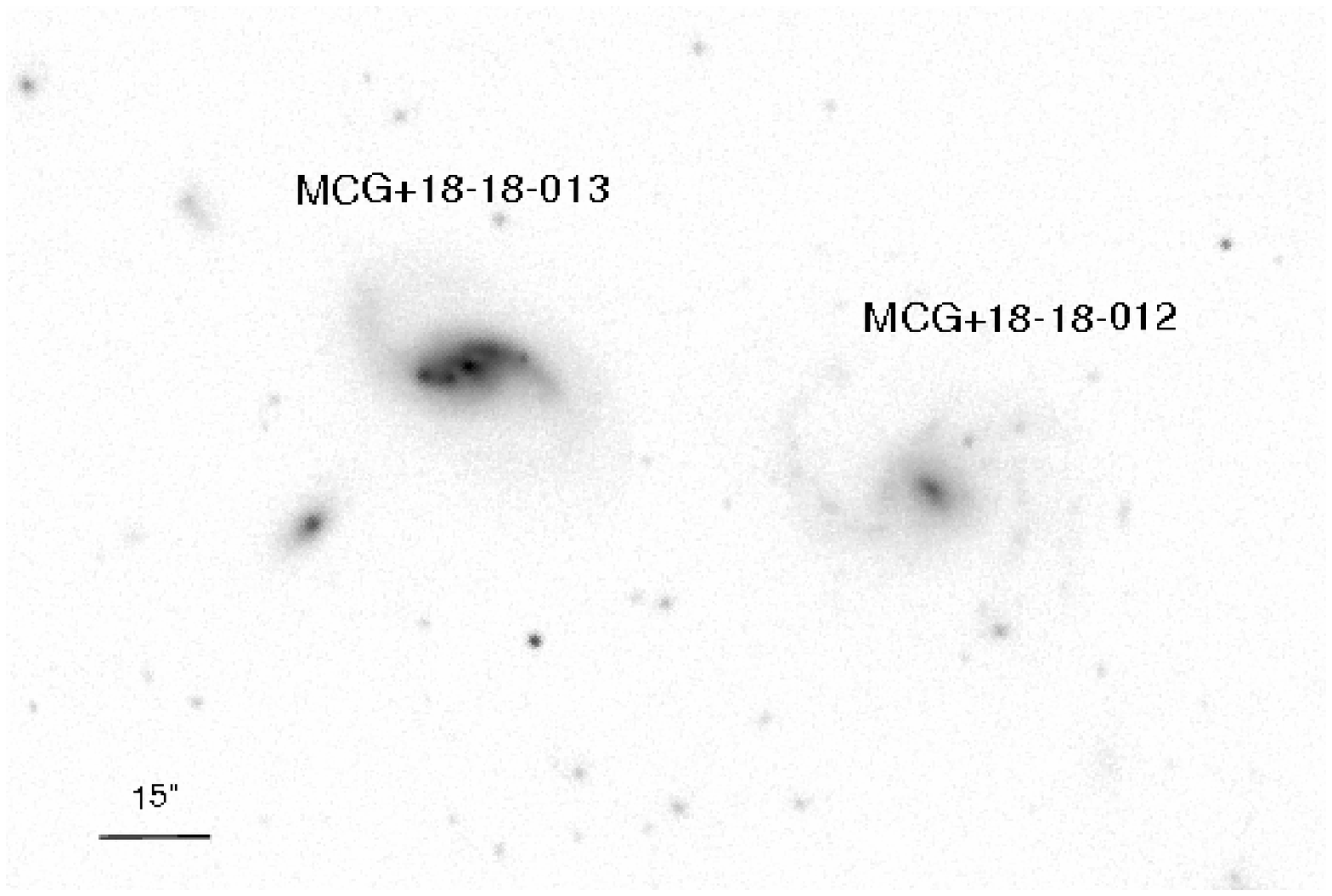}\end{minipage}
\begin{minipage}{.48\textwidth}
\centering
\textbf{IRAS\,10015$-$0614}\par\smallskip
\includegraphics[width=9cm]{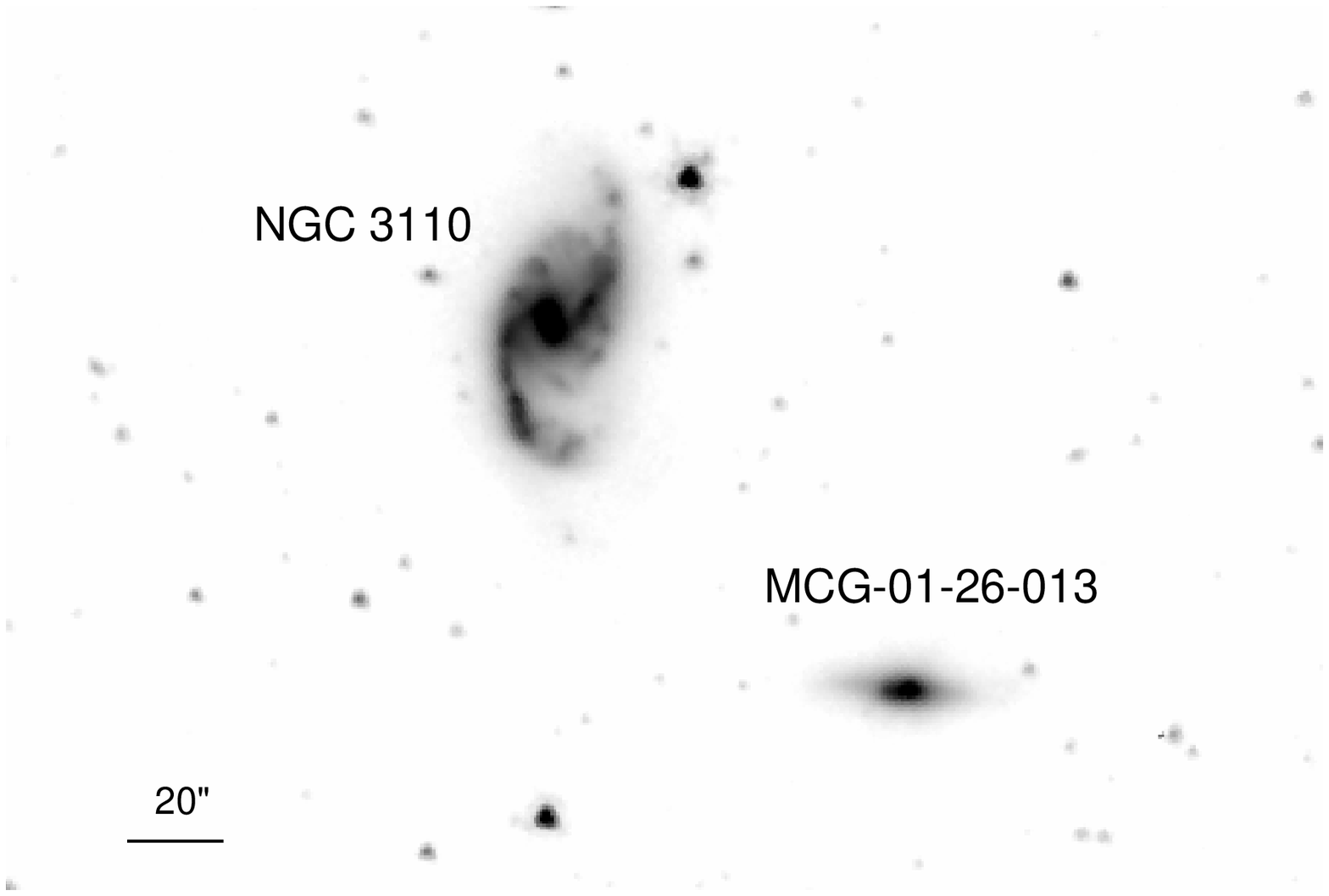}\end{minipage}
\par\smallskip
\par\smallskip
\par\smallskip
\par\smallskip
\par\smallskip
\begin{minipage}{.48\textwidth}
\centering
\textbf{IRAS\,10257$-$4339}\par\smallskip
\includegraphics[width=9cm]{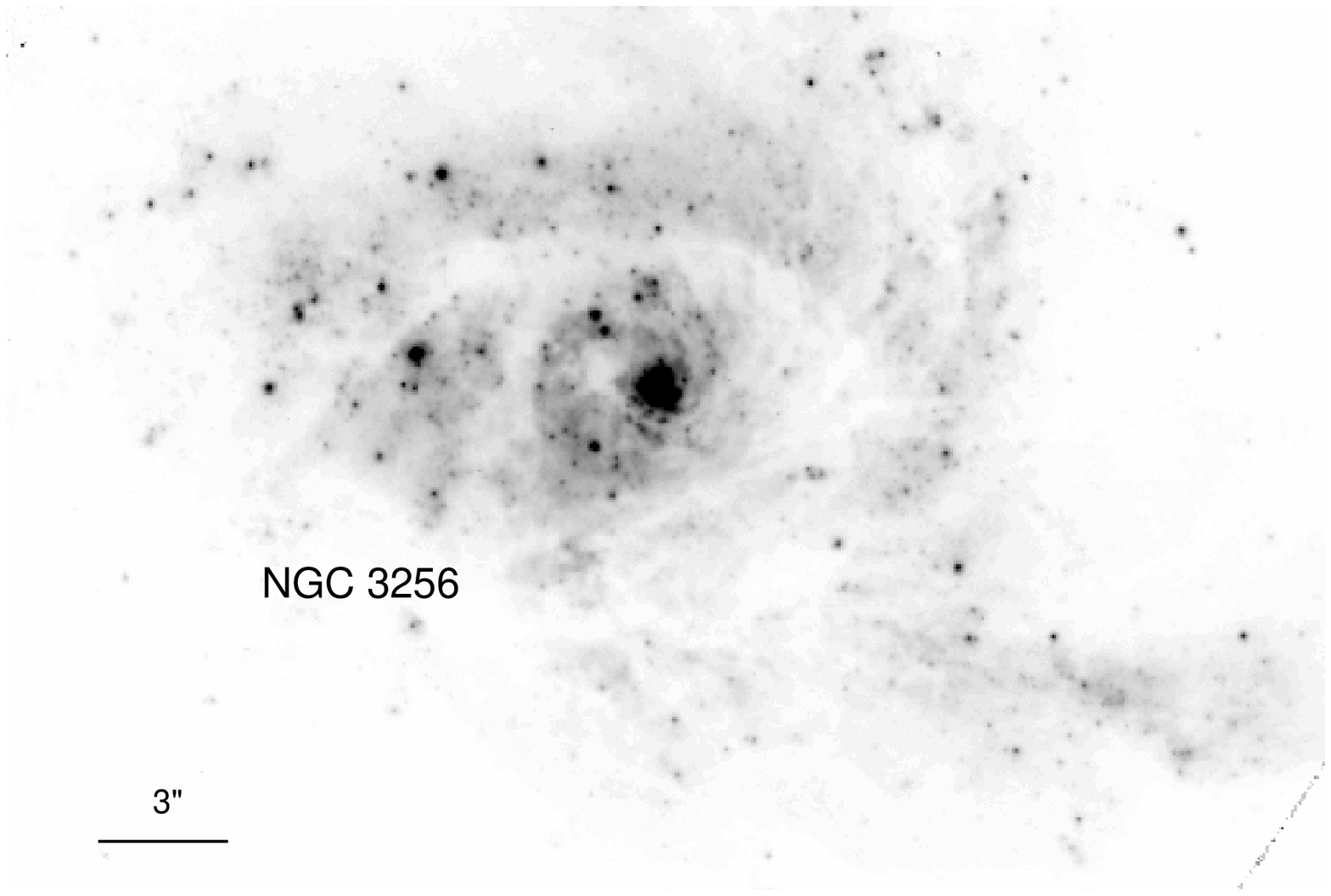}\end{minipage}
\begin{minipage}{.48\textwidth}
\centering
\textbf{IRAS\,10565$+$2448}\par\smallskip
\includegraphics[width=9cm]{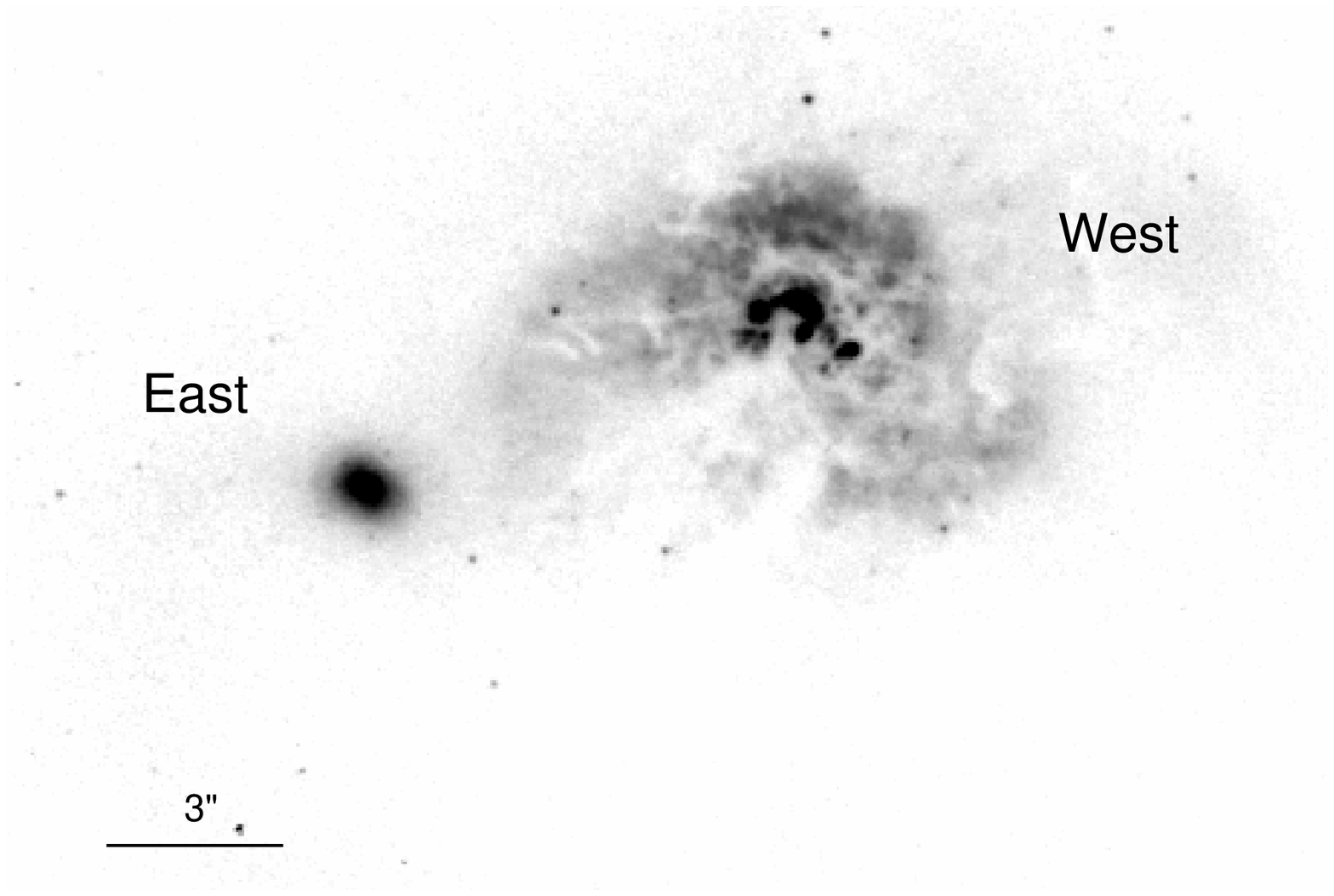}\end{minipage}
%
% %% caption
 \begin{minipage}{1\textwidth}
  \caption{Fields around IRAS\,08572$+$3915 ({\it HST} F435W; $3\arcsec\simeq3.8$\,kpc),  IRAS\,F09320+6134 ({\it HST} F435W; $5\arcsec\simeq4.1$\,kpc), IRAS\,F09333+4841 (SDSS $r$ band; $15\arcsec\simeq8.0$\,kpc), IRAS\,F10015$-$0614 ({\it Spitzer}/IRAC 3.6$\mu$m; $20\arcsec\simeq6.9$\,kpc), IRAS\,10257$-$4339 ({\it HST} F814W; $3\arcsec\simeq0.6$\,kpc), and IRAS\,10565$+$2448 ({\it HST} F435W; $3\arcsec\simeq2.7$\,kpc). In all images North is to the top and East is to the left.}
\label{fig:images2}
 \end{minipage}
\end{figure*}

\begin{figure*}
\centering
\begin{minipage}{.48\textwidth}
\centering
\textbf{IRAS\,11257$+$5850}\par\smallskip
\includegraphics[width=9cm]{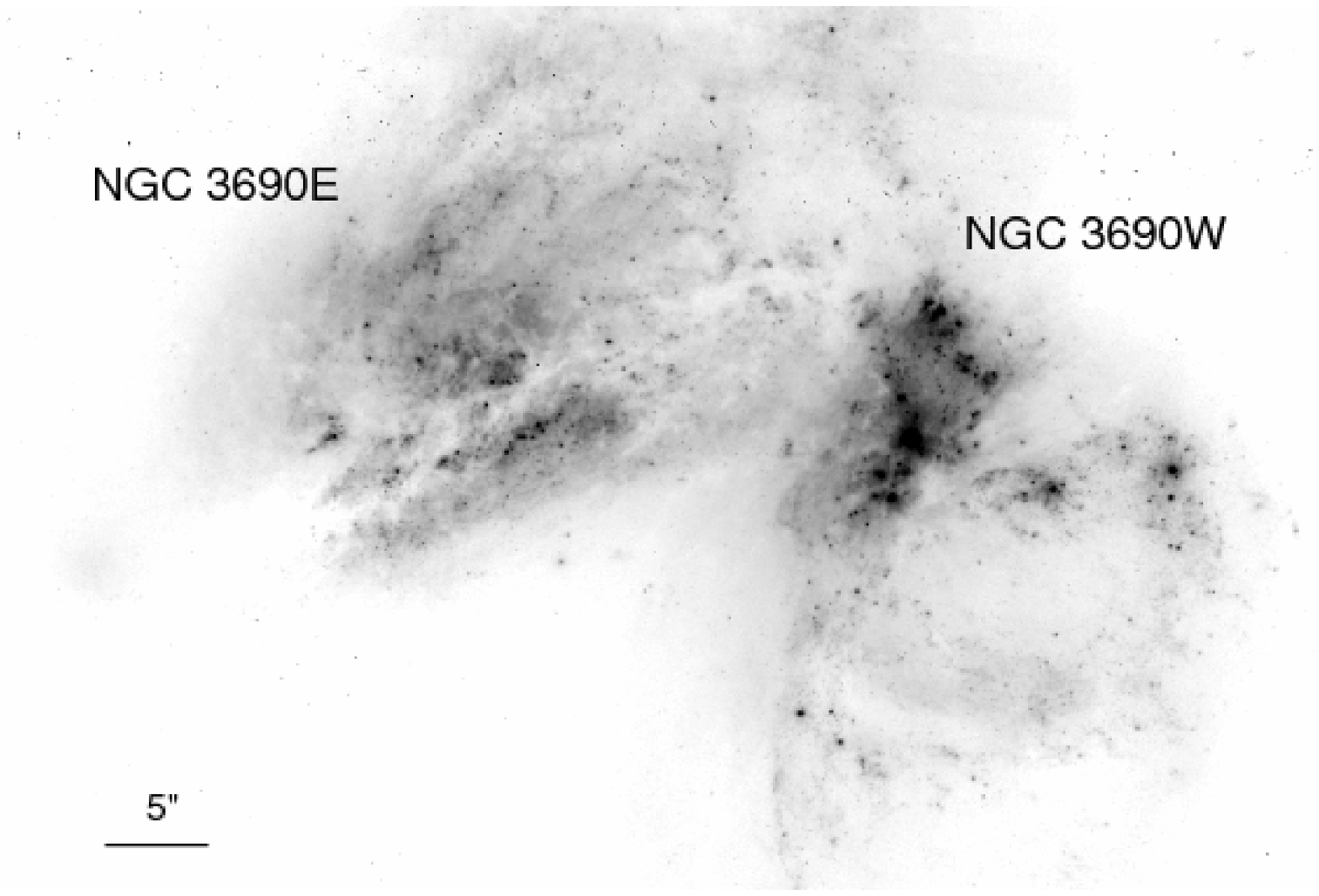}\end{minipage}
\begin{minipage}{.48\textwidth}
\centering
\textbf{IRAS\,12043$-$3140}\par\smallskip
\includegraphics[width=9cm]{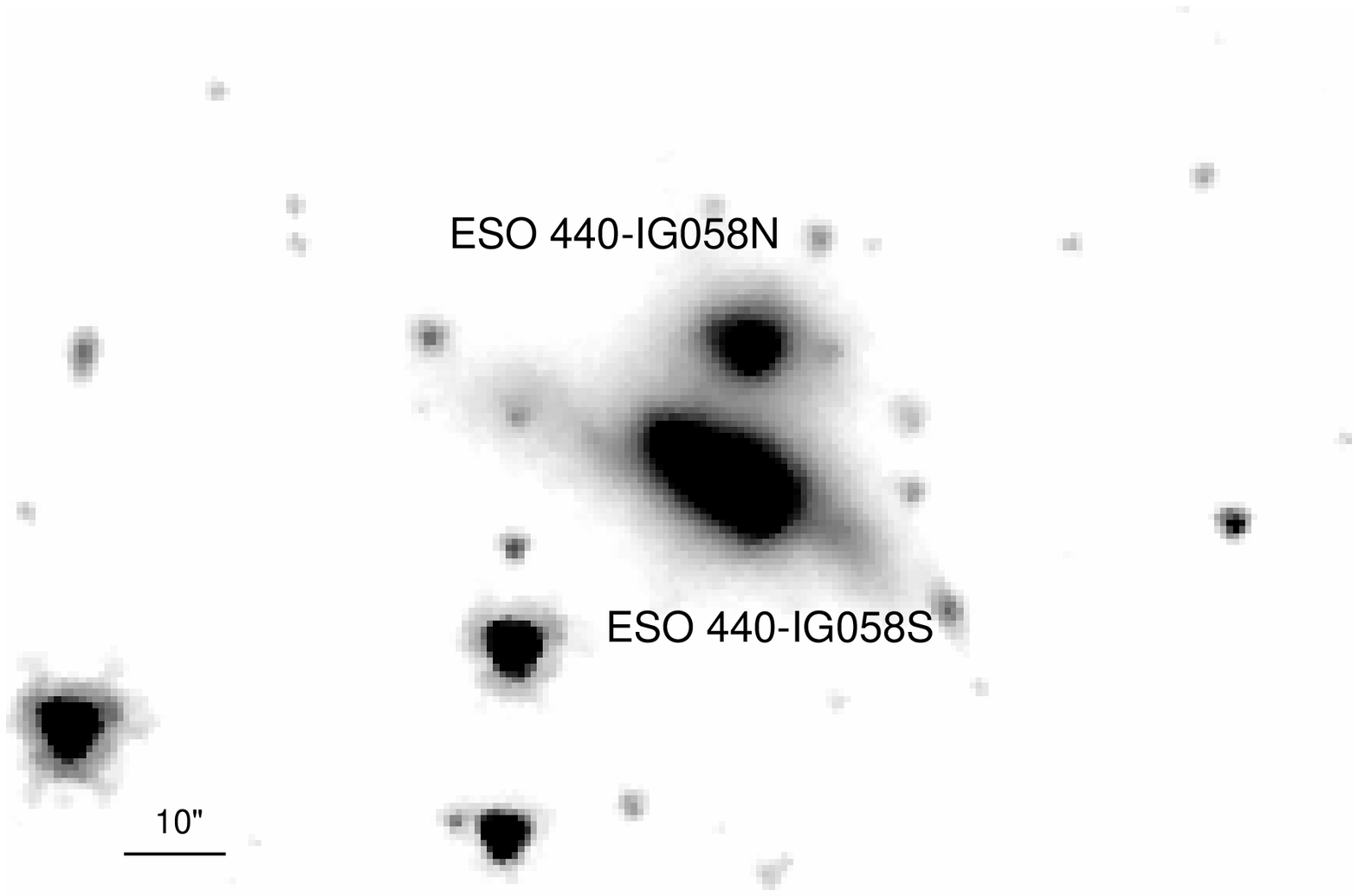}\end{minipage}
\par\smallskip
\par\smallskip
\par\smallskip
\par\smallskip
\par\smallskip
\begin{minipage}{.48\textwidth}
\centering
\textbf{IRAS\,F12540$+$5708}\par\smallskip
\includegraphics[width=9cm]{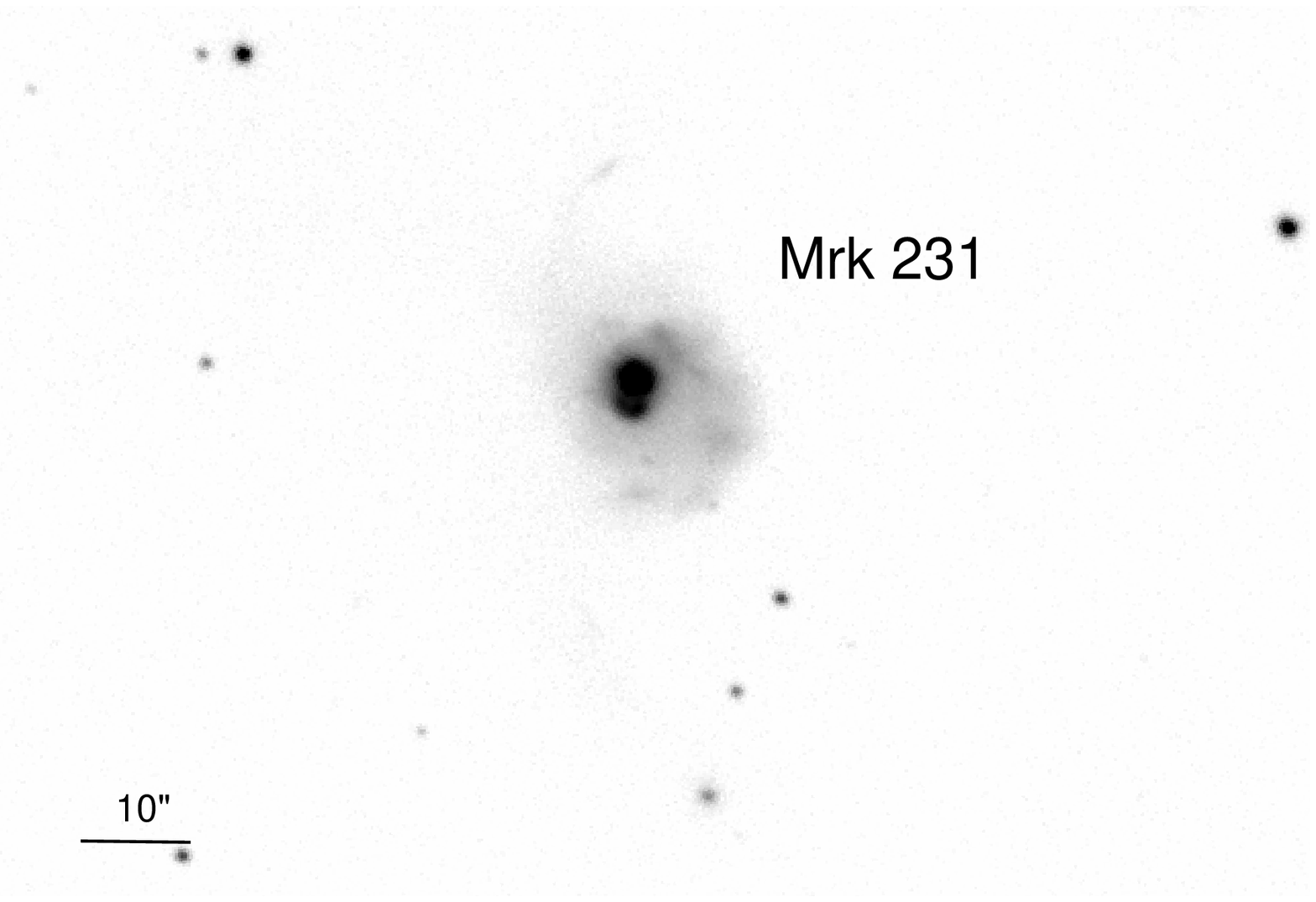}\end{minipage}
\begin{minipage}{.48\textwidth}
\centering
\textbf{IRAS\,F12590$+$2934}\par\smallskip
\includegraphics[width=9cm]{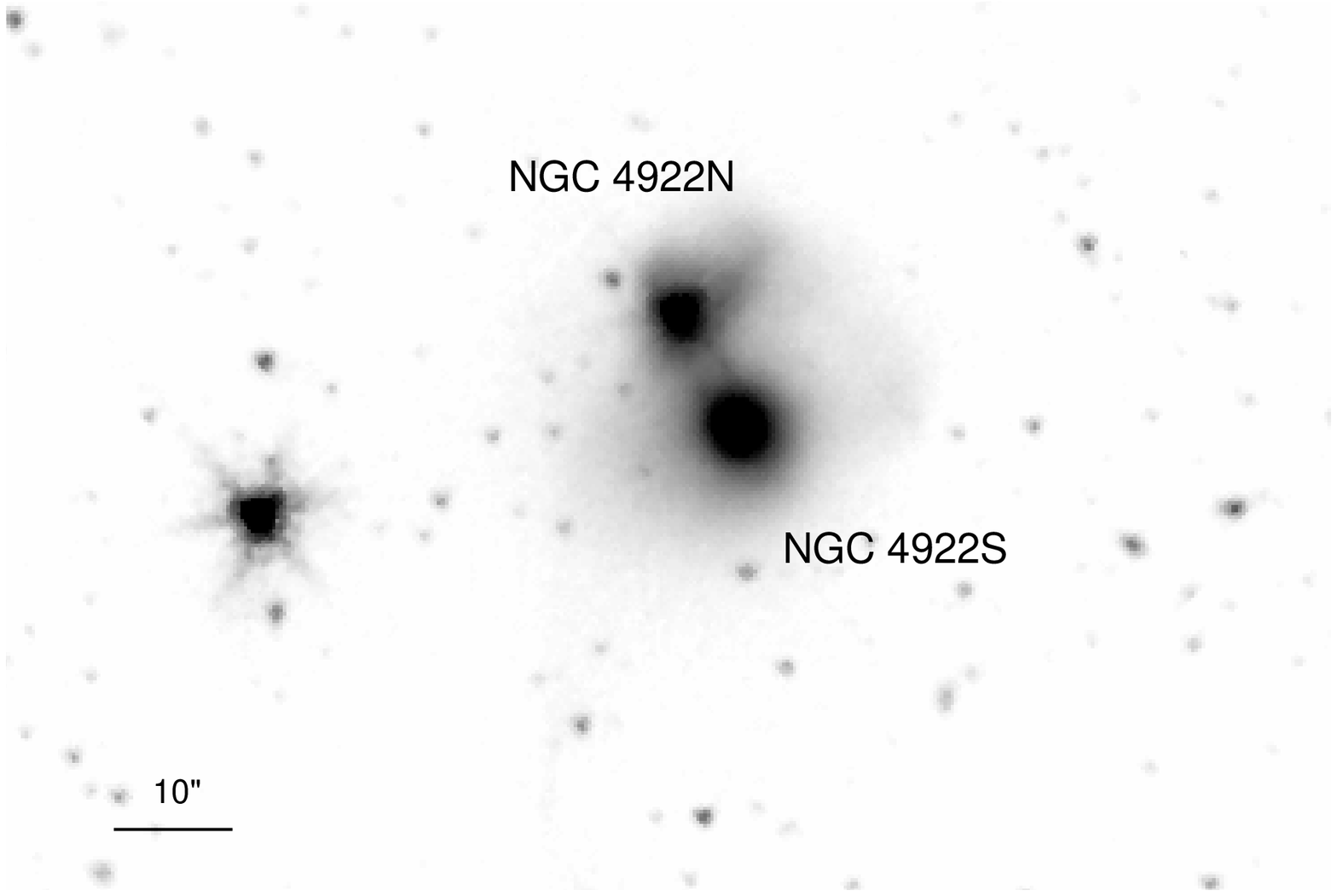}\end{minipage}
\par\smallskip
\par\smallskip
\par\smallskip
\par\smallskip
\par\smallskip
\begin{minipage}{.48\textwidth}
\centering
\textbf{IRAS\,13120$-$5453}\par\smallskip
\includegraphics[width=9cm]{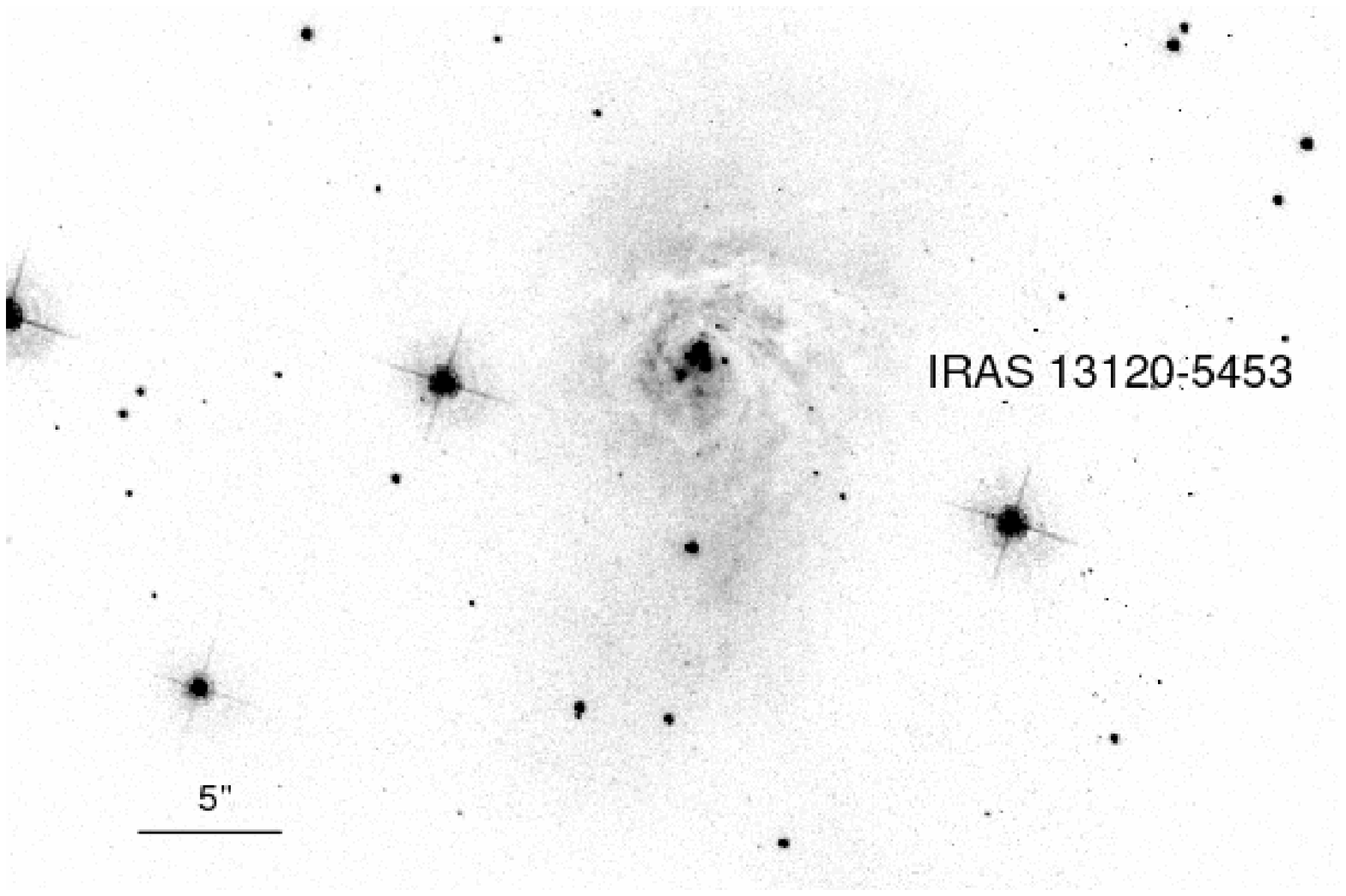}\end{minipage}
\begin{minipage}{.48\textwidth}
\centering
\textbf{IRAS\,F13197$-$1627}\par\smallskip
\includegraphics[width=9cm]{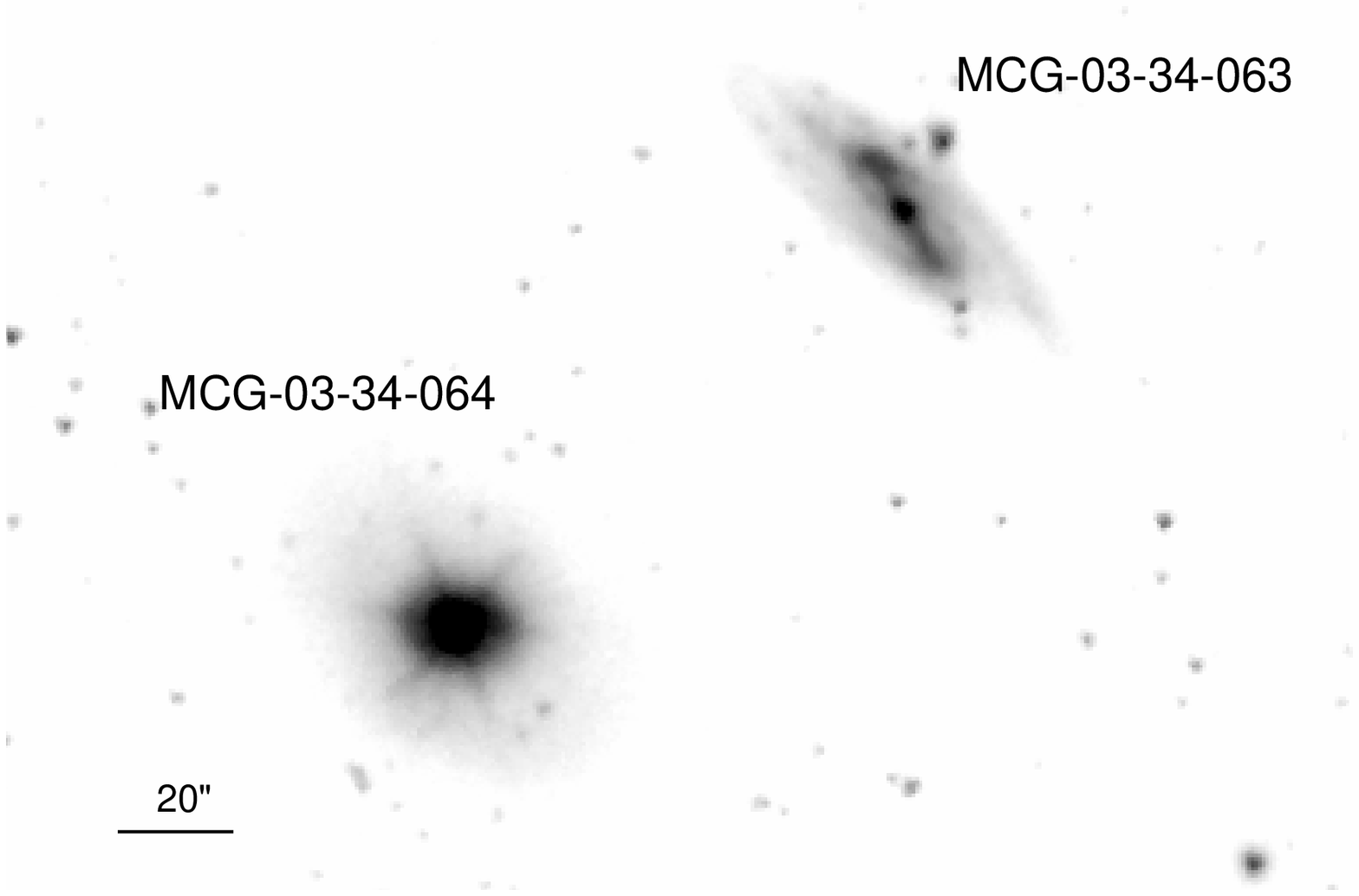}\end{minipage}
%
%
%%
% %% caption
 \begin{minipage}{1\textwidth}
  \caption{Fields around IRAS\,11257$+$5850 ({\it HST} F435W; $5\arcsec\simeq1.1$\,kpc), IRAS\,12043$-$3140 ({\it Spitzer}/IRAC 3.6$\mu$m; $10\arcsec\simeq4.9$\,kpc), IRAS\,F12540$+$5708 (SDSS $g$ band; $10\arcsec\simeq8.8$\,kpc), IRAS\,F12590$+$2934 ({\it Spitzer}/IRAC 3.6$\mu$m; $10\arcsec\simeq5.1$\,kpc), IRAS\,13120$-$5453 ({\it HST} F435W; $5\arcsec\simeq3.2$\,kpc) and IRAS\,F13197$-$1627 ({\it Spitzer}/IRAC 3.6$\mu$m; $20\arcsec\simeq7.1$\,kpc). In all images North is to the top and East is to the left.} 
\label{fig:images3}
 \end{minipage}
\end{figure*}

\begin{figure*}
\centering
\begin{minipage}{.48\textwidth}
\centering
\textbf{IRAS\,13428$+$5608}\par\smallskip
\includegraphics[width=9cm]{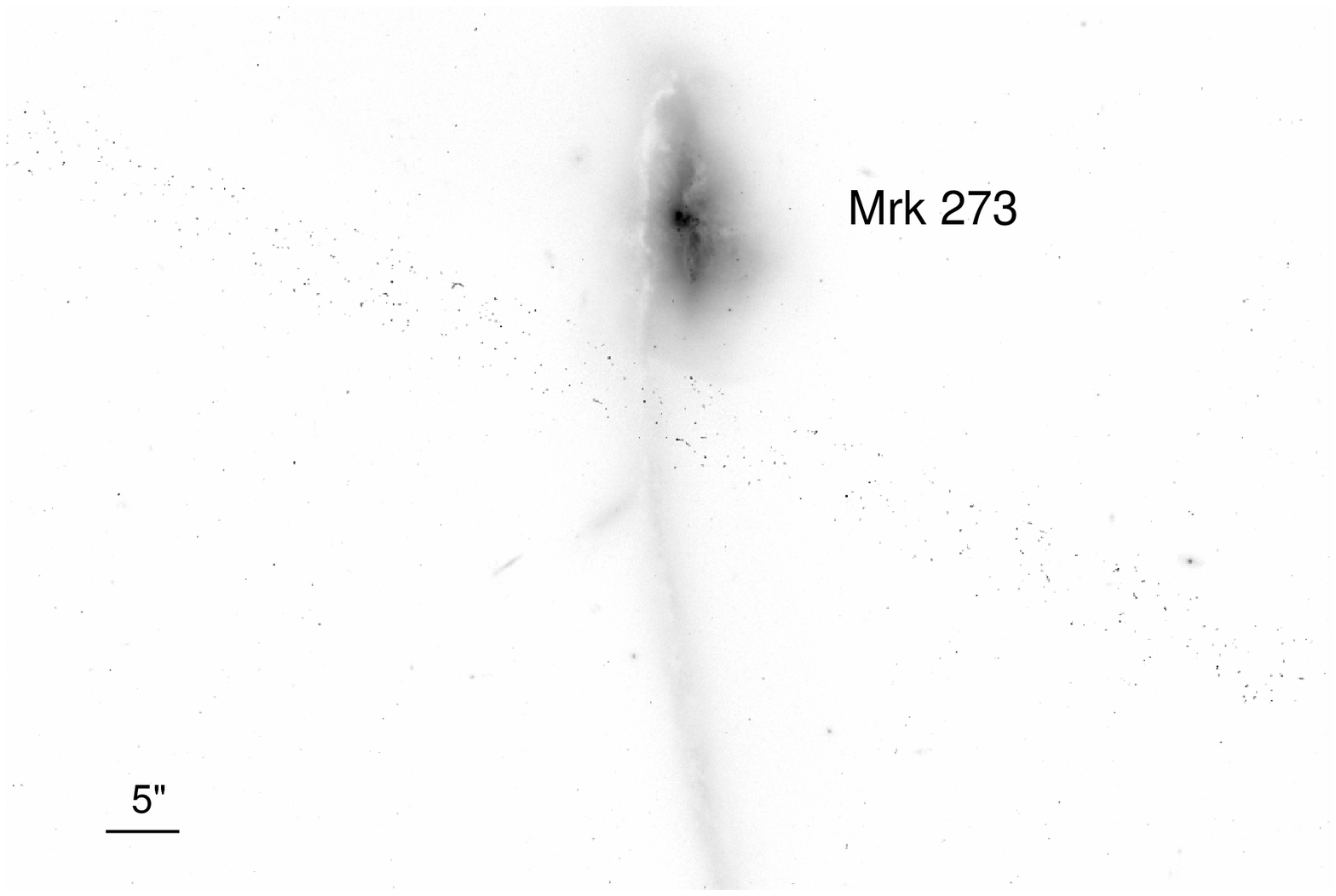}\end{minipage}
\begin{minipage}{.48\textwidth}
\centering
\textbf{IRAS\,14378$-$3651}\par\smallskip
\includegraphics[width=9cm]{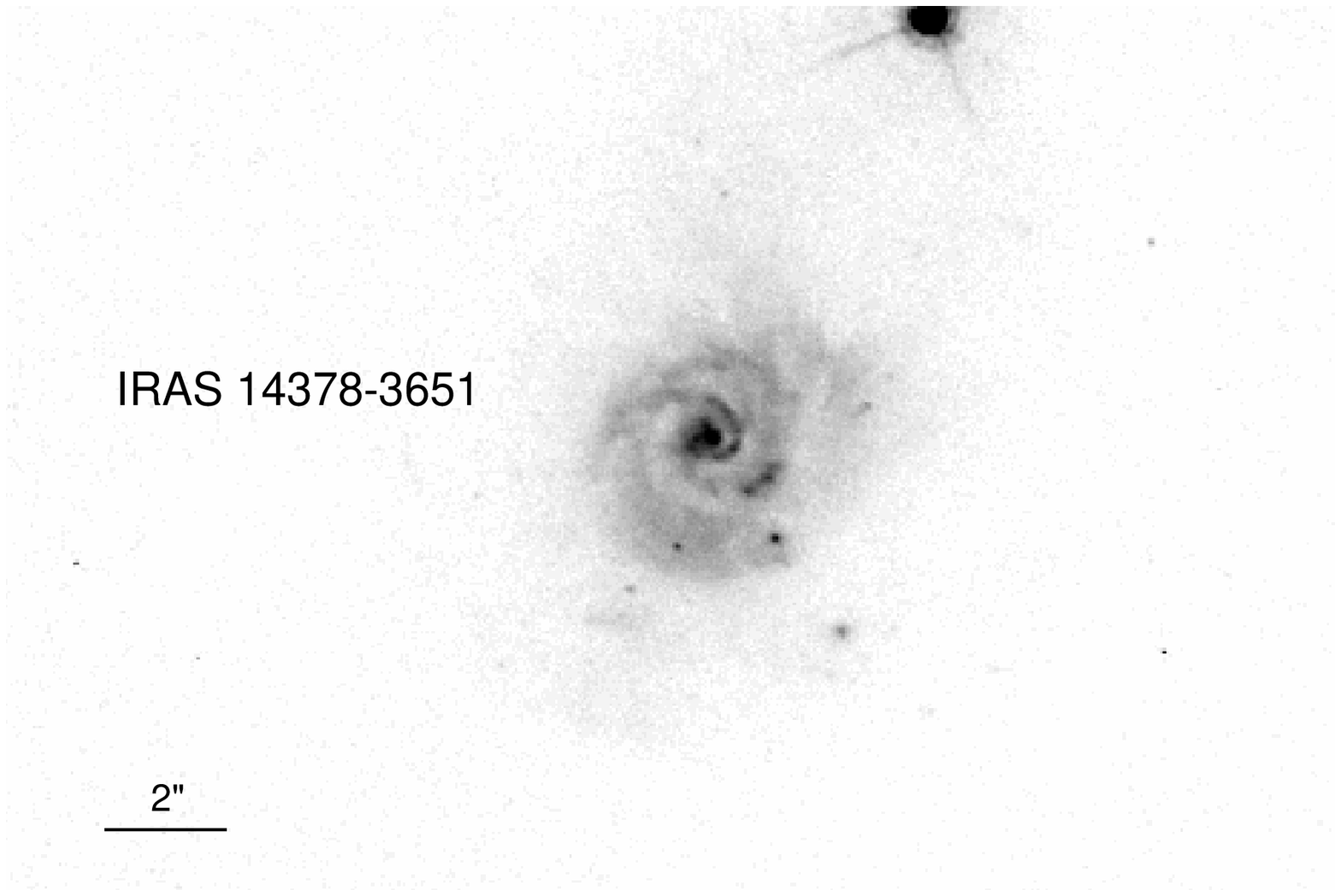}\end{minipage}
\par\smallskip
\par\smallskip
\par\smallskip
\par\smallskip
\par\smallskip
\begin{minipage}{.48\textwidth}
\centering
\textbf{IRAS\,14544$-$4255}\par\smallskip
\includegraphics[width=9cm]{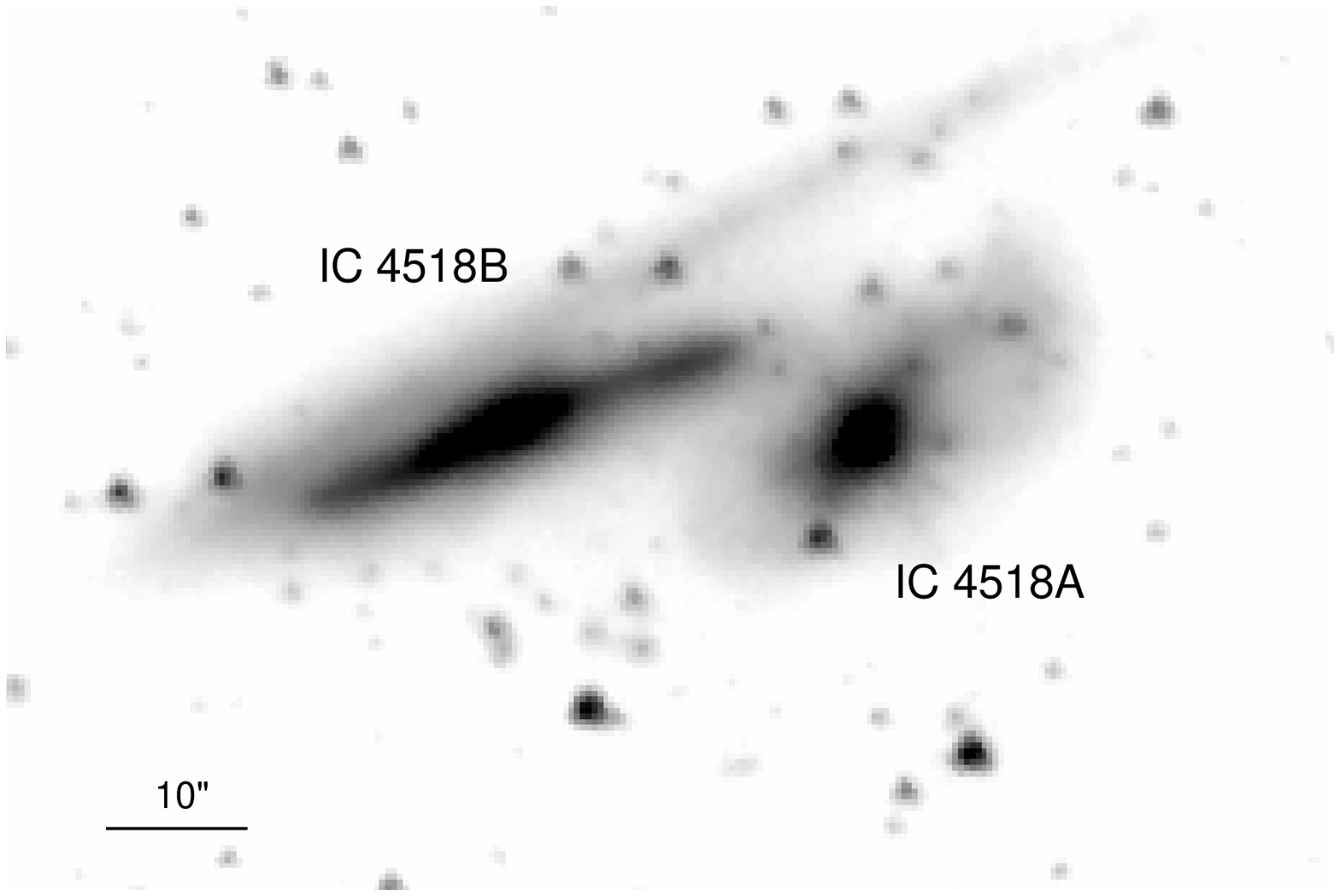}\end{minipage}
\begin{minipage}{.48\textwidth}
\centering
\textbf{IRAS\,15327$+$2340}\par\smallskip
\includegraphics[width=9cm]{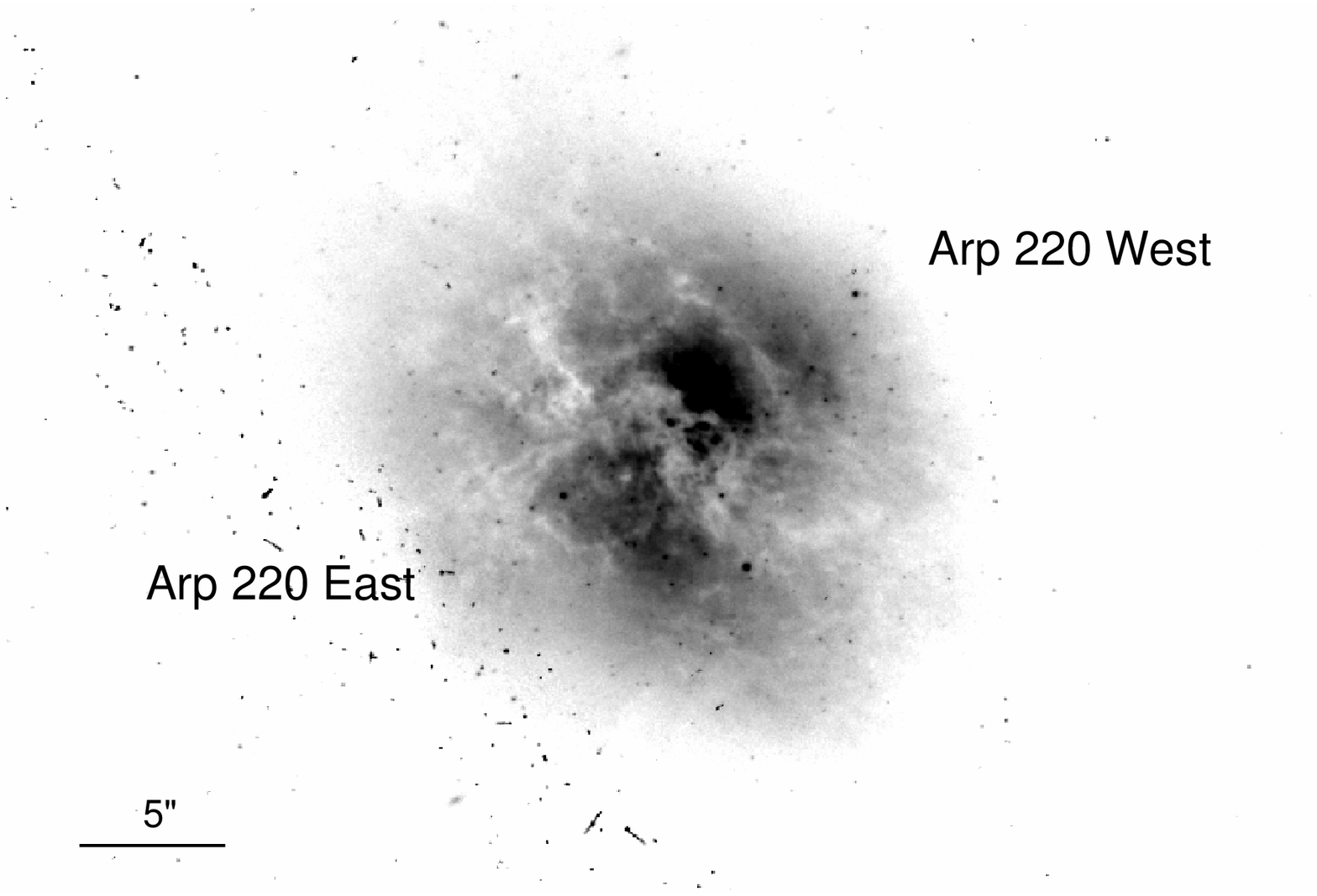}\end{minipage}
\par\smallskip
\par\smallskip
\par\smallskip
\par\smallskip
\par\smallskip
\begin{minipage}{.48\textwidth}
\centering
\textbf{IRAS\,16504$+$0228}\par\smallskip
\includegraphics[width=9cm]{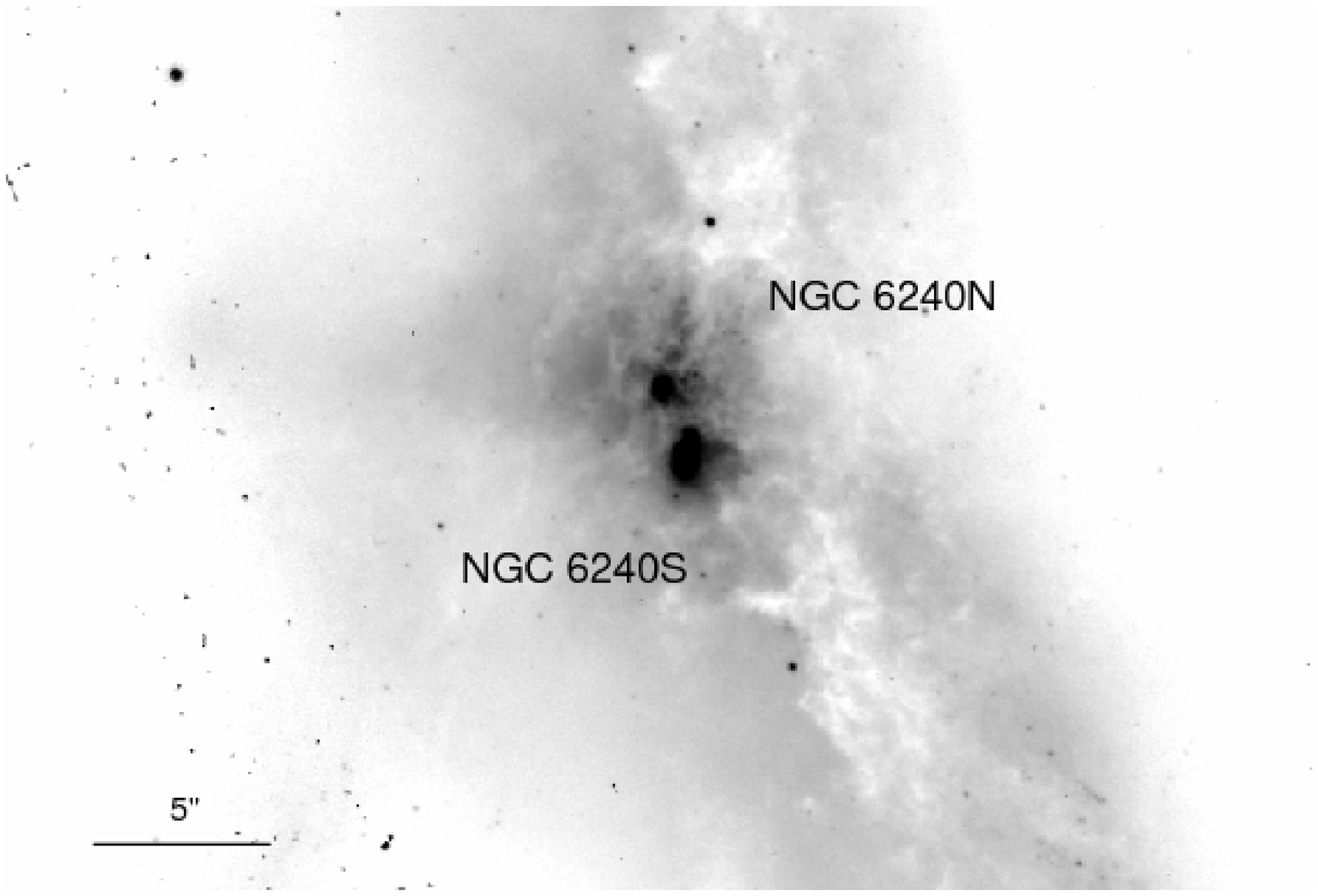}\end{minipage}
\begin{minipage}{.48\textwidth}
\centering
\textbf{IRAS\,F16577+5900}\par\smallskip
\includegraphics[width=9cm]{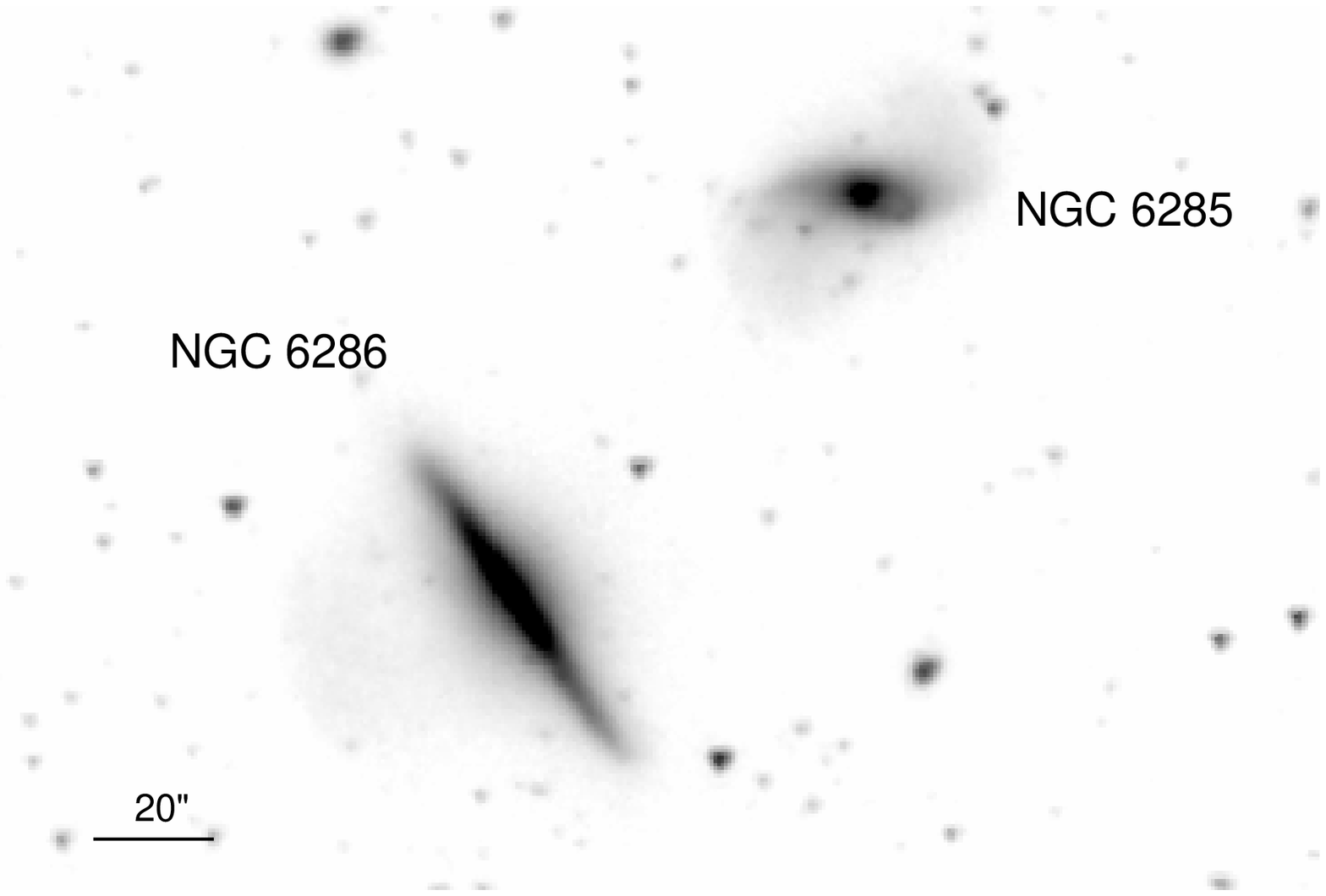}\end{minipage}
%
%
%%
% %% caption
 \begin{minipage}{1\textwidth}
  \caption{Fields around IRAS\,13428$+$5608 ({\it HST} F814W; $5\arcsec\simeq3.9$\,kpc), IRAS\,14378$-$3651 ({\it HST} F435W; $2\arcsec\simeq2.9$\,kpc), IRAS\,14544$-$4255 ({\it Spitzer}/IRAC 3.6$\mu$m; $10\arcsec\simeq3.3$\,kpc), IRAS\,15327$+$2340 ({\it HST} F814W; $5\arcsec\simeq2$\,kpc), IRAS\,16504+0228  ({\it HST} F814W; $5\arcsec\simeq2.5$\,kpc), IRAS\,F16577+5900 ({\it Spitzer}/IRAC 3.6$\mu$m; $20\arcsec\simeq7.9$\,kpc). In all images North is to the top and East is to the left.} 
\label{fig:images4}
 \end{minipage}
\end{figure*}

\begin{figure*}
\centering
\begin{minipage}{.48\textwidth}
\centering
\textbf{IRAS\,F17138$-$1017}\par\smallskip
\includegraphics[width=9cm]{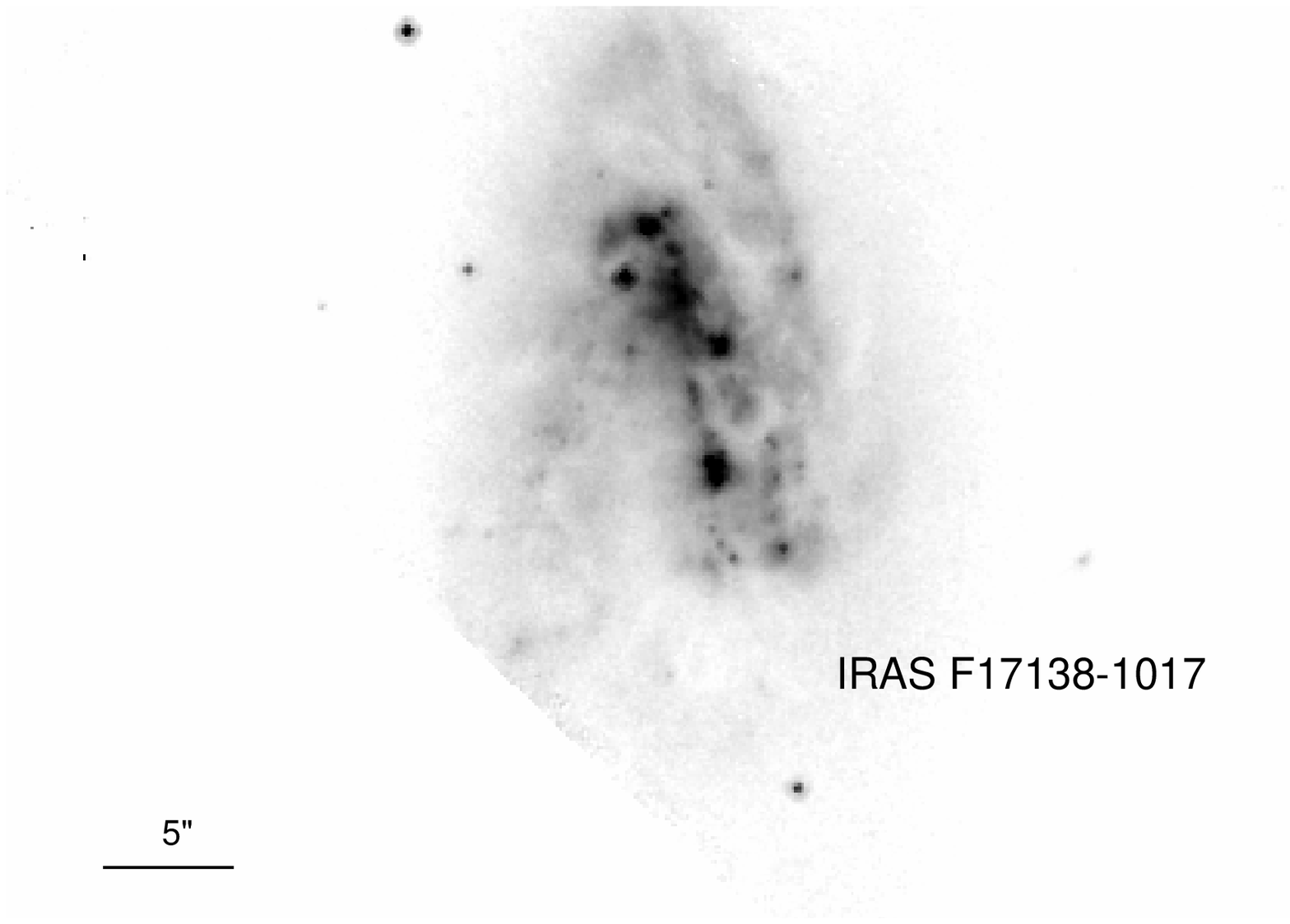}\end{minipage}
\begin{minipage}{.48\textwidth}
\centering
\textbf{IRAS\,20264+2533}\par\smallskip
\includegraphics[width=9cm]{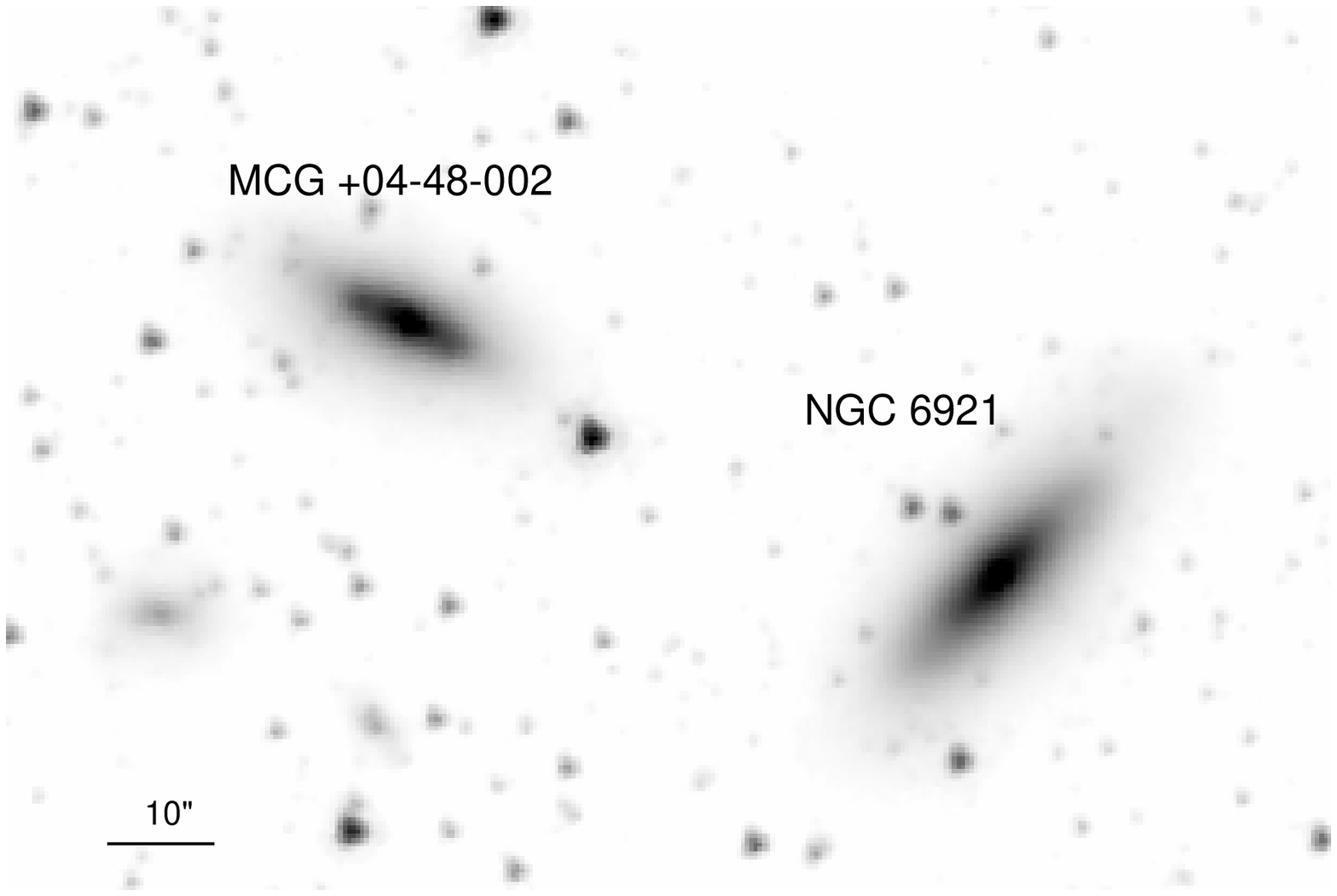}\end{minipage}
\par\smallskip
\par\smallskip
\par\smallskip
\par\smallskip
\par\smallskip
\begin{minipage}{.48\textwidth}
\centering
\textbf{IRAS\,21453$-$3511}\par\smallskip
\includegraphics[width=9cm]{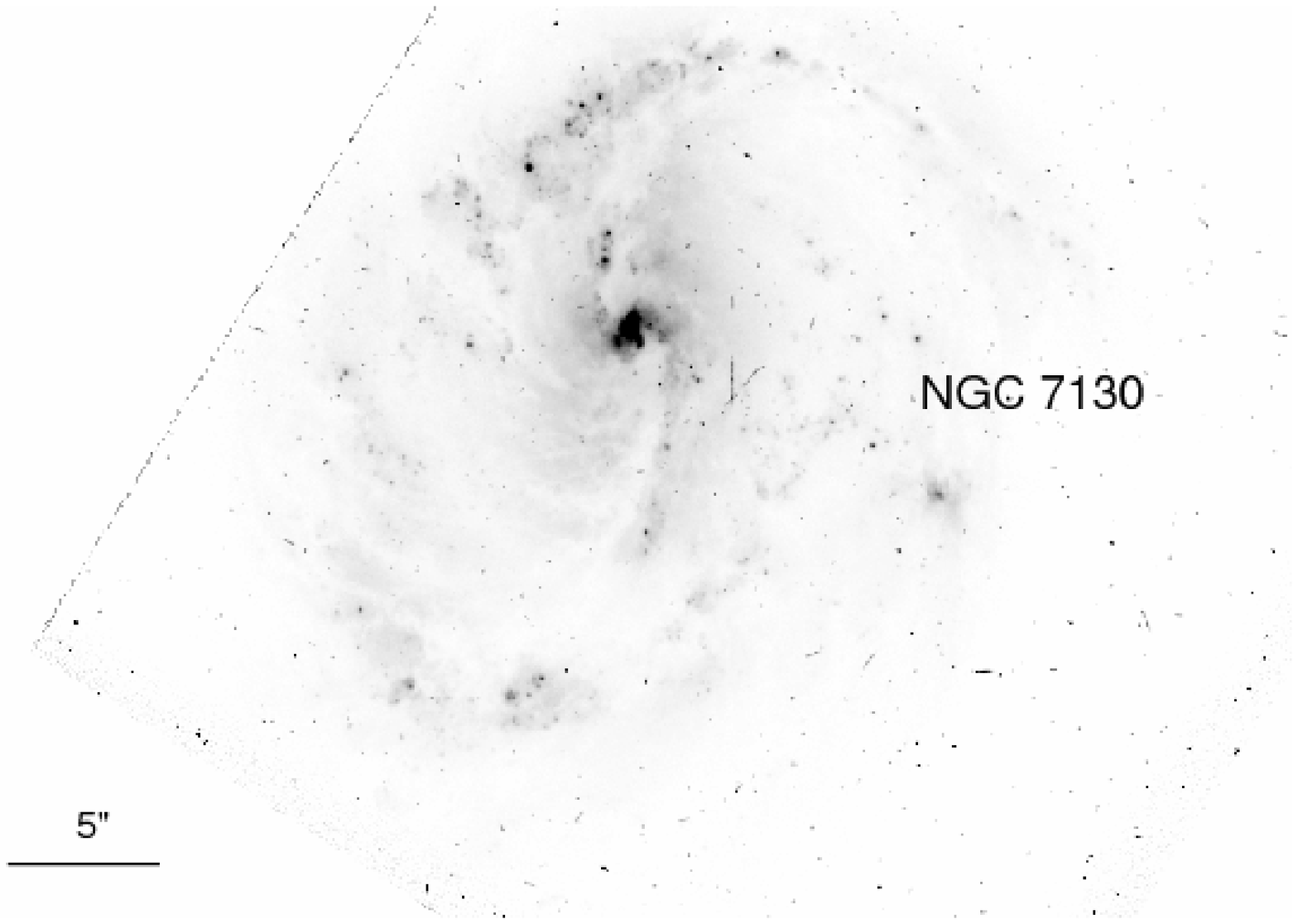}\end{minipage}
\begin{minipage}{.48\textwidth}
\centering
\textbf{IRAS\,23007$+$0836}\par\smallskip
\includegraphics[width=9cm]{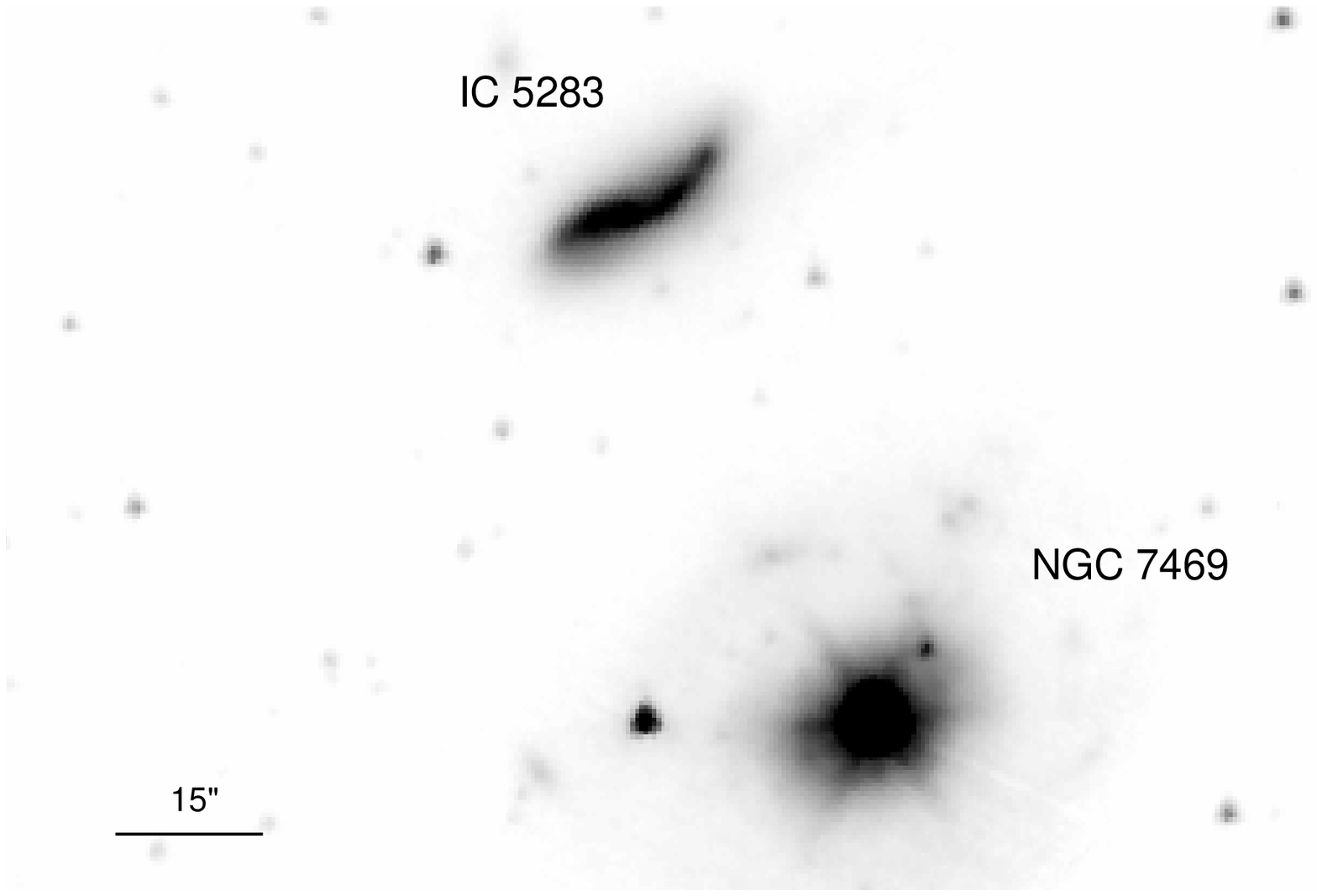}\end{minipage}
\par\smallskip
\par\smallskip
\par\smallskip
\par\smallskip
\par\smallskip
\begin{minipage}{.48\textwidth}
\centering
\textbf{IRAS\,23254$+$0830}\par\smallskip
\includegraphics[width=9cm]{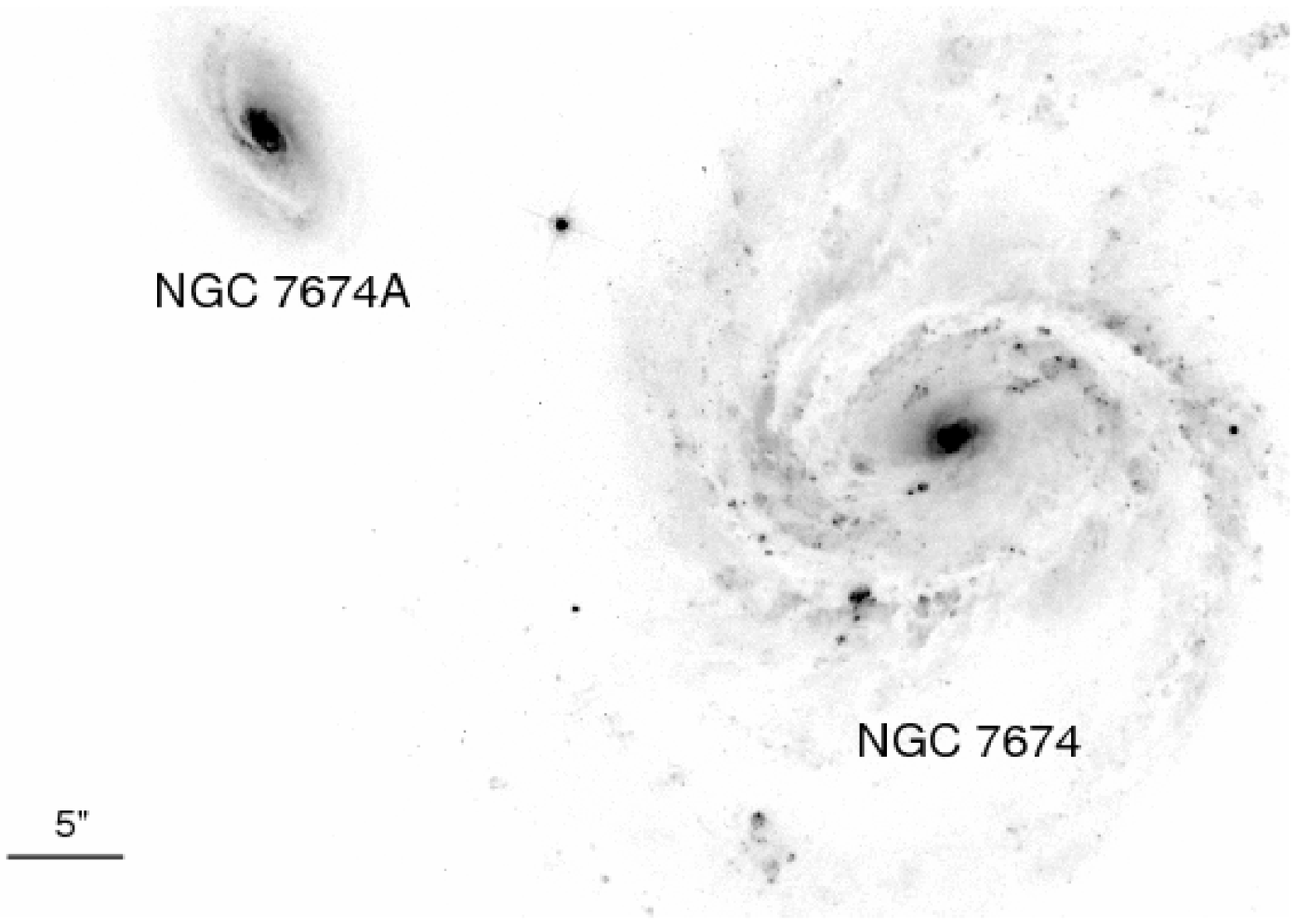}\end{minipage}
\begin{minipage}{.48\textwidth}
\centering
\textbf{IRAS\,23262$+$0314}\par\smallskip
\includegraphics[width=9cm]{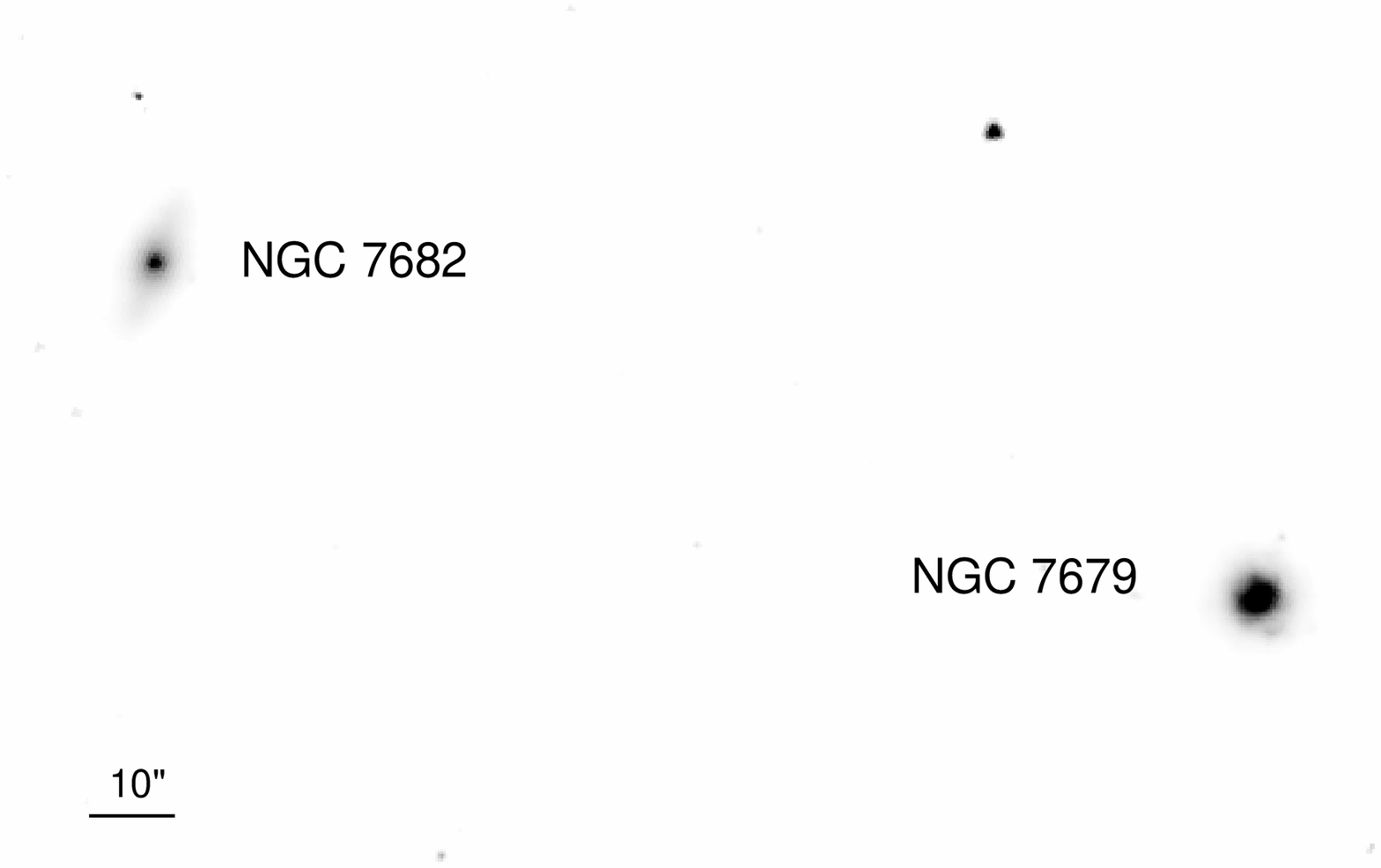}\end{minipage}
%
% %% caption
 \begin{minipage}{1\textwidth}
  \caption{Fields around IRAS\,F17138$-$1017 ({\it HST} F110W; $5\arcsec\simeq1.8$\,kpc), IRAS\,20264+2533 ({\it Spitzer}/IRAC 3.6$\mu$m; $10\arcsec\simeq3.0$\,kpc), IRAS\,21453$-$3511 ({\it HST} F606W; $5\arcsec\simeq1.5$\,kpc), IRAS\,23007$+$0836 ({\it Spitzer}/IRAC 3.6$\mu$m; $15\arcsec\simeq5.0$\,kpc), IRAS\,23254$+$0830 ({\it HST} F435W; $5\arcsec\simeq3.0$\,kpc) and IRAS\,23262$+$0314 ({\it Spitzer}/IRAC 3.6$\mu$m; $10\arcsec\simeq3.6$\,kpc). In all images North is to the top and East is to the left.} 
\label{fig:images5}
 \end{minipage}
\end{figure*}

The sample objects, together with their merger stage and the distance between the two nuclei are reported in Table\,\ref{tab:sample}. The images of the objects of our samples are shown in Figs.\,\ref{fig:images}-\ref{fig:images5}. {\it NuSTAR} observations of 15\,\,systems have been studied before, and for these objects we report the results from the literature in Table\,\ref{tab:results_other}.

\section{X-ray observations and data reduction}
\label{SM_datared}

We analysed the X-ray observations of 16\,\,systems, using {\it XMM-Newton} ($\S$\ref{sect:XMM}), {\it Chandra} ($\S$\ref{sect:Chandra}), {\it NuSTAR} ($\S$\ref{sec:Nustar}), {\it Swift}/BAT and {\it Swift}/XRT data ($\S$\ref{sect:Swift}). The observation log of our sample is reported in Table\,\ref{tab:obslog}. Details about the analysis of all sources are reported in Appendix\,\ref{SM_individual}. The angular separation between the nuclei and the relation with the extraction regions used for the different instruments are discussed in detail in Appendix\,\ref{SM_individual} for every source analysed here.

\subsection{{\it NuSTAR}}\label{sec:Nustar}

We studied here the 13 observations carried out by the {\it Nuclear Spectroscopic Telescope Array} ({\it NuSTAR}, \citealp{Harrison:2013lq}) not studied before. Of these, nine were observed as a part of a dedicated campaign awarded to our group during {\it NuSTAR} AO-1 (PI F. Bauer). The {\it NuSTAR} data were processed using the {\it NuSTAR} Data Analysis Software \textsc{nustardas}\,v1.4.1 within \textsc{heasoft\,v6.16}, adopting the calibration files released on UT 2015 March 16 \citep{Madsen:2015dz}. A circular region of 45\,arcsec was used for the sources, while the background was extracted from an annulus centred on the X-ray source, with an inner and outer radius of 50 and 100 arcsec, respectively. 
In none of the {\it NuSTAR} observations analysed here are two AGN confused or within the same extraction region. This is due to the fact that either i) no nucleus was detected by {\it NuSTAR}; ii) one nucleus only was detectable by {\it NuSTAR}; iii) if both nuclei were detectable, they were at a distance sufficient to avoid confusion.

\begin{table*}
\centering
\caption{The table reports the values obtained from the X-ray spectral analysis of the sources of our sample. For each source we list (1) the IRAS name of the source, (2) the counterparts, (3) the column density of X-ray emission related to star formation, (4) the temperature of the collisionally-ionised plasma, (5) the photon index of the soft X-ray emission due to X-ray binaries or to the scattered emission from the AGN, (6) the column density and (7) the photon index of the AGN, and (8) the value of the Cash [C] or $\chi^2$ statistics and the number of degrees of freedom (DOF). Objects in which both statistics were used to fit different spectra are reported as [C/$\chi^{2}$], and the value of the statistic is the combination of the two. }\label{tab:Xray_results1}
\begin{center}
\begin{tabular}{llcccccc}
\noalign{\smallskip}
\hline \hline \noalign{\smallskip}
\multicolumn{1}{c}{(1)}  & \multicolumn{1}{c}{(2)} & (3) & (4) & (5) & (6) & (7) & (8)  \\
\noalign{\smallskip}
{\it IRAS} name & Source & 	$N_{\rm\,H}^{\rm SF}$ & 	kT & $\Gamma_{\rm\,bin.}$ & $N_{\rm\,H}$ & $\Gamma$  & $\chi^2/DOF$\\
\noalign{\smallskip}
 & 	&($10^{21}\rm\,cm^{-2}$) &  (keV) &   & ($10^{22}\rm\,cm^{-2}$) &    \\
\noalign{\smallskip}
\hline \noalign{\smallskip}
\noalign{\smallskip}
F00085$-$1223 & 	NGC\,34				&  -- 	& $0.78^{+0.07}_{-0.08}$ &  $1.57^{A}$   &  $53\pm11$    & $1.57^{+0.12}_{-0.14}$   &  240.7/288 [C/$\chi^{2}$]  \\   
\noalign{\smallskip}
\noalign{\smallskip}
 F00163$-$1039 & 	Arp\,256				&  $3.4^{+2.4}_{-2.3}$ 	&  $0.30^{+0.45}_{-0.10}$&   $2.47^{+0.38}_{-0.32}$ &  --    & --   & 167.9/191 [C] \\
\noalign{\smallskip}
 & 	MCG$-$02$-$01$-$052	&  -- 	& $0.61\pm0.27$ & $1.93^{+0.56}_{-0.60}$   &  --    & --   & 62.0/91 [C] \\
\noalign{\smallskip}
\noalign{\smallskip}
  F00506+7248  & 	MCG+12$-$02$-$001	&  $5.4^{+2.4}_{-2.0}$ 	& -- & $3.07^{+0.57}_{-0.49}$   &  --    & --   & 131.8/144 [C] \\   
\noalign{\smallskip}
\noalign{\smallskip}
F05054+1718 & 	CGCG\,468$-$002W		&  -- 	& -- & --   &  $1.50\pm0.09$    & $1.69\pm0.04$   &  604.0/545 [$\chi^{2}$] \\    
\noalign{\smallskip}
\noalign{\smallskip}
F09333+4841 & 	MCG+08$-$18$-$013	&  $3.0^{+2.6}_{-2.1}$  	& --   &  $2.26^{+0.27}_{-0.21}$   &  --    & --   & 57.6/71 [C] \\     
\noalign{\smallskip}
\noalign{\smallskip}
F10015$-$0614 & 	  NGC 3110			&   $0.7\pm0.4$  	& $0.63^{+0.10}_{-0.06}$   &  $2.18^{+0.27}_{-0.24}$  &  --    &  --  &  245.2/270 [C/$\chi^{2}$]  \\ 
\noalign{\smallskip}
 & 	  MCG$-$01$-$26$-$013	&  --  	&  $0.36^{+0.31}_{-0.11}$  &   $1.82^{+0.57}_{-0.71}$  &  --    &  --  & 44.9/53 [C/$\chi^{2}$]      \\ 
\noalign{\smallskip}
\noalign{\smallskip}
F12043$-$3140 & 	ESO440$-$IG058N		&  -- 	& -- & $1.76^{+0.44}_{-0.46}$   &  --    & --   &40.1/45 [C] \\    
\noalign{\smallskip}
 & 	ESO440$-$IG058S		&  -- 	& $0.97^{+0.21}_{-0.17}$ & $2.32^{+0.21}_{-0.17}$   &  --    & --   &  66.7/71 [C] \\    
\noalign{\smallskip}
\noalign{\smallskip}
F12590+2934 & 	NGC 4922N			&  $7.7_{-2.8}^{+4.0}$ 	& $0.62_{-0.33}^{+0.27}$ & $2.11^{A}$   &  $\geq 427$    & $2.11^{+0.44}_{-0.46}$   & 56.3/108  [C/$\chi^{2}$] \\    
\noalign{\smallskip}
 \noalign{\smallskip}
F13197$-$1627 & 	MCG$-$03$-$34$-$064$^{\rm\,B}$		&  $-$ 	& $0.85^{+0.04}_{-0.02}/1.67_{-0.06}^{+0.17}$ & $2.80^{\rm\,A}$   &  $54.2^{+0.7}_{-0.9}$  & $ 2.80^{+0.18 }_{-0.05 }$   & 1805.1/1396  [$\chi^{2}$] \\   
\noalign{\smallskip}
\noalign{\smallskip}
F14544$-$4255 & 	IC\,4518A				&  -- 	& $0.77^{+0.04}_{-0.05}$ & $1.72^{\rm\,A}$&  $24\pm2$    & $1.72^{+0.09}_{-0.05}$   &  399.2/331  [$\chi^{2}$] \\    
\noalign{\smallskip}
 & 	IC\,4518B				&  $3.1^{+2.9}_{-2.5}$ 		& $0.24^{+0.20}_{-0.06}$ & $1.9^{\rm\,A}$ & $32_{-14}^{+41}$  & $1.9^{\rm\,C}$       &      165.7/188 [C]  \\    
\noalign{\smallskip}
\noalign{\smallskip}
F16577+5900 & 	NGC\,6285	&  -- 	& -- &  $1.64\pm0.62$   &   --  & --  &  16.8/16 [C]  \\    
\noalign{\smallskip}
\noalign{\smallskip}
F17138$-$1017 & 	IRASF17138$-$1017	&  -- 	& -- & $1.13^{+0.17}_{-0.16}$   &  --    & --   & 108.2/131 [C/$\chi^{2}$]   \\    	
\noalign{\smallskip}
\noalign{\smallskip}
20264+2533 & 	MCG +04$-$48$-$002	&  -- 	& $0.76_{-0.25}^{+0.23}$  & $1.50^{\rm\,A}$   &  $58^{+7}_{-4}$    & $1.50^{+0.12}_{-0.14}$   &  185.5/169 [$\chi^{2}$]  \\    
\noalign{\smallskip}
 & 	 NGC 6921			&  -- 	& $0.63_{-0.52}^{+0.34}$ & $2.08^{\rm\,A}$   &  $178^{+30}_{-53}$    & $2.08^{+0.16}_{-0.39}$   &  64.7/82 [$\chi^{2}$]  \\    
\noalign{\smallskip}
\noalign{\smallskip}
F21453$-$3511 & 	NGC\,7130			&  $5.0\pm1.2$ 	& $0.24^{+0.07}_{-0.06}/0.79^{+0.25}_{-0.09}$ & $2.36^{\rm\,A}$   &  $407_{-91}^{+152}$    & $2.36_{-0.12}^{+0.20}$   & 330.5/346 [C/$\chi^{2}$]   \\
\noalign{\smallskip}
\noalign{\smallskip}
F23007+0836 & 	NGC\,7469 		&  -- 	& -- & --   &  $0.006\pm0.002$    & $2.12^{+0.14}_{-0.09}$   &   1923.5/1766 [$\chi^{2}$]  \\
\noalign{\smallskip}
\noalign{\smallskip}
23262+0314  & 	NGC\,7679 		&  -- 	& $0.63^{+0.12}_{-0.10}$ & --   &   $\leq 0.02$   & $1.66\pm0.04$   &  278.9/259  [$\chi^{2}$]   \\
\noalign{\smallskip}
 & 	NGC\,7682		&  -- 	& $0.26^{+0.08}_{-0.05}$ & $2.27^{A}$      & $243^{+60}_{-44}$  &  $2.27^{+0.18}_{-0.17}$  &  876.2/914 [C]  \\
\noalign{\smallskip}
\hline
\noalign{\smallskip}
\multicolumn{8}{l}{{\bf Notes.} $^{\rm\,A}$: value of $\Gamma_{\rm\,bin.}$ fixed to that of the AGN continuum ($\Gamma$); $^{\rm\,B}$: additional plasma ($kT=0.11_{-0.06}^{+0.01}$\,keV); $^{\rm\,C}$: value fixed} \\
\end{tabular}
\end{center}
\end{table*}

\subsection{{\it XMM-Newton}}\label{sect:XMM}

{\it XMM-Newton} \citep{Jansen:2001vn} observations of seven systems of our sample were available. We analysed EPIC/PN \citep{Struder:2001uq} and MOS \citep{Turner:2001fk} data by reducing the original data files (ODFs) using {\it XMM-Newton} Standard Analysis Software (SAS) version 12.5.0 \citep{Gabriel:2004fk}. The raw PN and MOS data files were then reduced using the \texttt{epchain} and \texttt{emchain} tasks, respectively.  

The observations were then filtered for periods of high background activity by analysing the EPIC/PN and MOS background light curves in the 10--12 keV band and above 10\,keV, respectively. For both cameras we extracted the spectra of the sources using a circular region of 25\,arcsec radius, while the background was extracted from a circular region of 40\,arcsec radius, located on the same CCD of the source and in a zone where no other source was found. 

For all the {\it XMM-Newton} observations analysed here, with the exception of IRAS\,F14544-4255, the two nuclei are at a distance larger than the size of the extraction region used. In the case of IRAS\,F14544-4255 the two galaxies are $\sim 36\arcsec$ from each other, and we used a radius of 5\,arcsec (20\,arcsec) for the spectral extraction of IC\,4518B (IC\,4518A).

\subsection{{\it Chandra}}\label{sect:Chandra}

{\it Chandra} \citep{Weisskopf:2000vn} ACIS \citep{Garmire:2003kx} observations are available for ten sources in our sample. Most of these observations were carried out as a part of the campaign aimed at following up GOALS sources (PI: D. Sanders, see \citealp{Iwasawa:2011fk,Torres:2016hp} for details). We reduced {\it Chandra} ACIS data following standard procedures, using \textsc{CIAO} v.4.6. All data were reprocessed using the \textsc{chandra\_repro} task. For the extraction we used a circular region with a radius of 10\,arcsec, which included all the X-ray emission associated to the objects. A circular region with the same radius, selected in region where no other source was detected was used for the background.

\subsection{{\it Swift} XRT and BAT}\label{sect:Swift}

Data from the X-ray telescope (XRT, \citealp{Burrows:2005vn}) on board {\it Swift} were used only for IRAS\,F05054+1718. {\it Swift}/XRT data analysis was performed using the \textsc{xrtpipelinev0.13.0} within \textsc{heasoft\,v6.16} following the standard guidelines. {\it Swift}/BAT time-averaged spectra were used for two systems (IRAS\,F23007+0836 and IRAS\,F13197$-$1627), and were taken from the latest release of the {\it Swift}/BAT catalog \citep{Baumgartner:2013ek}.

\section{X-ray spectral analysis}
\label{SM_spectral}

The X-ray spectral analysis was carried out within \textsc{xspec}\,v.12.8.2 \citep{Arnaud:1996kx}. Galactic absorption in the direction of the source was added to all models using the Tuebingen-Boulder interstellar matter absorption model \textsc{TBabs} \citep{Wilms:2000vn}, fixing the column density ($N_{\rm\,H}^{\rm\,Gal}$) to the value reported by \citet{Kalberla:2005fk} for the coordinates of the source. Abundances were fixed to solar values. Spectra were typically rebinned to 20\,counts per bin in order to use $\chi^2$ statistics. Cash statistics \citep{Cash:1979fk} were used to fit {\it Chandra} spectra, and the source spectra were binned to have one count per bin, in order to avoid issues related to empty bins in \textsc{xspec}.

We used a variety of spectral models to reproduce the X-ray emission from the galaxies in our sample. To model the underlying stellar processes, which can lead to X-ray emission in U/LIRGs, we used a thermal plasma (\textsc{apec} in \textsc{xspec}) and a power-law component (\textsc{zpowerlaw}) to reproduce X-ray emission from hot plasma gas and an underlying population of X-ray binaries, respectively.
The free parameters of the \textsc{apec} model are the temperature ($kT$) and the normalization ($n_{\rm\,apec}$), while for the power-law component they are the photon index ($\Gamma$) and the normalization ($n_{\rm\,po}$). If required by the data we added photoelectric absorption (\textsc{zphabs}) to the thermal plasma and the X-ray binary emission. The only free parameter of this component is the column density ($N_{\rm\,H}^{\rm\,SF}$).

From X-ray spectroscopy and multi-wavelength properties evidence of AGN emission is found in 25 nuclei, of which 13 are early mergers and 12 are late mergers. While we cannot completely exclude the presence of low-luminosity or heavily obscured AGN in the systems for which no evidence of AGN activity is found, the X-ray spectra of all the sources analysed here (with the exception of IRAS\,F17138$-$1017) have characteristics consistent with those of star-forming regions in the X-ray band. We refer the reader to Appendix\,\ref{SM_individual} for a detailed discussion on the presence of accreting SMBHs and on the X-ray and multi-wavelength properties of all sources of our sample.

In order to self-consistently reproduce absorbed and reprocessed X-ray radiation from the circumnuclear material of the AGN we used the torus model developed by \citet{Brightman:2011fe}, which considers an X-ray source surrounded by a spherical-toroidal structure (\textsc{atable\{torus1006.fits\}}). This model was developed from ray-tracing simulations and reproduces the main features arising from the reprocessing of the primary X-ray radiation from neutral material (e.g., \citealp{Matt:1991ly,Murphy:2009hb}): a narrow Fe\,K$\alpha$ line at 6.4\,keV (e.g., \citealp{Shu:2010tg,Ricci:2014qe,Ricci:2014fr}) and the Compton hump at $\sim 20-30$\,keV (e.g., \citealp{Koss:2016kq}). These features are particularly prominent in CT AGN, due to the fact that most of the primary X-ray emission is obscured. The free parameters of the torus model we used here are the column density ($N_{\rm\,H}$), the photon index ($\Gamma$) and the normalization ($n$) of the primary X-ray emission. In this model $N_{\rm\,H}$ does not vary with the inclination angle, which we set to the maximum value permitted ($\theta_{\rm\,i}=87.1$\,deg). The model allows also to vary the half-opening angle of the torus ($\theta_{\rm\,OA}$). This component was fixed to $\theta_{\rm\,OA}=60$\,deg whenever it could not be constrained. In obscured AGN a second, unabsorbed, power-law component is often observed emerging at energies lower than the photoelectric cutoff. This component has often been associated to Thomson scattering of the primary X-ray radiation from ionised gas (e.g., \citealp{Ueda:2007th}). To take into account this feature, which could be confused with the emission arising from a population of X-ray binaries, when an AGN is present we fixed the photon index of the second power-law component to the value of the AGN primary X-ray emission. The amount of scattered radiation is parametrised in terms of $f_{\rm\,scatt}$, which is the ratio between the normalizations of the scattered component and of the primary X-ray emission.

\begin{figure}
\centering
\includegraphics[width=0.49\textwidth]{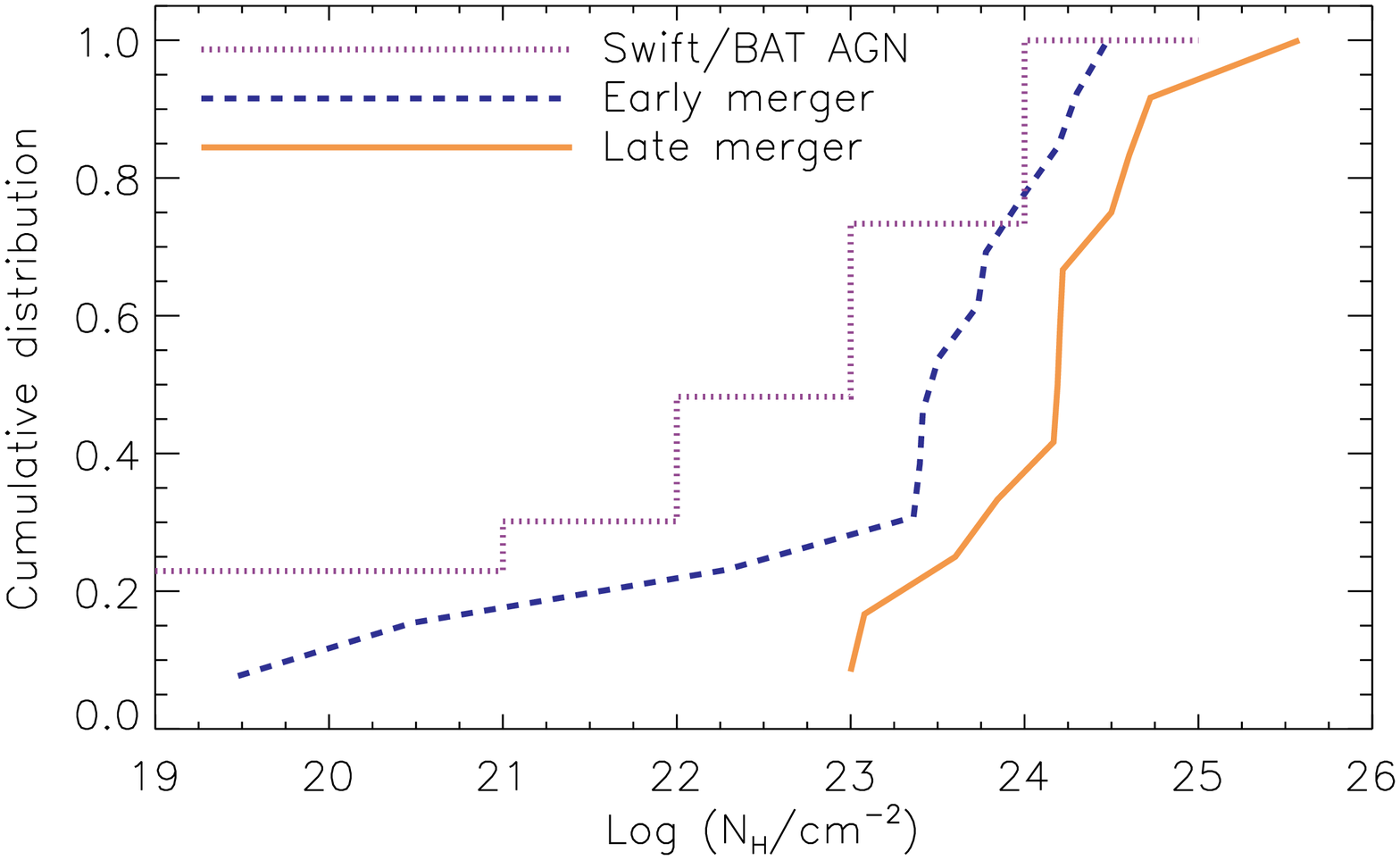} 
\caption{ Cumulative $N_{\rm\,H}$ distribution for AGN in early (blue dashed line) and late-merger (orange continuous line) galaxies. For comparison we also show the cumulative intrinsic $N_{\rm\,H}$ distribution of local, mostly non-merging {\it Swift}/BAT AGN (purple dotted line) from \citet{Ricci:2015tg}. The plot shows that: i) AGN in late mergers are systematically more obscured than those in early mergers and {\it Swift}/BAT AGN; ii) all AGN in late mergers have $N_{\rm\,H}\geq 10^{23}\rm\,cm^{-2}$, which implies that they are almost completely surrounded by material with $N_{\rm\,H}\geq 10^{23}\rm\,cm^{-2}$. Following the approach described in \citet{Cameron:2011cl} we found that the obscuring material covers $95^{+4}_{-8}\%$ of the X-ray source, where the uncertainties represent the 16th and 84th quantiles of a binomial distribution.}
\label{fig:cdf}
\end{figure}

The results obtained by our spectral analysis are summarised in Table\,\ref{tab:Xray_results1}. Details about the spectral fitting and the multi-wavelength properties of the galaxies of our sample are reported in Appendix\,\ref{SM_individual}, while the X-ray spectra are illustrated in Appendix\,\ref{appendix:xrayspec}.

\section{Discussion}\label{sect:discussion}

\begin{figure}
\centering
\includegraphics[width=0.49\textwidth]{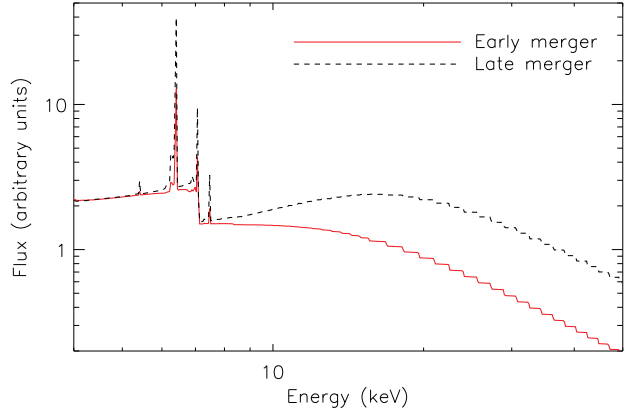}
% %% caption
 \begin{minipage}{0.49\textwidth}
  \caption{Average best-fit spectral model for the AGN in galaxies in early (red continuous line) and late (black dashed line) merger stage. Each individual spectral model was normalised to have the same flux in the 4--50\,keV range. The two average spectral models were then normalised to have the same flux in the 4--5\,keV band for visualization purposes. The figure clearly illustrates the different mean spectra of the two subsamples, with AGN in late-merger galaxies showing more prominent features of heavy absorption (i.e. Fe\,K$\alpha$ line at 6.4\,keV and Compton hump at $\sim 20-30$\,keV) than those in early-merger galaxies.}
\label{fig:averagespectra}
 \end{minipage}
\end{figure}

\begin{figure}
\centering
\includegraphics[width=0.49\textwidth]{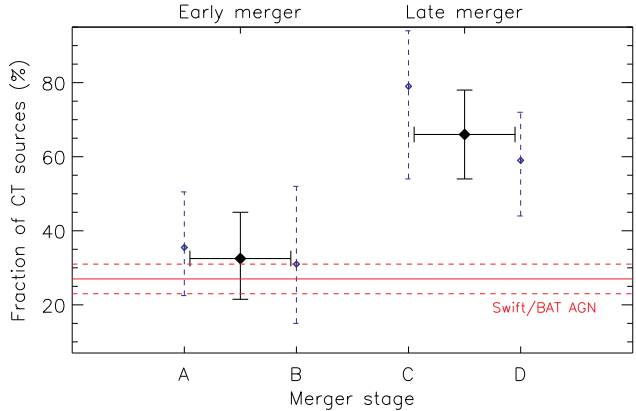} 
\caption{
Fraction of Compton-thick (CT, i.e., $N_{\rm\,H}\geq 10^{24}\rm\,cm^{-2}$) AGN versus merger stage. The empty blue diamonds represent the values for the four merger stages separately, while the filled black diamonds are the values for early and late mergers. The uncertainties represent the 16th/84th quantiles of a binomial distribution. The red continuous line represents the intrinsic fraction of CT AGN measured by {\it Swift}/BAT \citep{Ricci:2015tg} for local hard X-ray selected AGN, while the dashed lines are the 1$\sigma$ uncertainty associated with this value. The figure shows that AGN in the late stages of mergers are more likely to be CT than those in isolated galaxies, which confirms the idea that the amount of dense material around the SMBH is larger in advanced mergers.
}
\label{fig:CTmergerstatge}
\end{figure}

The broad-band X-ray spectral analysis of the objects in our sample shows that AGN in advanced merger stages are typically more obscured than AGN in isolated galaxies or in galaxies in the early stages of mergers (Fig.\,\ref{fig:cdf}). In Fig.\,\ref{fig:averagespectra} we illustrate the average normalised spectral models of AGN in galaxies in early (red continuous line) and late (black dashed line) merger stage. The image clearly illustrates that the main features indicating heavy obscuration, the Fe\,K$\alpha$ line and Compton hump, are stronger in AGN in late merger stage galaxies. We find that AGN in late-merger stage galaxies host a significantly ($99\%$ confidence) larger fraction of CT AGN ($65^{+12}_{-13}\%$) compared to local hard X-ray selected AGN ($27\pm 4\%$, \citealp{Ricci:2015tg}; see also \citealp{Burlon:2011dk}), which are mostly found in non-merging systems\footnote{Only $8\pm2\%$ of {\it Swift}/BAT AGN are found in close mergers \citep{Koss:2010rc}.}. The fraction of CT AGN in early-merger stage galaxies ($35^{+13}_{-12}\%$) is marginally lower than in late-merger stage galaxies, and is consistent with that of local hard X-ray selected AGN (Fig.\,\ref{fig:CTmergerstatge}). All the uncertainties reported on the fractions of AGN represent the 16th and 84th quantiles of a binomial distribution, computed with the beta function \citep{Cameron:2011cl}.

\begin{figure}
\centering
\includegraphics[width=0.49\textwidth]{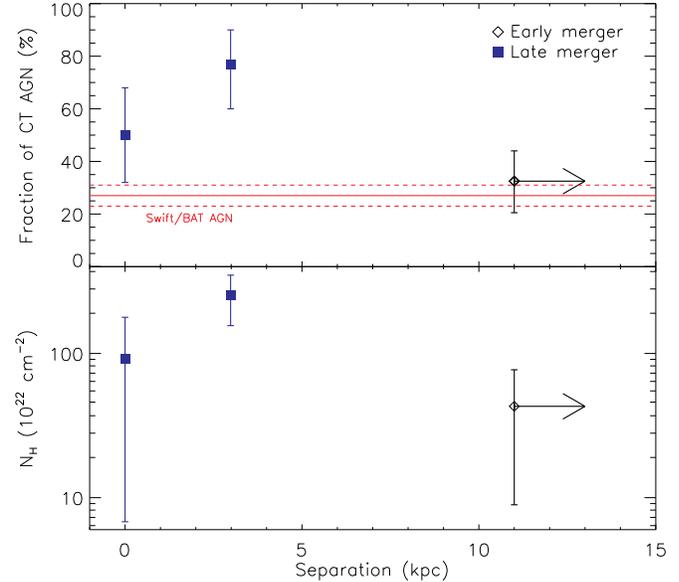} 
\caption{{\it Top panel:} fraction of CT AGN versus the separation between the two galactic nuclei. U/LIRGs in the late stages of galaxy mergers are divided into objects with two nuclei and those with a single nucleus. The separation for U/LIRGs in late merger stages showing two nuclei is set to the mean value of the distance, while for objects with a single nucleus we set it to zero kpc. The red continuous line represents the intrinsic fraction of CT AGN measured by {\it Swift}/BAT \citep{Ricci:2015tg} for local hard X-ray selected AGN. The uncertainties on $f_{\rm\,CT}$ represent the 16th/84th quantiles of a binomial distribution. {\it Bottom panel:} median value of the line-of-sight column density versus the separation between the two galactic nuclei. The error bars on $N_{\rm\,H}$ show the 1$\sigma$ interval. The three objects (NGC\,4922N, Arp 220W and NGC\,7674) for which only a lower limit of $N_{\rm\,H}$ could be constrained were assigned the minimum value. The two plots show that nuclear obscuration seems to peak when the two nuclei are at a distance of $0.4-10.8$\,kpc.
}
\label{fig:ctdistatnce}
\end{figure}

Dividing the sample based on the distance between the nuclei of the merging galaxies (top panel of Fig.\,\ref{fig:ctdistatnce}), we find that the fraction of CT AGN peaks ($77_{-17}^{+13}\%$) when the sources are in late merger stages and at a projected distance of 0.4--10.8 kiloparsecs. The median column density (bottom panel of Fig.\,\ref{fig:ctdistatnce}) also reaches its maximum value at this stage [$N_{\rm\,H}=(2.53\pm0.97)\times 10^{\rm\,24}\rm\,cm^{-2}$], and is larger than for early mergers at a distance $\geq11$\,kpc [$N_{\rm\,H}=(4.3\pm3.4)\times 10^{\rm\,23}\rm\,cm^{-2}$]. 
In agreement with these X-ray based results, a further four nuclei (the NW and SE nuclei of IRAS\,08572+3915, NGC\,3690E and IRAS\,14378$-$3651) in the sample have indications for the presence of accreting SMBHs from other multi-wavelength tracers, although they are only weakly or not detected by {\it NuSTAR} (and are therefore very likely heavily obscured). All are in late stage of mergers, with three of them being in systems in which the two galactic nuclei are separated by a few kiloparsecs (see Appendix\,\ref{SM_individual}).

The increase of the fraction of CT AGN along the merger sequence is in strong agreement with the long-standing paradigm that galaxy mergers are able to trigger the inflow of material onto the close environment of SMBHs. Recent numerical simulations of mergers \citep{Blecha:2016aa} show that an increase in $N_{\rm\,H}$ caused by the merger dynamics is expected as the distance between the two nuclei decreases. Because these simulations can only probe column densities on resolved scales ($\gtrsim 48$\,pc), they provide a lower limit on the total line-of-sight obscuration. On scales of $\gtrsim 48$\,pc, the median value of $N_{\rm\,H}$ predicted by the simulations ($N_{\rm\,H}\simeq 3\times10^{23}\rm\,cm^{-2}$) is significantly lower than what we found here, and the simulations shows that AGN can be obscured by CT material only for a very brief phase around coalescence. This could imply that the material responsible for most of the obscuration is located on scales smaller than those probed by the simulations (i.e., $\lesssim 48$\,pc).
Our findings are consistent with recent observations carried out in the submm band, which found evidence of Compact Obscuring Nuclei (CON; \citealp{Aalto:2015rw}) in U/LIRGS (e.g., \citealp{Sakamoto:2010db,Costagliola:2013ul,Martin:2016dq}). These CON typically have sizes of tens of parsecs (e.g., \citealp{Aalto:2015rw,Scoville:2015fy}), and the high column densities ($N_{\rm\,H}>10^{24}\rm\,cm^{-2}$) inferred from the submm observations may be responsible for the bulk of the obscuration traced by the X-rays.

\paragraph*{}
\label{sec:discussion}

According to the classical unification scheme of AGN \citep{Antonucci:1993fu}, obscuration is only due to anisotropic material distributed in the form of a torus, and the sole difference between obscured and unobscured objects is the inclination of the observer with respect to the torus. The fraction of CT AGN in mergers vastly exceeds that expected considering a random viewing angle with respect to the torus, and this enhanced obscuration is indicative of additional material on pc-scales. Moreover, as illustrated in Fig.\,\ref{fig:cdf}, AGN in late mergers are almost completely surrounded by material with $N_{\rm\,H}\geq 10^{23}\rm\,cm^{-2}$, which covers $95^{+4}_{-8}\%$ of the solid angle of the accreting SMBH\footnote{The covering factor was calculated following the approach described in \citet{Cameron:2011cl}, and the uncertainties represent the 16th and 84th quantiles of a binomial distribution.}. As a comparison, the covering factor of material with $\log (N_{\rm\,H}/\rm cm^{-2})=23-25$ in {\it Swift}/BAT AGN is $52\pm2\%$ \citep{Ricci:2015tg}. This implies that the classical unification scenario is not sufficient to describe the structure of obscuration in galaxies undergoing mergers. 
Our results can be interpreted in the framework of the AGN evolutionary model first proposed by \citet{Sanders:1988ys} for U/LIRGS. This scenario starts with a merger of galaxies, which triggers accretion and heavy obscuration, along with strong starbursts, and is then followed by a phase in which feedback from the AGN removes the obscuring material and the source can be observed as a red quasar (e.g., \citealp{Urrutia:2008qq,Banerji:2012mz,LaMassa:2016ly}). The object eventually ends its life as an unobscured quasar (e.g., \citealp{Treister:2010lo}). Here we find a hint of a decrease in the fraction of CT AGN and in the median $N_{\rm\,H}$ with the separation (for $D_{12}<11$\,kpc). In particular closely-separated nuclei seem to have a larger fraction of CT AGN than objects in the single-nuclei phase, which in the evolutionary scheme would be the stage when the AGN is starting to clean up its environment. This behaviour, if confirmed, would also be in agreement with recent numerical simulations \citep{Blecha:2016aa}. 

Several studies over the past years have shown that the fraction of AGN in mergers increases with the AGN luminosity (e.g., \citealp{Treister:2012kk,Glikman:2015lk}), reaching $\sim 40-80\%$ at bolometric luminosities of $\log (L_{\rm\,Bol}/\rm erg\,s^{-1})\sim 46-48$, in agreement with the idea that mergers of galaxies are able to effectively fuel SMBHs. This picture is consistent with what has been recently found for mid-IR selected Hot Dust-obscured galaxies (e.g., \citealp{Eisenhardt:2012ve,Wu:2012bh,Assef:2015zr}), which host the most luminous obscured AGN known (e.g., \citealp{Stern:2014kx,Piconcelli:2015uq,Assef:2016qf,Ricci:2017xe}), show an excess of submillimiters galaxies as neighbours \citep{Jones:2015lq}, and tend to be found in mergers of galaxies \citep{Fan:2016sf}. Some recent studies of galaxies at $z\sim 0.6-0.7$ (e.g., \citealp{Villforth:2014sf,Villforth:2016mz}) have argued that the fraction of AGN in mergers is significantly lower, and it does not increase with the AGN bolometric luminosity. However, the samples used for these studies were selected in the soft X-ray band, which is strongly biased against heavily obscured sources. In particular, \cite{Villforth:2016mz} used a {\it ROSAT}-selected sample, which, in the rest-frame of the sources they considered, corresponds to a selection in the $\sim 0.2-3.8$\,keV band. Our work shows that the circumnuclear environment of AGN in mergers is different (i.e. richer in gas and dust) from that of AGN in isolated galaxies, which implies that soft X-ray selection would fail to detect most of the AGN in mergers. Considering the median $N_{\rm\,H}$ of sources in advanced stages of merger with two distinct nuclei [$N_{\rm\,H}=(2.53\pm0.97)\times 10^{\rm\,24}\rm\,cm^{-2}$], which would be the sources easier to distinguish as mergers at $z\sim 0.6$, the AGN would lose $\sim 99.9\%$ of the $\sim 0.2-3.8$\,keV flux.

\cite{Treister:2012kk} showed that, while 90\% of AGN are triggered by secular processes (i.e., stellar bars, supernova winds, etc.), $\sim 50-60\%$ of the SMBH growth is caused by mergers. Since mergers are predicted to be ubiquitous and play a fundamental role in the formation and evolution of galaxies \citep{Springel:2005eq}, accretion triggered by tidal torques could contribute significantly to the growth of SMBHs, and could produce the observed relation between the mass of the SMBH and the velocity dispersion of the galaxy bulge (e.g., \citealp{Di-Matteo:2005qv,Blecha:2011pb}). With the {\it NuSTAR} observations analysed here, we show that mergers of galaxies are able to drive material into the proximity of the SMBH, thus strengthening the idea that interactions between galaxies are critical to understand the link between accreting SMBHs and their hosts.

\section{Summary and conclusions}\label{sect:summary}

In this work we studied the relation between AGN obscuration and galaxy mergers. This was done analysing the broad-band X-ray emission of a sample of 30 luminous and ultra-luminous IR galaxies in different merger stages from the GOALS sample (for a total of 52 individual galactic nuclei). In a forthcoming paper (Ricci et al. in prep.) we will discuss the multi-wavelength properties of the U/LIRGs of our sample, and how they relate to the bolometric luminosity of the AGN. In the following we summarise our findings.

\begin{itemize}

\item All AGN of our sample in late merger galaxies have $N_{\rm\,H}\geq 10^{23}\rm\,cm^{-2}$, which implies that the obscuring material covers $95^{+4}_{-8}\%$ of the X-ray source (see Fig.\,\ref{fig:cdf}). The close environment of these objects is therefore very different from what would be foreseen by the unification model of AGN.

\item The fraction of CT AGN in late merger galaxies is higher ($f_{\rm\,CT}=65^{+12}_{-13}\%$) than in local hard X-ray selected AGN ($f_{\rm\,CT}=27\pm 4\%$, \citealp{Ricci:2015tg}), which are mostly found in isolated galaxies, and marginally higher than AGN in early merger galaxies (see Figs.\,\ref{fig:averagespectra} and \ref{fig:CTmergerstatge}).

\item A peak in the fraction of CT AGN is found when the nuclei of the two merging galaxies are at a projected distance of $\simeq0.4-10.8$\, kiloparsecs ($f_{\rm\,CT}=77_{-17}^{+13}\%$, see Fig.\,\ref{fig:ctdistatnce}). This is also the stage at which the maximum of the median $N_{\rm\,H}$ is observed [$N_{\rm\,H}=(2.53\pm0.97)\times 10^{\rm\,24}\rm\,cm^{-2}$]. We also find a hint of a decrease both in $f_{\rm\,CT}$ and in the median $N_{\rm\,H}$ when only a single nucleus is observed. If confirmed, this decrease of the obscuring material might be related to feedback from the AGN clearing out its environment.

\end{itemize}

Our results confirm the long-standing idea that galaxy mergers are able to trigger the inflow of material onto the inner tens of parsecs, and that the close environment of AGN in late-merger galaxies is richer in gas and dust as compared to AGN in isolated galaxies.

\appendix

\section{Notes on individual sources}
\label{SM_individual}

In the following we report the details on the individual sources, including previous literature studies on the presence of an accreting SMBH using multi-wavelength tracers of AGN activity. The proxies of AGN activity we use are the following:
\begin{itemize}
\item The presence of high-excitation mid-IR (MIR, 5--40\,$\mu$m) emission lines (e.g., \citealp{Sturm:2002tw}), and in particular [Ne\,V]\,$14.32\mu$m and [Ne\,V]\,$24.32\mu$m, indicate of the presence of an AGN, since the ionization potential of [Ne\,V] is 97\,eV, which is considered too high to be produced by star formation (e.g., \citealp{Weedman:2005cr}).
\item The slope of the near IR continuum, with a very red continuum ($\Gamma>1$, with $F_{\nu}\propto \lambda^{\Gamma}$, \citealp{Imanishi:2010uq}) suggesting the presence of a buried AGN \citep{Risaliti:2006kx,Sani:2008wq,Imanishi:2008ly}. Similarly, we also used the {\it Wide-field Infrared Survey Explorer} satellite ({\it WISE}, \citealp{Wright:2010fk}) colours, adopting $W1-W2>0.8$ as threshold for the presence of an AGN \citep{Stern:2012fk}. It should be however remarked that this tracer can be problematic for low-luminosity AGN in strongly star-forming galaxies (e.g., \citealp{Griffith:2011fj,Hainline:2016nr}).
\item The equivalent width (EW) of the 3.3$\mu$m and 6.2$\mu$m polycyclic aromatic hydrocarbon (PAH) features. PAH features are destroyed by the radiation field of the AGN or diluted by the strong MIR continuum, thus low values of the EW typically indicate the presence of an AGN [EW$(3.3\mu\rm m)<40$\,nm, \citealp{Imanishi:2010uq,Ichikawa:2014db}; EW$(6.2\mu\rm m)<0.27\,\mu$m, \citealp{Stierwalt:2013eu}].
\item Optical line ratios. The U/LIRGs of our sample are optically classified as objects dominated by star formation in the optical band (i.e. HII-regions; e.g., \citealp{Ho:1997kk}), objects that might contain both star-formation and AGN activity (i.e., composite objects; e.g. \citealp{Yuan:2010ye}), and objects that are clearly dominated by AGN activity (i.e., Seyfert\,1s or Seyfert\,2s).
\item The spatial variations of the radio spectral index (i.e., radio-spectral-index maps; \citealp{Vardoulaki:2015mi}) allow objects to be classified as radio-AGN,
composite AGN/starburst (SB), and radio-starburst (radio-SB).
 \end{itemize}

%################################################################################################################

\subsection{IRAS\,F00085$-$1223 (NGC\,34)}\label{sect:ngc34}

The LIRG NGC\,34 is in merger stage\,\,D, and shows a single nucleus (Fig.\,\ref{fig:images}). The object is reported to be a Seyfert\,2 by \citet{Yuan:2010ye}. Using the radio spectral index maps, \citet{Vardoulaki:2015mi} also confirmed the presence of an AGN.

The source is detected by {\it NuSTAR} in the 3--10 keV (10-24\,keV) band at 5.7$\sigma$ (6.3$\sigma$) and 6.8$\sigma$ (4.4$\sigma$) for FPMA and FPMB, respectively. The source is also detected by {\it Chandra} and by {\it XMM-Newton}. The X-ray spectrum of NGC\,34 (Fig.\,\ref{fig:spectra1}) shows a clear Fe K$\alpha$ feature at $6.48^{+0.06}_{-0.05}$\,keV, which also indicates the presence of an AGN. We therefore applied a model including both thermal emission and an AGN component (\textsc{atable\{torus1006.fits\} + apec + zpowerlaw}). We added a cross-calibration constant between the different, non-simultaneous, observations. We found that the flux varied by a factor $\sim$2 between the {\it NuSTAR}, {\it Chandra} and {\it XMM-Newton} observations. Our spectral analysis shows that the AGN is heavily obscured, with a column density of $N_{\rm\,H}=5.3\pm1.1\times 10^{23}\rm\,cm^{-2}$. The value of the half-opening angle could not be constrained for this observation, and was therefore fixed to $\theta_{\rm\,OA}=60$\,deg. The scattered radiation has a flux of $6.2^{+7.9}_{-4.3}\%$ of the primary X-ray emission in the X-ray energy band probed here.

%################################################################################################################

\subsection{IRAS\,F00163$-$1039 (Arp\,256 \& MCG$-$02$-$01$-$052)}\label{sect:IRASF00163-1039}

The two objects are in stage\,\,B, and their projected separation is 33.1\,kpc (Fig.\,\ref{fig:images}). While Arp\,256 is a LIRG, the IR luminosity of MCG$-$02$-$01$-$052 is $\log (L_{\rm\,IR}/L_{\odot})=10.36$. Optically, Arp\,256 is classified by \citet{Yuan:2010ye} as a HII-region, and none of the tracers detect an AGN in Arp\,256 or MCG$-$02$-$01$-$052.

The two galactic nuclei are at a distance (56.1\arcsec) that allows them to be well separated by {\it Chandra}. While the two galaxies are detected by {\it Chandra}, they are not detected by {\it NuSTAR}, with both nuclei only exhibiting significances of $\sim 3\sigma$ in the 3--10\,keV band and of $<2\sigma$ in the 10--24\,keV band. The {\it Chandra} spectra of both sources are soft, and can be well fit by a starburst model. For Arp\,256 a thermal plasma and a power law component, both of which appear obscured [\textsc{zphabs$\times$(zpo+apec)}], are necessary to best reproduce the spectrum. The same model, with the exception of the absorbing component, was used for MCG$-$02$-$01$-$052. The spectra and the fit are shown in Fig.\,\ref{fig:spectra1}.

%################################################################################################################

\subsection{IRAS\,F00506+7248 (MCG+12$-$02$-$001)}\label{sect:IRASF00506+7248}

This LIRG is in merger stage\,\,C, and the projected separation of the two nuclei is 0.3\,kpc (Fig.\,\ref{fig:images}). The system is classified in the optical as a composite \citep{Alonso-Herrero:2009kl}, and none of the tracers analysed suggest the presence of an AGN. 

While the system is detected by {\it Chandra}, it is not detected by {\it NuSTAR}, and the significance is $<3\sigma$ for each camera in the 3--10\,keV and 10--24\,keV band. Given the very small separation between the two nuclei (0.9\arcsec) the X-ray emission observed by {\it Chandra} could come from any of the two nuclei, or from both of them. The {\it Chandra} spectrum was fitted with a simple absorbed power law model (\textsc{zphabs$\times$zpowerlaw}). Using the relation of \citet{Ranalli:2003kb}, we find that the star formation rate obtained for the source ($\sim 54.5\rm\,M_{\odot}\,yr^{-1}$, \citealp{Howell:2010ty}) is able to account for the totality of the 2--10\,keV luminosity observed. The steep photon index obtained ($\Gamma \sim 3 $) also suggests that the X-ray emission is related to star formation. The spectrum of the source is illustrated in Fig.\,\ref{fig:spectra1}.

%################################################################################################################

\subsection{IRAS\,F02069$-$1022 (NGC\,833 \& NGC\,835)}\label{sect:IRASF02069-1022}

IRAS\,F02069$-$1022 (also called Arp\,318) is composed of two galaxies at a projected distance of 15.7\,kpc and is in merger stage\,\,A (Fig.\,\ref{fig:images}). The multi-wavelength tracers fail to detect evidence of AGN activity, with the exception of optical spectroscopy for NGC\,835, which is classified as a Seyfert\,2 \citep{Veron-Cetty:2010tg}. NGC\,835 hosts a low-luminosity AGN which appears responsible for most of the 2--10\,keV emission \citep{Gonzalez-Martin:2016mb}. NGC\,833 is classified as a LINER \citep{Veron-Cetty:2010tg}, but it has been found to be AGN dominated, with a very low star-formation rate ($\lesssim 3\rm\,M_{\odot}\,yr^{-1}$) over the past few hundreds of Myr \citep{OSullivan:2014ng}. 

The two galaxies are separated by 55.9\arcsec, and could be therefore well resolved by the {\it NuSTAR}, {\it XMM-Newton}/EPIC and {\it Chandra} observations studied by \cite{Oda:2017aa}, who found that both sources are strongly detected by {\it NuSTAR}. \cite{Oda:2017aa} found that the analysis of the combined {\it XMM-Newton} EPIC, {\it Chandra}/ACIS and {\it NuSTAR} spectra of NGC\,833 results in a column density of $N_{\rm\,H}=2.8\pm0.3\times 10^{23}\rm\,cm^{-2}$. Variability of the line-of-sight column density between the {\it Chandra} and {\it XMM-Newton} observations of NGC\,835 was found by \citet{Gonzalez-Martin:2016mb}. \cite{Oda:2017aa} find that the {\it NuSTAR} spectrum is very different from the {\it XMM-Newton} observation carried out in January 2000, while it is consistent with the spectrum of the more recent (July 2013) {\it Chandra} observation. Fitting separately the combined {\it Chandra}/{\it NuSTAR} spectra and the {\it XMM-Newton} spectrum, \cite{Oda:2017aa} found that the column density varies from $N_{\rm\,H}^a=5.5^{+1.5}_{-1.0}\times 10^{23}\rm\,cm^{-2}$ ({\it XMM-Newton}) to $N_{\rm\,H}^b=3.0\pm0.2\times 10^{23}\rm\,cm^{-2}$ ({\it Chandra}/{\it NuSTAR}). Through the rest of the paper we use the mean value of the column density for this source ($N_{\rm\,H}^{\rm mean}=4.3^{+1.5}_{-1.0}\times 10^{23}\rm\,cm^{-2}$).

%################################################################################################################

\subsection{IRAS\,F05054+1718 (CGCG 468$-$002E \& CGCG 468$-$002W)}\label{sect:IRASF05054+1718}

This system in merger stage\,\,B is composed of an eastern (CGCG\,468$-$002E) and a western (CGCG\,468$-$002W) component (Fig.\,\ref{fig:images}). While the eastern object is a LIRG, the western component has a luminosity in the IR of $\log (L_{\rm\,IR}/L_{\odot})=10.74$. The western component shows a [Ne\,V] line at 14.32\,$\mu$m \citep{Inami:2013il}, which indicates the presence of an AGN. From the ratio of the [Ne\,V]/[Ne\,II] and [O\,IV]/[Ne\,II] lines the AGN is found to contribute to $\sim 25-30\%$ of the MIR emission \citep{Petric:2011zt}. The western component also shows a weak (EW$=0.12\mu$m) PAH 6.2$\mu$m feature, indicative of a significant AGN contribution to the MIR flux. None of the proxies of AGN activity find evidence of an AGN in the eastern component.

The two galaxies are separated by $29.6\arcsec$, and could therefore be distinguished both by the {\it Swift}/XRT and the {\it NuSTAR} observations studied here.
CGCG\,468$-$002W is clearly detected by {\it NuSTAR} in both detectors ($\sim 51\sigma$ in 0.3--10\,kev and $\sim 31\sigma$ in 10--24\,keV for both FPMA and FPMB). The source is also detected by {\it Swift}/XRT in the 0.3--10\,keV band. The {\it Swift}/XRT spectrum of this source is comprised from the integration of several observations carried out within a month in late 2014. The model we used to fit the spectra is similar to that used for NGC\,34, with the exception of the thermal plasma, which is not required by the data. We also added a line at $6.47\pm0.07$\,keV (EW$=211^{+61}_{-66}$\,eV) to account for the observed excess at $\sim 6.4\,$keV. The model we used is \textsc{atable\{torus1006.fits\} + zgauss + zpowerlaw}, and shows that the level of obscuration of the source is low [$N_{\rm\,H}= (1.50\pm0.09)\times 10^{22}\rm\,cm^{-2}$]. A cross-calibration constant was also added, and showed that the flux of {\it NuSTAR} is $\sim 1.5$ times higher than that of the XRT observation. The ratio between the scattered power-law and the primary X-ray emission is $1.5\pm0.5\%$. The broad-band spectrum is illustrated in Fig.\,\ref{fig:spectra1}.
CGCG\,468$-$002E was detected by neither {\it Swift}/XRT (2--8\,keV) nor by {\it NuSTAR} (10--24\,keV).

%################################################################################################################

\subsection{IRAS\,05189$-$2524}

This merger stage\,\,D ULIRG shows a single nucleus (Fig.\,\ref{fig:images}), which has been found to host an AGN using several proxies. The source shows strong 14.32\,$\mu$m \citep{Inami:2013il} and 24.32\,$\mu$m \citep{Pereira-Santaella:2010kx} [Ne\,V] lines, and very weak PAH features at 3.3$\mu$m (EW$=10$\,nm, \citealp{Imanishi:2008ly}) and at 6.2$\mu$m (EW$=0.03$\,$\mu$m, \citealp{Stierwalt:2013eu}). Optically the source is classified as a Seyfert\,2 \citep{Yuan:2010ye}. The [Ne\,V]/[Ne\,II] and [O\,IV]/[Ne\,II] ratios indicate that the AGN contributes $\sim 50\%$ of the MIR flux.

Previous studies in the X-ray band have also confirmed the presence of an AGN. The source has been detected by {\it XMM-Newton} \citep{Imanishi:2004bx,Teng:2010df}, by {\it Chandra} \citep{Ptak:2003to}, and by the XIS camera on board {\it Suzaku} \citep{Teng:2009sh}. These X-ray spectral studies have shown that the AGN is obscured, with a line-of-sight column density $\sim 10^{23}\rm\,cm^{-2}$. The source has also been detected by {\it Swift}/BAT \citep{Koss:2013pi}.
IRAS\,05189$-$2524 was detected by {\it NuSTAR} in three observations carried out in 2013 for a total of $\sim 54$\,ks \citep{Teng:2015vn}. Analyzing the combined {\it XMM-Newton} and {\it NuSTAR} spectrum, \citet{Teng:2015vn} showed that the obscuration can be explained by considering two absorbers, with column densities of $5.2\pm0.2$ and $9.3^{+1.0}_{-0.7}\times 10^{22}\rm\,cm^{-2}$, covering $\simeq 98\%$ and $\simeq 74\%$ of the X-ray source, respectively. To account for the partial covering, for this object we considered $N_{\rm\,H}$ to be the sum of the two column densities weighted by the covering factor.

%################################################################################################################

\subsection{IRAS\,08572+3915}

IRAS\,08572+3915 is a ULIRG in a merger stage\,\,D showing two distinct nuclei (northwest and southeast), located at a distance of 5.6\,kpc (Fig.\,\ref{fig:images2}). While no [Ne\,V] emission line is detected \citep{Pereira-Santaella:2010kx,Inami:2013il}, the system shows a very strong silicate absorption feature at $9.7\mu$m ($\tau \sim -3.58$) and weak PAH features at 3.3$\mu$m (EW$<5$\,nm, \citealp{Imanishi:2008ly}) and at 6.2$\mu$m (EW$<0.03$\,$\mu$m, \citealp{Stierwalt:2013eu}). Using the radio spectral index, \citet{Vardoulaki:2015mi} also found results consistent with the presence of an AGN. Both the northwest (NW) and the southeast (SE) nuclei are classified as Seyfert\,2s by \citet{Yuan:2010ye}.

The {\it NuSTAR} observation of this object did not yield significant detections of either unresolved nuclei \citep{Teng:2015vn}. The source was also not detected by a previous {\it Suzaku} observation \citep{Teng:2009sh}. \citet{Teng:2015vn} argued that the source could be X-ray weak, although it cannot be excluded that it is heavily obscured, similar to what we find here for most of the sources showing two nuclei with a separation of a few kpc. Using CO, \citet{Evans:2002xz} estimated the average column density to be in the range of $N_{\rm\,H}\sim (3-10)\times 10^{24}$ for this system. The NW nucleus was detected by {\it Chandra}, and \citet{Teng:2009sh} reported an observed 0.5--10\,keV luminosity of $\sim 2\times 10^{41}\rm\,erg\,s^{-1}$. The hardness ratios points to a photon index of $\Gamma \sim -0.43$, also indicative of heavy obscuration. We conclude therefore that both nuclei are likely to host heavily obscured AGN, but do not use these sources since an estimation of $N_{\rm\,H}$ is lacking.

%################################################################################################################

\subsection{IRAS\,F09320+6134 (UGC\,5101)}

UGC\,5101 is a ULIRG with a single nucleus, classified as merger stage\,\,D (Fig.\,\ref{fig:images2}). Multi-wavelength studies have shown strong evidence for the presence of an AGN, such as 14.32$\mu$m and 24.32$\mu$m [Ne\,V] emission lines \citep{Pereira-Santaella:2010kx,Inami:2013il}, and weak 3.3$\mu$m (EW$=33$\,nm, \citealp{Imanishi:2008ly}) and 6.2$\mu$m (EW$=0.13$\,$\mu$m, \citealp{Stierwalt:2013eu}) PAH features. The {\it WISE} $W1-W2$ colour of UGC\,5101 is $1.697$ ($W2=8.34$\,mag), consistent with a significant AGN MIR emission \citep{Stern:2012fk}. In the optical the source is reported as being a Seyfert\,2 \citep{Yuan:2010ye}. 

{\it Chandra} and {\it XMM-Newton} observations of this object were studied by \citet{Imanishi:2003kr}, who found that the AGN is CT. UGC\,5101 was also reported as CT by \citet{Ricci:2015tg} analyzing {\it XMM-Newton} and {\it Swift}/BAT data. This in agreement with the depth of the 9.7$\mu$m silicate feature ($\tau_{9.7\mu\rm\,m}$=-0.78, \citealp{Stierwalt:2013eu}), which could also suggest the presence of a buried AGN. A recent study carried out by \citet{Oda:2016aa}, who studied in detail observations of this object carried out by {\it Chandra}, {\it XMM-Newton}, {\it Suzaku} and {\it NuSTAR}, confirmed the CT obscuration of the source ($N_{\rm\,H} =1.32^{+0.32}_{-0.37}\times10^{24}\rm\,cm^{-2}$).

%################################################################################################################

\subsection{IRAS\,F09333+4841 (MCG+08$-$18$-$013  \& MCG+08$-$18$-$012)}\label{sect:IRASF09333+4841}

These two merger stage\,\,A galaxies have a projected separation of 35.4\,kpc (Fig.\,\ref{fig:images2}). While MCG+08$-$18$-$013 is classified as a LIRG, the IR luminosity of MCG+08$-$18$-$012 is $\log (L_{\rm\,IR}/L_{\odot})=9.93$. MCG+08$-$18$-$013 is classified as a composite galaxy by \citet{Yuan:2010ye}, and none of the multi-wavelength tracers identify an AGN in the system. 

The two galaxies are separated by 66.5\arcsec, and could therefore be clearly resolved by both {\it Chandra} and {\it NuSTAR}.
In each of the 0.3--2\,keV and 3--8\,keV {\it Chandra} images, only MCG+08$-$18$-$013 is detected, showing extended emission in both bands. Neither nucleus is detected by {\it NuSTAR} in the 3--10\,keV or 3--24\,keV bands. The X-ray spectral analysis was carried out using the {\it Chandra} data alone. The X-ray spectrum of MCG+08$-$18$-$013 was fitted with an obscured power-law (\textsc{zphabs$\times$ po}) model. We found that the level of obscuration is low ($N_{\rm\,H}\sim 3\times 10^{21}\rm\,cm^{-2}$) and the X-ray continuum is rather steep ($\Gamma\sim 2.3$). No additional thermal component is necessary, and the X-ray spectrum is fully consistent with a scenario in which star formation is the only mechanism producing X-ray emission. The X-ray spectrum of MCG+08$-$18$-$013 is shown in Fig.\,\ref{fig:spectra1}.

%################################################################################################################

\subsection{IRAS\,F10015$-$0614 (NGC\,3110 \& MCG$-$01$-$26$-$013)}\label{sect:F10015-0614}

This system includes two galaxies in a merger stage\,\,A, with a projected separation of 37.7\,kpc (Fig.\,\ref{fig:images2}). Of the two galaxies, only NGC\,3110 is a LIRG. NGC\,3110 is classified as a HII-region by \citet{Yuan:2010ye}, and no trace of an AGN is evident for NGC\,3110 and MCG$-$01$-$26$-$013 from any of the multi-wavelength properties.

The two galaxies are separated by 108.9\arcsec, and could therefore be clearly resolved by both {\it Chandra} and {\it NuSTAR}.
Neither nucleus is detected by {\it NuSTAR} in the 3--10\,keV or 10--24\,keV band. The combined  {\it Chandra}/ACIS and {\it XMM-Newton}/EPIC X-ray spectrum of NGC\,3110 could be well fit by an absorbed collisionally ionised plasma plus a power law [\textsc{zphabs*(zpowerlaw+apec)}]. We applied a similar model to MCG$-$01$-$26$-$013, although this object did not require the presence of absorbing material along the line of sight (\textsc{zpowerlaw+apec}). For both sources the X-ray spectral analysis does not require the presence of an AGN. The X-ray spectra of these two sources are illustrated in Fig.\,\ref{fig:spectra2}.

%################################################################################################################

\subsection{IRAS\,F10257$-$4339 (NGC\,3256)}

NGC\,3256 is a LIRG in stage\,\,D (Fig.\,\ref{fig:images2}), with two nuclei detected in the radio \citep{Norris:1995ys} with a projected separation of 1\,kpc. None of the multi-wavelength proxies of AGN activity indicate the presence of an accreting SMBH, and the optical spectrum has been classified as that of an HII-region \citep{Lipari:2000ij}.

The {\it Chandra} spectrum was studied by \citet{Lira:2002fk} who could detect both nuclei (separated by 5.1\arcsec), and found X-ray emission consistent with being purely from star formation. The spectrum is in fact very soft, and could be modelled by the superposition of three thermal plasma components or by a steep power law with a photon index $\Gamma\sim 3$. NGC\,3256 was observed by {\it NuSTAR}, and the results were reported in \citet{Lehmer:2015ys}, who concluded that the emission at $>3$\,keV detected by {\it NuSTAR} is produced by a population of 5--10 ultra-luminous X-ray sources. 

%################################################################################################################

\subsection{IRAS\,10565+2448}

IRAS\,10565+2448 is a ULIRG in stage\,\,D (Fig.\,\ref{fig:images2}) showing two nuclei (east and west components) with a projected separation of 6.7\,kpc \citep{Scoville:2000zr}. The west component was classified as a composite galaxy in the optical by \citet{Yuan:2010ye}, and no evidence of AGN activity has been found from the multi-wavelength properties of the object. 

The two nuclei are separated by 7.4\arcsec, and {\it Chandra} observations could be able to resolve both of them. The source was detected in the X-rays by {\it Chandra} and {\it XMM-Newton} \citep{Teng:2010df,Iwasawa:2011fk}, and the spectra were found to be consistent with a star-formation origin of the X-ray emission. The galaxy was not detected in a 25\,ks {\it NuSTAR} observation \citep{Teng:2015vn}.

%################################################################################################################

\subsection{IRAS\,F11257+5850 (Arp\,299)}

Arp\,299 is a LIRG in merger stage\,\,C, with two nuclei (NGC\,3690W and NGC\,3690E) at a projected separation of 4.6\,kpc (Fig.\,\ref{fig:images3}). The eastern nucleus is often confused with IC\,0694, which is a lenticular galaxy nearby. Both nuclei show signatures of AGN activity in different energy bands. NGC\,3690W has a weak 3.3$\mu$m PAH feature (EW$=16$\,nm; \citealp{Imanishi:2010uq}) and a red 2.5-5\,$\mu$m continuum ($\Gamma_{2.5-5\,\mu\rm m}=1.9$; \citealp{Imanishi:2010uq}). A red near-IR continuum is also found for NGC\,3690E ($\Gamma_{2.5-5\,\mu\rm m}=1.05$; \citealp{Imanishi:2010uq}). The {\it WISE} $W1-W2$ colours of both nuclei are also consistent with AGN activity, being $W1-W2\sim 1.5$ ($W2=6.26$\,mag) and $W1-W2\sim 1.0$ ($W2=8.28$\,mag) for NGC\,3690W and NGC\,3690E, respectively. In the MIR the 6.2$\mu$m PAH feature is rather weak for both NGC\,3690W (EW$=0.12\,\mu$m; \citealp{Stierwalt:2013eu}) and NGC\,3690E (EW$=0.38\,\mu$m; \citealp{Stierwalt:2013eu}), and indicates AGN contributions to the MIR luminosities of each nucleus of $\sim 75\%$ and $\sim 20\%$ \citep{Stierwalt:2013eu}, respectively. In the optical \citet{Yuan:2010ye} classified NGC\,3690W as a Seyfert\,2 and NGC\,3690E as a HII-region. 

The two nuclei are separated by 21.3\arcsec, and could be resolved in the X-ray band by several studies carried out in the past decade.
The presence of a buried AGN in NGC\,3690W was first found by \citet{Della-Ceca:2002wm} using {\it BeppoSAX} data. Combining {\it BeppoSAX} with {\it Chandra} and {\it XMM-Newton} observations, \citet{Ballo:2004im} confirmed the heavily obscured nature of the western component, detecting a prominent Fe K$\alpha$ line. \citet{Ballo:2004im} also argued for the presence of a CT AGN in NGC\,3690E, which shows a strong Fe\,XXV line at 6.7\,keV. \citet{Alonso-Herrero:2013cr}, using the CanariCam instrument on the 10.4\,m Gran Telescopio Canarias found evidence of AGN activity in both nuclei. For NGC\,3690E they estimated that the AGN is $\sim 5$ times less luminous than NGC\,3690W and the material surrounding the SMBH has an extinction of $A_{\rm\,V}\sim 24$\,mag.
More recently, studying the simultaneous {\it NuSTAR} and {\it Chandra} spectra of NGC\,3690W in the $3-40$\,keV range, \citet{Ptak:2015nx} found a column density of $N_{\rm\,H}\sim 4\times 10^{24}\rm\,cm^{-2}$. No evidence of X-ray emission above 10\,keV from NGC\,3690E was found, and \citet{Ptak:2015nx} concluded that the AGN is heavily obscured and/or significantly less luminous than NGC\,3690W. Both NGC\,3690W and NGC\,3690E show deep silicate features ($\tau_{9.7\mu\rm m}=-0.77$ and $-1.65$, respectively; \citealp{Stierwalt:2013eu}), consistent with a buried AGN scenario.

%################################################################################################################

\subsection{IRAS\,F12043$-$3140 (ESO\,440$-$IG058N \& ESO\,440$-$IG058S)}\label{sect:F12043-3140}

The two objects are in merger stage\,\,B, and their projected separation is 5.9\,kpc (Fig.\,\ref{fig:images3}). The northern source is a LIRG, while the southern source has an IR luminosity of $\log (L_{\rm\,IR}/L_{\odot})=10.54$. In the optical ESO\,440$-$IG058N has been classified as a HII-region, while ESO\,440$-$IG058S as a composite system \citep{Yuan:2010ye}. Both galaxies show strong $6.2\mu$m PAH features (EW$=0.56$ and 0.66\,$\mu$m for ESO\,440$-$IG058N \& ESO\,440$-$IG058S, respectively; \citealp{Stierwalt:2013eu}), and none of the tracers considered demonstrate the presence of an AGN.

The two sources are at 12\arcsec, and both have been detected by {\it Chandra}.
No source was detected by {\it NuSTAR} at the position of the system, and therefore the spectral analysis was carried out using only {\it Chandra}. The X-ray spectrum of ESO\,440$-$IG058N was fitted using a simple power law model, with no absorption required (\textsc{zpowerlaw}). The model used for ESO\,440$-$IG058S takes into account thermal emission and a power-law component (\textsc{zpowerlaw+apec}). Both sources are consistent with no AGN contribution in the X-ray band, and their X-ray spectra are shown in Fig.\,\ref{fig:spectra2}.

%################################################################################################################

\subsection{IRAS\,F12540+5708 (Mrk\,231)}

Mrk\,231 is an ULIRG with a single nucleus in merger stage\,\,D (Fig.\,\ref{fig:images3}), and notably is the nearest broad-absorption line quasar. While the MIR spectrum of the source does not show evidence of [Ne\,V] emission lines \citep{Pereira-Santaella:2010kx,Inami:2013il}, the presence of a luminous AGN is inferred from the weak 3.3$\mu$m (EW$=8$\,nm, \citealp{Imanishi:2010uq}) and 6.2$\mu$m (EW$=0.01$\,$\mu$m, \citealp{Stierwalt:2013eu}) PAH features and from the MIR colours ($W1-W2\sim 1.1$). In the optical band the source is classified as a Seyfert\,1 \citep{Yuan:2010ye}. Studying several different tracers of AGN emission, \citet{Veilleux:2009qo} found that the AGN contributes to $\sim 71\%$ of the bolometric output of the system. A lower value ($\sim 34\%$) was obtained by \citet{Nardini:2010dz} using spectral decomposition. Mrk\,231 is also known to have a strong kpc-scale outflow \citep{Rupke:2013tg,Feruglio:2015ek}, which has been interpreted as proof of quasar-mode feedback \citep{Feruglio:2010xe}.

The broad-band 0.5--30\,keV {\it NuSTAR} and {\it Chandra} spectrum of the source was studied by \citet{Teng:2014oq}, who found that the primary X-ray continuum is flat, and the X-ray source is obscured by Compton-thin material. \citet{Teng:2014oq} found that the source is X-ray weak, with the 2--10\,keV to bolometric luminosity ratio being $\sim 0.03\%$ (compared to a typical value for local Seyferts of $\sim5\%$, e.g., \citealp{Vasudevan:2007qt}). The values of the column density were found to vary between the 2003 {\it Chandra} ($19.4^{+5.7}_{-4.4}\times 10^{22}\rm\,cm^{-2}$) and the 2012 {\it Chandra}/{\it NuSTAR} ($9.5^{+2.3}_{-1.9}\times 10^{22}\rm\,cm^{-2}$) observations. Since we are interested in studying the global properties of AGN in mergers, here we used the average of the two values of the column density.

%################################################################################################################

\subsection{IRAS\,F12590+2934 (NGC\,4922N \& NGC\,4922S)}\label{sect:ngc4922}

This merger stage\,\,C system includes a LIRG (NGC\,4922N) and a nucleus significantly weaker in the IR [NGC\,4922S, $\log (L_{\rm\,IR}/L_{\odot})=8.87 $]. The two nuclei have a projected separation of 10.8\,kpc (Fig.\,\ref{fig:images3}). Significant [Ne\,V] 14.32$\mu$m emission is observed from the system \citep{Inami:2013il}, while only an upper limit was reported for the 24.32$\mu$m line \citep{Pereira-Santaella:2010kx}. The presence of an AGN in the northern nucleus is suggested by the {\it WISE} colours ($W1-W2=1.26$, $W2=9.42$\,mag), while the southern nucleus does not present evidence of a dominating AGN component in the MIR ($W1-W2=-0.04$). The system is classified as a Seyfert\,2 \citep{Yuan:2010ye}, and shows a weak 6.2$\mu$m PAH feature (EW$=0.16\,\mu$m; \citealp{Stierwalt:2013eu}). The depth of the 9.7$\mu$m silicate feature is $\tau_{9.7\mu\rm m}=-0.60$ \citep{Stierwalt:2013eu}, possibly indicating a buried AGN.

The two nuclei are at a distance of 21.2\arcsec, and could be well resolved by {\it Chandra}.
While in the optical NGC\,4922S is significantly brighter than NGC\,4922N, the galaxy is detected by {\it Chandra} only in the 0.3--2\,keV band. NGC\,4922N is detected both in the 0.3--2 and 2--10\,keV band. A source at a position coincident with NGC\,4922N is detected by {\it NuSTAR} at 4.2$\sigma$ and 4.6$\sigma$ in the 3--10\,keV band for FPMA and FPMB, respectively. In the 10--24\,keV band the system is detected at 3.3$\sigma$ in both detectors. Considering the position of the {\it NuSTAR} source, the fact that NGC\,4922S is not detected in the 2--10\,keV, and the similar flux level of the {\it NuSTAR} source with NGC\,4922N in the overlapping energy band, we conclude that the source detected by {\it NuSTAR} is NGC\,4922N. The X-ray spectrum of NGC\,4922N shows a prominent Fe K$\alpha$ line at $6.48^{+0.07}_{-0.07}$\,keV. Fitting the combined 2--24\,keV {\it Chandra}/{\it NuSTAR} spectrum with a power-law plus a Gaussian line, we find that the line has an equivalent width of $3.0_{-1.3}^{+1.4}$\,keV and the X-ray continuum is very hard ($\Gamma=0.2\pm0.4$). Using the Torus model together with a scattered component and a thermal plasma (\textsc{zpowerlaw+atable\{torus1006.fits\}} + \textsc{zphabs$\times$apec}) we find that the X-ray source is CT ($N_{\rm\,H}\geq 4.27\times 10^{24}\rm\,cm^{-2}$), in agreement with the flat X-ray spectrum and the strong Fe K$\alpha$ line. The torus half-opening angle could not be constrained, and was therefore fixed to $\theta_{\rm\,OA}=60$\,deg in the model. The fraction of scattered radiation is $f_{\rm\,scatt}\lesssim 0.5\%$. The broad-band X-ray spectrum of NGC\,4922N is shown in Fig.\,\ref{fig:spectra2}.

%################################################################################################################

\subsection{IRAS\,13120$-$5453}

This ULIRG in merger stage\,\,D shows a single nucleus (Fig.\,\ref{fig:images3}). The presence of an AGN is inferred from the {\it WISE} colours ($W1-W2=0.86$, $W2=8.85$\,mag) and from the Seyfert\,2 optical classification \citep{Veron-Cetty:2010tg}. The spectral decomposition study of \citet{Nardini:2010dz} found that the AGN does not contribute significantly to the bolometric output of the system ($<1.1\%$).

The system was detected by {\it Chandra} \citep{Iwasawa:2011fk} and more recently by {\it NuSTAR} \citep{Teng:2015vn} up to 20\,keV. The X-ray spectral analysis of the combined {\it Chandra} and {\it NuSTAR} spectrum of the source found that the AGN is CT, with a line-of-sight column density of $N_{\rm\,H}\sim 3.2\times10^{24}\rm\,cm^{-2}$ \citep{Teng:2015vn}.

%################################################################################################################

\subsection{IRAS\,F13197$-$1627 (MCG$-$03$-$34$-$064 \& MCG$-$03$-$34$-$063)}\label{sect:13197}

This system is composed of the LIRG MCG$-$03$-$34$-$064, and the normal galaxy MCG$-$03$-$34$-$063. The two galaxies are located at a projected distance of 37.7\,kpc, and are reported as being in merger stage\,\,A. The presence of an AGN in MCG$-$03$-$34$-$064 is confirmed by the detection of the 14.32$\mu$m and 24.32$\mu$m [Ne\,V] emission lines \citep{Pereira-Santaella:2010kx,Inami:2013il}, and by the very weak (EW$<0.01\mu$m) 6.2$\mu$m PAH feature \citep{Stierwalt:2013eu}. The optical spectrum of this source is consistent with that of an HII-region \citep{Yuan:2010ye}.

The two galaxies are separated by 106.2\arcsec, and could be clearly resolved by {\it XMM-Newton}/EPIC. While, MCG$-$03$-$34$-$063 is not detected in the X-rays, MCG$-$03$-$34$-$064 is clearly detected by {\it XMM-Newton}, and it was associated to the {\it Swift}/BAT source.
The X-ray spectrum of MCG$-$03$-$34$-$064 is dominated by the emission from the AGN \citep{Miniutti:2007ys}. We fitted the combined {\it XMM-Newton} EPIC and {\it Swift}/BAT spectra with a model that includes both absorption and reflection from a torus, a scattered component, three thermal plasmas and two Gaussian lines (\textsc{atable\{torus1006.fits\} + 3$\times$apec + 2$\times$zgauss + zpowerlaw}; Fig.\,\ref{fig:spectra2}). We find that, in agreement with the results of \citet{Miniutti:2007ys}, the X-ray source is obscured by Compton-thin material ($N_{\rm\,H}=5.42^{+0.07}_{-0.09}\times 10^{23}\rm\,cm^{-2}$).
The half-opening angle of the torus is $\theta_{\rm\,OA}=79.1^{+0.1}_{-2.3}$\,deg, while only an upper limit is obtained for the fraction of scattered radiation ($\leq 0.2$\%). The two Gaussian lines have energies of $1.86_{-0.01}^{+0.01}$ and $6.62_{-0.01}^{+0.01}$\,keV, with an equivalent width of $40^{+4}_{-30}$\,eV and $197^{+21}_{-40}$\,eV, respectively.

%################################################################################################################

\subsection{IRAS\,13428$+$5608 (Mrk\,273)}

The ULIRG Mrk\,273 is a late merger system (stage\,\,D, Fig.\,\ref{fig:images4}) composed of two nuclei located separated by 0.7\,kpc from each other \citep{Scoville:2000zr}. The MIR spectrum of the source shows a 14.32$\mu$m [Ne\,V] emission line \citep{Inami:2013il}. The presence of an AGN is confirmed by the {\it WISE} colour ($W1-W2=1.182$, $W2=9.23$\,mag), by radio spectral index maps \citep{Vardoulaki:2015mi} and by the weak 6.2$\mu$m PAH feature (EW$=0.12$\,$\mu$m, \citealp{Stierwalt:2013eu}). The source is classified as a Seyfert\,2 in the optical \citep{Veilleux:1999hq}.

\cite{Iwasawa:2011kq} discussed the analysis of the {\it Chandra} observations of the two nuclei (separated by 0.9\arcsec), and their analysis indicated that the AGN coincides with the south-western nucleus, while the northern nucleus contains a powerful starburst which dominates the far infrared emission. However, the extended emission in the 6--7\,keV range in direction of the northern nucleus might imply the presence of a second obscured AGN. Mrk\,273 was clearly detected by {\it NuSTAR} in a $\sim 70$\,ks observation carried out in late 2013. The broad-band {\it XMM-Newton}/{\it NuSTAR} X-ray spectrum of this object was recently studied by \citet{Teng:2015vn}, who found that the X-ray source is obscured by material with a column density of $N_{\rm\,H}=4.4\pm0.1\times 10^{23}\rm\,cm^{-2}$.

%################################################################################################################

\subsection{IRAS\,14378$-$3651}

IRAS\,14378$-$3651 is a ULIRG in merger stage\,\,D with a single nucleus (Fig.\,\ref{fig:images4}). The 14.32$\mu$m and 24.32$\mu$m [Ne\,V] lines are not detected \citep{Pereira-Santaella:2010kx,Inami:2013il}. The {\it WISE} $W1-W2$ colour is above the threshold for AGN activity ($W1-W2=0.87$, $W2=11.44$\,mag), while from the EW of the 6.2$\mu$m feature ($0.39\,\mu$m) \citet{Stierwalt:2013eu} estimate a contribution of the AGN to the MIR flux of $\sim 20\%$. The study of \citet{Nardini:2010dz}, based on spectral decomposition, finds a significantly lower contribution from the AGN to the bolometric luminosity ($<1.3\%$). In the optical band IRAS\,14378$-$3651 has been classified as a LINER by \citet{Kim:1998bh} and as a Seyfert\,2 by \citet{Duc:1997jt}.

The source was detected by {\it Chandra}, and \citet{Iwasawa:2011fk} argued that the hard X-ray colour implies the presence of an AGN with a relatively low 2--10\,keV luminosity. From the small ratio between the 2--10\,keV and the IR luminosity, \citet{Iwasawa:2011fk} proposed that the AGN is a CT candidate. The source was observed by {\it NuSTAR} for 24.5\,ks, and was weakly detected only in the 3--10\,keV band \citep{Teng:2015vn}, thereby providing no additional constraints beyond those obtained by {\it Chandra}.

%################################################################################################################

\subsection{IRAS\,F14544$-$4255 (IC\,4518A \& IC\,4518B)}\label{sect:F14544-4255}

IRAS\,F14544$-$4255 is a LIRG which is composed of two galaxies (IC\,4518A \& IC\,4518B) at a projected separation of 12.0\,kpc (Fig.\,\ref{fig:images4}). The system is classified as an early merger (stage\,\,B). Both the 14.32$\mu$m and the 24.32$\mu$m [Ne\,V] emission lines are observed in the MIR spectrum of IC\,4518A, while they are not detected in that of IC\,4518B \citep{Pereira-Santaella:2010kx,Inami:2013il}. The 6.2$\mu$m PAH feature of IC\,4518A is very weak (EW$=0.05\,\mu$m), but is significantly stronger in IC\,4518B (EW$=0.47\,\mu$m). The {\it WISE} colour confirms the presence of an AGN in IC\,4518A ($W1-W2=1.228$, $W2=8.80$\,mag), and the galaxy is also classified as a Seyfert\,2 in the optical \citep{Masetti:2008bs}. From spectral decomposition, \citet{Hernan-Caballero:2015uq} found that the contribution from the AGN in IC\,4518A to the MIR flux is $\sim 79\%$.

The two galaxies are both detected by {\it XMM-Newton}/EPIC and are at $\sim 36\arcsec$ from each other, so that we used a radius of 5\,arcsec (20\,arcsec) for the spectral extraction of IC\,4518B (IC\,4518A).
IC\,4518A was significantly detected both by EPIC/PN and MOS1 on board {\it XMM-Newton} and by {\it NuSTAR}. The EPIC/MOS2 spectrum could not be extracted since the inner chip where the source was located was not available during this observation. The spectrum was fitted with the same model we used for other AGN (\textsc{atable\{torus1006.fits\} + apec + zpowerlaw}). The model reproduces well the X-ray spectrum, and we find a column density of $N_{\rm\,H}=(2.4\pm 0.2)\times 10^{23}\rm\,cm^{-2}$ in the direction of the X-ray source. The fraction of scattered radiation is $f_{\rm\,scatt}=2.1^{+0.4}_{-0.3}\%$, and only an upper limit is obtained for the half-opening angle of the torus ($\theta_{\rm\,tor}\leq 34$\,deg). We found that the cross-correlation constant for the EPIC/PN and MOS1 spectra is $\sim 0.4$, which implies that the source was $\sim 2.5$ times dimmer than at the time of the {\it NuSTAR} observation. IC\,4518B is detected only by EPIC/PN. Due to the low signal-to-noise ratio the spectrum was rebinned to 1 count per bin and we applied Cash statistics. The X-ray spectrum shows a strong Fe K$\alpha$ line at $\sim 6.4$\,keV (EW$=554_{-240}^{+319}$\,eV), and a very flat 2--10\,keV continuum ($\Gamma=0.23^{+0.61}_{-0.62}$), both clear indications of the presence of a buried AGN. We applied the same model as for IC\,4518A, with the addition of an absorption component for the thermal plasma and the scattered component [\textsc{atable\{torus1006.fits\} + zphabs$\times$(apec + zpowerlaw})]. Due to the low signal-to-noise ratio the photon index was fixed to $\Gamma=1.9$ and the half-opening angle of the torus to 60\,deg. We found that the source is obscured by material with a column density of $N_{\rm\,H}=3.2^{+4.1}_{-1.4}\times 10^{23}\rm\,cm^{-2}$, and the scattered flux is $6.8^{+6.9}_{-3.6}\%$ of the primary X-ray emission. The observed flux of IC\,4518B is $\sim 2\%$ of that of IC\,4518A in the 3--10 and 10--24\,keV bands, so that its contribution to the {\it NuSTAR} spectrum can be safely ignored. The spectra of IC\,4518A and IC\,4518B are shown in Fig.\,\ref{fig:spectra3}.

%################################################################################################################

\subsection{IRAS\,F15327+2340 (Arp\,220W \& Arp\,220E)}

Arp\,220 is the nearest ULIRG, and is in merger stage\,\,D with two nuclei (east and west) separated by only 0.4\,kpc (Fig.\,\ref{fig:images4}). The presence of at least one AGN in this system has been confirmed by several pieces of observational evidence. The western nucleus is classified as a Seyfert\,2 by optical spectroscopy \citep{Yuan:2010ye}. The system shows a weak 6.2$\mu$m PAH feature (EW$=0.17\,\mu$m, \citealp{Stierwalt:2013eu}), which could imply an AGN contribution of $\sim 30\%$ to the MIR flux. Using six multi-wavelength tracers, \citet{Veilleux:2009qo} found that the AGN contributes to $\sim 18.5\%$ of the bolometric flux. A similar value ($17^{+2}_{-2}\%$) was obtained by \citet{Nardini:2010dz} through spectral decomposition. Studies carried out using CO have confirmed that the western nucleus hosts a deeply buried AGN, with a total column density of $\sim 10^{25}\rm\,cm^{-2}$ \citep{Downes:2007ad,Scoville:2015fy}. Consistent with this, \citet{Stierwalt:2013eu} found that this source exhibits a deep 9.7$\mu$m silicate absorption feature ($\tau_{9.7\mu\rm m}=-2.26$).

The two nuclei are separated by 1\arcsec, and {\it Chandra} observations have shown that the peak of the 2--7\,keV emission is in the western nucleus \citep{Iwasawa:2011fk}, while no emission is observed in the same band where the eastern nucleus lies.
In the X-ray band Arp\,220 has been shown to have a flat X-ray continuum ($\Gamma\sim 1$, \citealp{Ptak:2003to,Iwasawa:2005lr}) and a strong emission line at 6.7\,keV (EW$\sim 1.9$\,keV, \citealp{Iwasawa:2005lr}). {\it NuSTAR} detected X-ray emission from Arp\,220 up to 20\,keV \citep{Teng:2015vn}, and using a torus model yields only a lower limit on the column density of $N_{\rm\,H}\geq 5.3\times 10^{24}\rm\,cm^{-2}$, consistent with the results obtained by CO studies. \citet{Teng:2015vn} discuss that the X-ray spectrum could also be well reproduced by a combination of thermal plasma models. However, given the strong evidence for an AGN obtained at other wavelengths, we consider here that the hard X-ray emission is produced by the AGN in Arp\,220W.

%################################################################################################################

\subsection{IRAS\,16504$+$0228 (NGC\,6240N and NGC\,6240S)}

IRAS\,16504$+$0228 is a LIRG in merger stage\,\,D with two nuclei (NGC\,6240N and NGC\,6240s) separated by 0.7\,kpc (Fig.\,\ref{fig:images4}). The system shows a 14.32$\mu$m [Ne\,V] emission line, and in the optical is classified as a LINER \citep{Yuan:2010ye}. The contribution of AGN activity to the bolometric luminosity has been estimated to be $\sim 26\%$ \citep{Veilleux:2009qo}. 

The two nuclei are separated by 1.4\arcsec, and were both detected by {\it Chandra} observations \citep{Komossa:2003xu}.
The presence of an AGN has been confirmed by several studies carried out in the X-ray band \citep{Iwasawa:1998ys,Vignati:1999qc,Lira:2002wu}. {\it Chandra} observations have shown that NGC\,6240S is brighter than NGC\,6240N, and that both nuclei host CT AGN \citep{Komossa:2003xu}. Combining {\it Chandra}, {\it XMM-Newton} and {\it BeppoSAX} data with recent {\it NuSTAR} observations, \citet{Puccetti:2016cj} found that both NGC\,6240S ($N_{\rm\,H}= 1.47^{+0.21}_{-0.17}\times 10^{24}\rm\,cm^{-2}$) and NGC\,6240N ($N_{\rm\,H}= 1.55^{+0.72}_{-0.23}\times 10^{24}\rm\,cm^{-2}$) host CT AGN.

%################################################################################################################

\subsection{IRAS\,F16577+5900 (NGC\,6286 \& NGC\,6285)}\label{sect:F16577+5900}

IRAS\,F16577+5900 is a system composed of a LIRG (NGC\,6286) and its companion (NGC\,6285), which is significantly less luminous in the IR [$\log (L_{\rm\,IR}/L_{\odot})=10.72$]. The two galaxies are in merger stage\,\,B and are separated by 35.8\,kpc (Fig.\,\ref{fig:images4}). None of the tracers of AGN activity show evidence of an accreting SMBH in either nucleus, with the possible exception of faint 14.32$\mu$m and 24.32$\mu$m [Ne\,V] emission lines in NGC\,6286 \citep{Dudik:2009cq}. The detection of these lines, however, was questioned by \citet{Inami:2013il}. Optically, NGC\,6286 is classified as a composite \citep{Yuan:2010ye}. Studying the multi-wavelength SED of NGC\,6286, \citet{Vega:2008kh} found that an AGN might contribute to $\sim 5\%$ of the MIR emission of the galaxy.

Given the separation between the two sources (91.1\arcsec), they could be well resolved by {\it XMM-Newton}/EPIC, {\it Chandra} and {\it NuSTAR} observations.
A joint analysis of {\it XMM-Newton}, {\it Chandra} and {\it NuSTAR} observations of this system has shown that NGC\,6286 contains a buried AGN \citep{Ricci:2016zr}, with a column density of $N_{\rm\,H}= 1.1^{+1.1}_{-0.4}\times 10^{24}\rm\,cm^{-2}$. NGC\,6285 was not detected by {\it NuSTAR}, and the {\it Chandra} spectrum is well reproduced by a single power-law model (\textsc{zpowerlaw}; Fig.\,\ref{fig:spectra3}).

%################################################################################################################

\subsection{IRAS\,F17138$-$1017}\label{sect:17138}

The LIRG IRAS\,F17138$-$1017 is a coalesced merger (stage\,\,D) showing only one nucleus (Fig.\,\ref{fig:images5}). None of the multi-wavelength tracers find any evidence of AGN activity in this system, which is optically classified as a composite \citep{Yuan:2010ye}.

The source was detected by {\it NuSTAR} at the 4.3 and 5$\sigma$ levels in the 3--10\,keV band for FPMA and FPMB, respectively, but remains undetected in the 10-24\,keV band. Given the non-detection above 10\,keV we modelled only the 3--10\,keV region of the {\it NuSTAR} spectrum. The combined {\it Chandra} and {\it NuSTAR} spectra can be well fit with a simple power-law model (\textsc{zpowerlaw}) with a photon index of $\Gamma=1.13_{-0.16}^{+0.17}$. This is harder than the typical X-ray emission observed for star-forming regions. While we cannot exclude the presence of a low-luminosity or heavily obscured AGN in this object, both the multi-wavelength tracers and the low 2--10\,keV X-ray luminosity [$\log (L_{\rm\,2-10}/\rm erg s^{-1})=40.99$] are consistent with a star-forming galaxy. The X-ray spectrum of the source is shown in Fig.\,\ref{fig:spectra3}.

%################################################################################################################

\subsection{IRAS\,20264+2533 (MCG +04$-$48$-$002 \& NGC\,6921)}\label{sect:20264+2533}

IRAS\,20264+2533 is an early merger (stage\,\,A) composed of two galaxies with a projected separation of 27.1\,kpc: NGC\,6921 and the LIRG MCG +04$-$48$-$002 (Fig.\,\ref{fig:images5}). The presence of an AGN in MCG +04$-$48$-$002 is suggested by the detection of [Ne\,V] 14.32$\mu$m, although the source is found to be an HII-region in the optical \citep{Masetti:2006fv}.

The two galaxies are separated by 91.4\arcsec, and could therefore be resolved by both {\it XMM-Newton}/EPIC and {\it NuSTAR} observations.
X-ray observations of these two sources were recently discussed by \citet{Koss:2016oz}, who found both sources to host obscured AGN. Both objects show Fe K$\alpha$ lines at 6.4\,keV, and this feature is particularly prominent in NGC\,6921. We studied the combined {\it XMM-Newton} and {\it NuSTAR} spectrum of each source using the same combination of models applied to other AGN (\textsc{zpowerlaw+atable\{torus1006.fits\}+apec}). For MCG +04$-$48$-$002 we found that the flux of the primary X-ray emission during the {\it NuSTAR} observation was about one fourth of that measured by the {\it XMM-Newton} observation (the cross-calibration constants were $0.27\pm0.03$ and $0.26\pm0.03$ for FPMA and FPMB, respectively). The X-ray source is obscured by material with a column density of $N_{\rm\,H}=5.8^{+0.7}_{-0.4}\times 10^{23}\rm\,cm^{-2}$, while the ratio between scattered radiation and primary X-ray emission is found to be $\leq 0.3\%$. We also constrain the half-opening angle of the torus to be $\theta_{\rm\,OA}=78^{+1.0}_{-0.4}$\,deg. Previous observations of NGC 6921 found that the galaxy hosts a CT AGN \citep{Ricci:2015tg,Koss:2016oz}. Our spectral analysis confirms this result. Applying the same model we used for MCG +04$-$48$-$002, we obtained a column density in the direction of the X-ray source of $N_{\rm\,H}=1.78^{+0.30}_{-0.53}\times 10^{24}\rm\,cm^{-2}$. The scattered radiation is $\leq 0.6\%$ of the primary component, while the half-opening angle of the torus is $\theta_{\rm\,OA}=55^{+16}_{-16}$\,deg. The broad-band X-ray spectra of these two sources are shown in Fig.\,\ref{fig:spectra3}.

%################################################################################################################

\subsection{IRAS\,F21453$-$3511 (NGC\,7130)}\label{sect:n7130}

The LIRG IRAS\,F21453$-$3511 is in merger stage  {\it D} with a single nucleus (Fig.\,\ref{fig:images5}). The MIR spectrum of the source shows [Ne\,V] 14.32$\mu$m and 24.32$\mu$m emission lines \citep{Pereira-Santaella:2010kx,Inami:2013il}, clear indications of AGN activity. The galaxy is classified as a Seyfert\,2 \citep{Yuan:2010ye}, and from the 6.2$\mu$m PAH feature \citet{Stierwalt:2013eu} suggested that the AGN contributes $\sim 40\%$ of the MIR flux. From spectral decomposition, \citet{Hernan-Caballero:2015uq} concluded that $\sim 51\%$ of the MIR luminosity is due to the AGN.

Studying a {\it Chandra} observation, \citet{Levenson:2005ff} found that the galaxy hosts a CT AGN. This was also confirmed by \citet{Gilli:2010rp} using the X-ray to [Ne\,V]3426 flux ratio. Our analysis of the combined {\it Chandra} and {\it NuSTAR} spectrum confirms the heavy obscuration of this source, which shows a very strong Fe\,K$\alpha$ line at 6.4\,keV (EW$=1.10_{-0.22}^{+0.27}$\,keV). Using the torus model plus two different thermal plasmas [\textsc{zpowerlaw+atable\{torus1006.fits\}} + \textsc{zphabs$\times$(apec+apec)}] we found a column density of $N_{\rm\,H}=4.07^{+1.52}_{-0.91}\times 10^{24}\rm\,cm^{-2}$ and that $0.3\pm0.2\%$ of the primary X-ray radiation is scattered. Only an upper limit for the half-opening angle of the torus could be obtained by our analysis ($\theta_{\rm\,OA}\leq 37$\,deg). The broad-band X-ray spectrum of NGC\,7130 is presented in Fig.\,\ref{fig:spectra4}.

%################################################################################################################

\subsection{IRAS\,F23007+0836 (NGC\,7469 and IC\,5283)}\label{sect:F23007}

IRAS\,F23007+0836 is a system composed of a pair of galaxies in an early merger (stage\,\,A) at a projected distance of 26\,kpc (Fig.\,\ref{fig:images5}). NGC\,7469 is a LIRG, while IC\,5283 has a lower IR luminosity ($\log L_{\rm\,IR}=10.79\,L_{\odot}$). The presence of an AGN in NGC\,7469 has been confirmed by several tracers. The MIR spectrum shows both 14.32$\mu$m and 24.32$\mu$m [Ne\,V] emission lines \citep{Pereira-Santaella:2010kx,Inami:2013il}, and weak 3.3$\mu$m (EW$=32$\,nm) and 6.2$\mu$m (EW$=0.23\,\mu$m) PAH features \citep{Yamada:2013ly,Stierwalt:2013eu}. The AGN is optically classified as a Seyfert\,1 \citep{Yuan:2010ye}, and it has been estimated that it contributes $\sim 64\%$ of the MIR flux.

The two galaxies are separated by 79.7\arcsec, and could therefore be resolved by {\it XMM-Newton}/EPIC, {\it NuSTAR} and {\it Swift}/XRT observations.
Previous X-ray studies of NGC\,7469 carried out using {\it XMM-Newton} \citep{Blustin:2003uq,Scott:2005kx} and {\it Suzaku} \citep{Walton:2013vn} have shown that the AGN is obscured only by ionised material, in agreement with the optical classification.  To constrain the level of obscuration of the source we studied the combined {\it XMM-Newton} and {\it Swift}/BAT spectrum. Data from EPIC/MOS were not used because the inner chips of MOS1 and MOS2 were missing during the observation.
We used a model that considers reprocessed X-ray emission from a slab [\textsc{pexrav}, \citep{Magdziarz:1995pi}], a Gaussian line for the Fe K$\alpha$ emission (\textsc{zgauss}), a soft excess in the form of a blackbody (\textsc{bbody}) and ionised absorption (\textsc{zxipcf}). The model we used is \textsc{zxipcf$\times$(bbody + zgauss + zxipcf$\times$pexrav)}. We find that the neutral column density in the direction of the source is extremely low [$\log (N_{\rm\,H}/\rm\,cm^{-2})\sim 19.8$], and the Fe K$\alpha$ line at $6.40\pm0.01$\,keV is relatively weak (EW$=81^{+17}_{-14}$\,eV). The two ionised absorbers have column densities of $N_{\rm\,H}^{\rm\,W,1}=1.7^{+6.6}_{-0.6}\times10^{21}\rm\,cm^{-2}$ and $N_{\rm\,H}^{\rm\,W,2}=1.8^{+0.5}_{-0.2}\times10^{23}\rm\,cm^{-2}$, ionization parameters of $\log (\xi^1/\rm erg\,cm\,s^{-1})=2.36\pm0.07$ and $\log (\xi^2/\rm erg\,cm\,s^{-1})\geq -0.20$, and cover a fraction of $f^1\geq 37\%$ and $f^2= 32_{-4}^{+2}\%$ of the X-ray source. The temperature of the blackbody component is $kT=0.106\pm0.002$\,eV, while the reflection parameter is $R=0.77^{+0.41}_{-0.34}$. The X-ray spectrum of NGC\,7469 is shown in Fig.\,\ref{fig:spectra4}. 

The EPIC/PN observation of IRAS\,F23007+0836 was carried out in small-window mode, and therefore IC\,5283 fell outside of the limited FOV of this {\it XMM-Newton} observation. We therefore analysed the {\it Swift}/XRT image of the field, but found that IC\,5283 is not detected either in the 0.3--2\,keV or in the 2--10\,keV band.

%################################################################################################################

\subsection{IRAS\,F23254+0830 (NGC\,7674 \& NGC\,7674A)}

IRAS\,F23254+0830 is a galaxy pair in merger stage\,\,A, with the two galaxies having a separation of $20.7$\,kpc (Fig.\,\ref{fig:images5}). NGC\,7674 is a LIRG, while NGC\,7674A has an IR luminosity of $\log (L_{\rm\,IR}/L_{\odot})=10.01$. The presence of an AGN in NGC\,7674 is confirmed by several proxies. Both [Ne\,V] 14.32$\mu$m and 24.32$\mu$m are significantly detected \citep{Pereira-Santaella:2010kx,Inami:2013il}. The $W1-W2$ {\it WISE} colour is above the threshold ($W1-W2=1.16$, $W2=8.11$\,mag), and the NIR slope shows a red continuum ($\Gamma_{2.5-5\mu\rm m}=1.4$, \citealp{Imanishi:2010uq}). Both the 3.3$\mu$m (EW$=21$\,nm) and the 6.2$\mu$m (EW$=0.02\,\mu$m) PAH features are weak \citep{Imanishi:2010uq,Stierwalt:2013eu}. Optical spectroscopy studies have classified NGC\,7674 as  a Seyfert\,2 \citep{Yuan:2010ye}.

NGC\,7674A, which is located at 34.1\arcsec  from NGC\,7674, is not detected in the X-ray band by observations carried out with {\it Swift}/XRT, which would be able to resolve the two sources \citep{Gandhi:2016bq}.
A {\it NuSTAR} observation of NGC\,7674, combined with {\it Suzaku}/XIS and {\it Swift}/XRT observations, was recently studied by \citet{Gandhi:2016bq}. In this work it was found that, although the Fe K$\alpha$ line is relatively weak (EW$=400$\,eV), the AGN is CT and possibly reflection dominated, with a line-of-sight column density of $N_{\rm\,H}\gtrsim 3\times 10^{24}\rm\,cm^{-2}$.

%################################################################################################################

\subsection{IRAS\,23262+0314 (NGC\,7679 and NGC\,7682)}\label{sect:23262}

IRAS\,23262+0314 is a system in an early merger (stage\,\,A) with the two galaxies (NGC\,7679 and NGC\,7682) separated by 97.3\,kpc (Fig.\,\ref{fig:images5}). NGC\,7679 is a LIRG which shows a 14.32\,$\mu$m [Ne\,V] emission line \citep{Inami:2013il} and is optically classified as a Seyfert\,2 \citep{Yuan:2010ye}. Spectral decomposition studies have shown that the AGN in this galaxy could contribute to $\sim 19\%$ of the MIR emission \citep{Hernan-Caballero:2015uq}. NGC\,7682 is not a luminous IR source and is optically classified as a Seyfert\,2 \citep{Huchra:1992hb}.

The two galaxies are separated by 269.7\arcsec, and could be well resolved by the {\it XMM-Newton} observations analysed here.
NGC\,7679 was found to be unobscured with a luminosity of $3.4\times 10^{42}\rm\,erg\,s^{-1}$ in the analysis of a {\it BeppoSAX} observation \citep{Risaliti:2002le,Dadina:2007tw}. The X-ray luminosity is above the threshold typically used to separate AGN from starbursts in the 2--10 keV band ($10^{42}\rm\,erg\,s^{-1}$; \citealp{Kartaltepe:2010qf}). In a subsequent {\it XMM-Newton} observation we found the X-ray source was about ten times dimmer in the 2--10\,keV band ($4\times 10^{41}\rm\,erg\,s^{-1}$), and the spectrum could be well modeled with a power-law component, plus a thermal component, a Gaussian line and neutral absorption [\textsc{zphabs$\times$(apec+po+zgauss)}]. We confirm that the source is unobscured and find that the Gaussian line has an energy of $E=6.57_{-0.09}^{+0.07}$\,keV and an equivalent width of $450^{+171}_{-200}$\,eV. 
The strong X-ray variability indicates the presence of an AGN. Using the relation of \citet{Ranalli:2003kb}, we find that a star formation rate of $\sim 80\rm\,M_{\odot}\,yr^{-1}$ is needed to account for the 2--10\,keV luminosity of the source at the time of the {\it XMM-Newton} observation. This is $\sim 4$ times larger than the value obtained from H$\alpha$ ($21.2\pm0.2\rm\,M_{\odot}\,yr^{-1}$, \citealp{Davies:2016fq}) and $\sim 8$ times that inferred from the 3.6$\mu$m and 8$\mu$m images  ($11.4\pm0.6\rm\,M_{\odot}\,yr^{-1}$, \citealp{Davies:2016fq}). The star formation rate needed to reproduce the X-ray luminosity found by \citet{Dadina:2007tw} would be $\sim 680\rm\,M_{\odot}\,yr^{-1}$, $\sim 30-60$ times larger than the value observed. We therefore conclude that most of the X-ray emission observed is most likely due to the AGN. NGC\,7682 is a known CT AGN \citep{Singh:2011xw,Ricci:2015tg}. We analysed the {\it XMM-Newton} of this object using a torus model plus scattered emission and a thermal plasma [\textsc{zpowerlaw+atable\{torus1006.fits\}+zphabs$\times$apec}]. Our analysis confirms that the X-ray source is CT, with a column density of $N_{\rm\,H}=2.43^{+0.60}_{-0.44}\times 10^{24}\rm\,cm^{-2}$. We find that the half-opening angle of the torus is $\theta_{\rm\,OA}\geq 68$\,deg and $\leq 0.04$\% of the primary radiation is scattered. The X-ray spectra of NGC\,7679 and NGC\,7682 are illustrated in Fig.\,\ref{fig:spectra4}.

%################################################################################################################
%################################################################################################################
%################################################################################################################
%################################################################################################################
%################################################################################################################
%################################################################################################################
%################################################################################################################
%################################################################################################################
%################################################################################################################

\section{X-ray spectra}\label{appendix:xrayspec}
\begin{figure*}
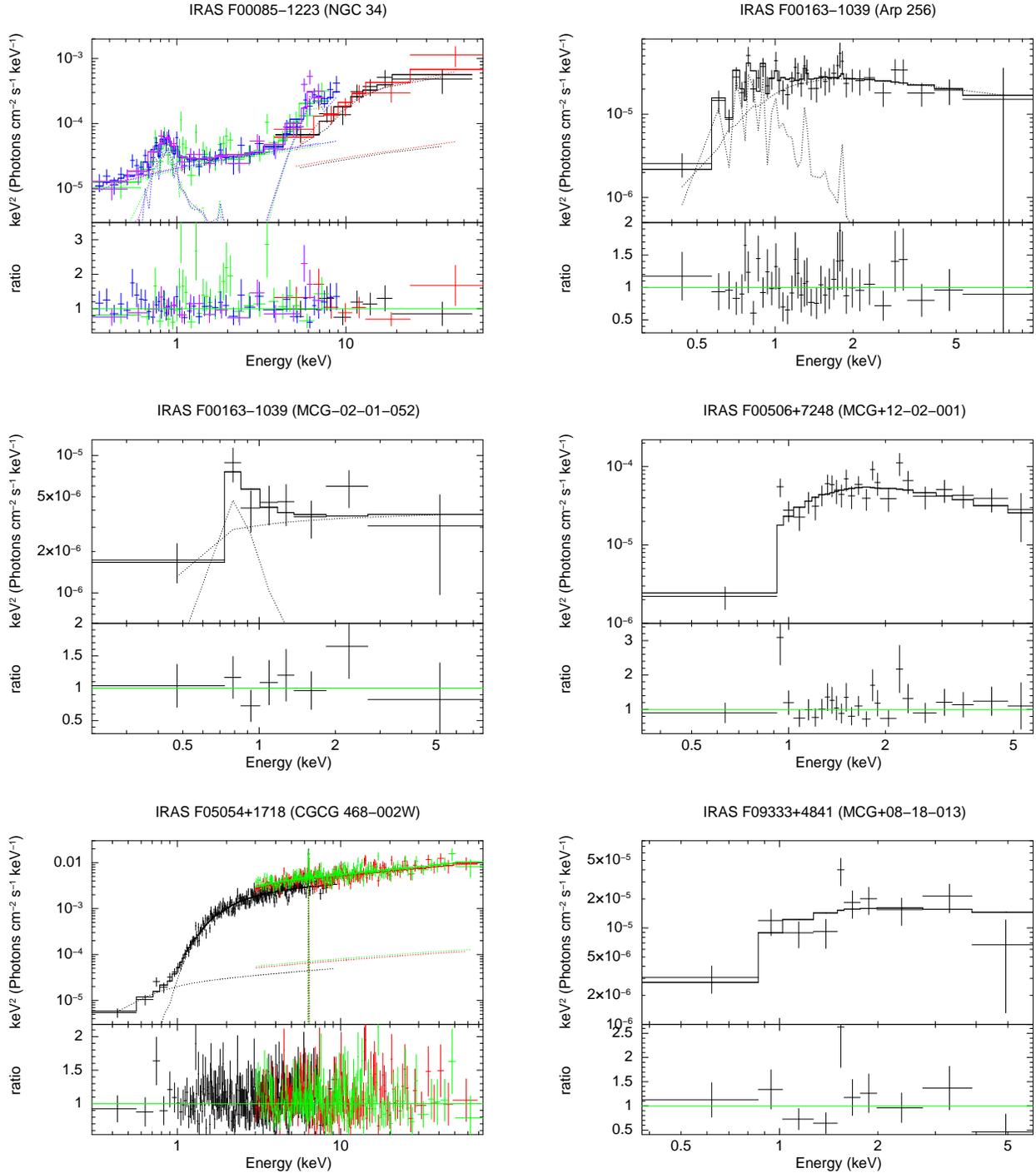

\centering
 \begin{minipage}{.49\textwidth}
\centering
\includegraphics[height=8cm, angle=270]{figures/NGC34_spec.ps}\end{minipage}
\begin{minipage}[!h]{.49\textwidth}
\centering
\includegraphics[height=8cm, angle=270]{figures/Arp256_spec.ps}\end{minipage}
\par\bigskip
\begin{minipage}[!h]{.49\textwidth}
\centering
\includegraphics[height=8cm, angle=270]{figures/MCG-02-01-052_spec.ps}\end{minipage}
\begin{minipage}{.49\textwidth}
\centering
\includegraphics[height=8cm, angle=270]{figures/MCG+12-02-001_spec.ps}\end{minipage}
\par\bigskip
\begin{minipage}{.49\textwidth}
\centering
\includegraphics[height=8cm, angle=270]{figures/CGCG468-002W_spec.ps}\end{minipage}
\begin{minipage}{.49\textwidth}
\centering
\includegraphics[height=8cm, angle=270]{figures/IRASF09333+4841_spec.ps}\end{minipage}
 \begin{minipage}{1\textwidth}
  \caption{X-ray spectra of NGC\,34 [{\it NuSTAR}, {\it Chandra} and {\it XMM-Newton}; $\S$\ref{sect:ngc34}, $N_{\rm\,H}=(5.3\pm1.1)\times10^{23}\rm\,cm^{-2}$], Arp\,256 ({\it Chandra}; $\S$\ref{sect:IRASF00163-1039}), MCG$-$02$-$01$-$052 ({\it Chandra}; $\S$\ref{sect:IRASF00163-1039}), MCG+12$-$02$-$001 ({\it Chandra}; $\S$\ref{sect:IRASF00506+7248}), CGCG 468$-$002W [{\it NuSTAR} and {\it Swift}/XRT; $\S$\ref{sect:IRASF05054+1718}, $N_{\rm\,H}=(1.50\pm0.09)\times10^{22}\rm\,cm^{-2}$], and MCG+08$-$18$-$013 ({\it Chandra}; $\S$\ref{sect:IRASF09333+4841}). The bottom panels show the ratio between the data and the models.}
\label{fig:spectra1}
 \end{minipage}
\end{figure*}

\begin{figure*}
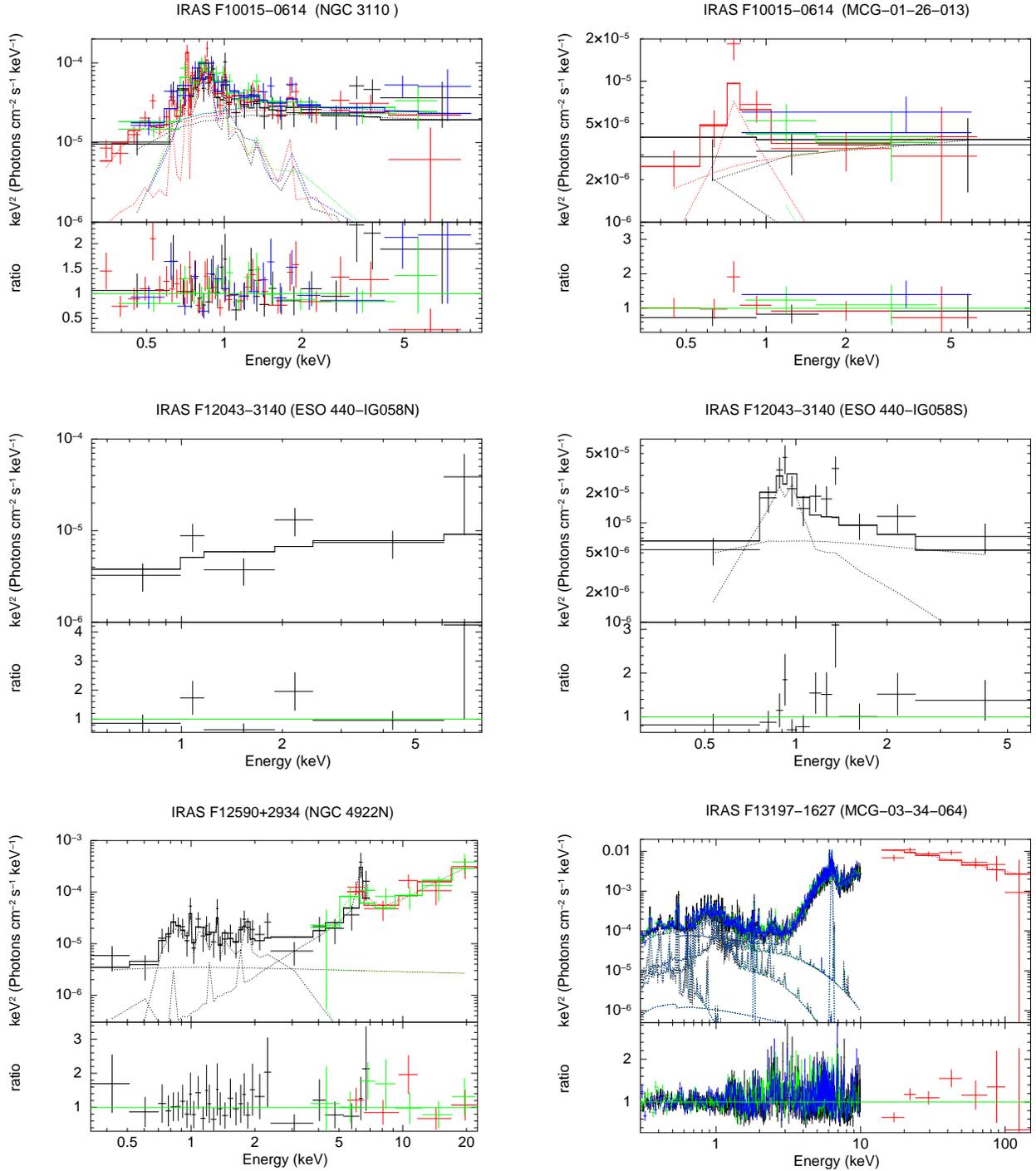

\centering
\begin{minipage}[!h]{.49\textwidth}
\centering
\includegraphics[height=8cm, angle=270]{figures/NGC3110_spec.ps}\end{minipage}
\begin{minipage}[!h]{.49\textwidth}
\centering
\includegraphics[height=8cm, angle=270]{figures/MCG-01-26-013_spec.ps}\end{minipage}
\par\bigskip
\begin{minipage}[!h]{.49\textwidth}
\centering
\includegraphics[height=8cm, angle=270]{figures/ESO440-IG058N_spec.ps}\end{minipage}
\begin{minipage}[!h]{.49\textwidth}
\centering
\includegraphics[height=8cm, angle=270]{figures/ESO440-IG058S_spec.ps}\end{minipage}
\par\bigskip
\begin{minipage}[!h]{.49\textwidth}
\centering
\includegraphics[height=8cm, angle=270]{figures/NGC4922N_spec.ps}\end{minipage}
\begin{minipage}[!h]{.49\textwidth}
\centering
\includegraphics[height=8cm, angle=270]{figures/IRASF13197.ps}\end{minipage}
 \begin{minipage}{1\textwidth}
  \caption{X-ray spectra of NGC\,3110 ({\it XMM-Newton} and {\it Chandra}; $\S$\ref{sect:F10015-0614}), MCG$-$01$-$26$-$013 ({\it XMM-Newton} and {\it Chandra}, $\S$\ref{sect:F10015-0614}), ESO\,440$-$IG058N ({\it Chandra}; $\S$\ref{sect:F12043-3140}), ESO\,440$-$IG058S ({\it Chandra}; $\S$\ref{sect:F12043-3140}), NGC\,4922N ({\it NuSTAR} and {\it Chandra}; $\S$\ref{sect:ngc4922}, $N_{\rm\,H}\geq4.27\times10^{24}\rm\,cm^{-2}$), and MCG$-$03$-$34$-$064 [{\it XMM-Newton} EPIC and {\it Swift}/BAT; $\S$\ref{sect:13197}, $N_{\rm\,H}=(54.2^{+0.7}_{-0.9})\times10^{22}\rm\,cm^{-2}$].}
\label{fig:spectra2}
\end{minipage}
\end{figure*}

\begin{figure*}
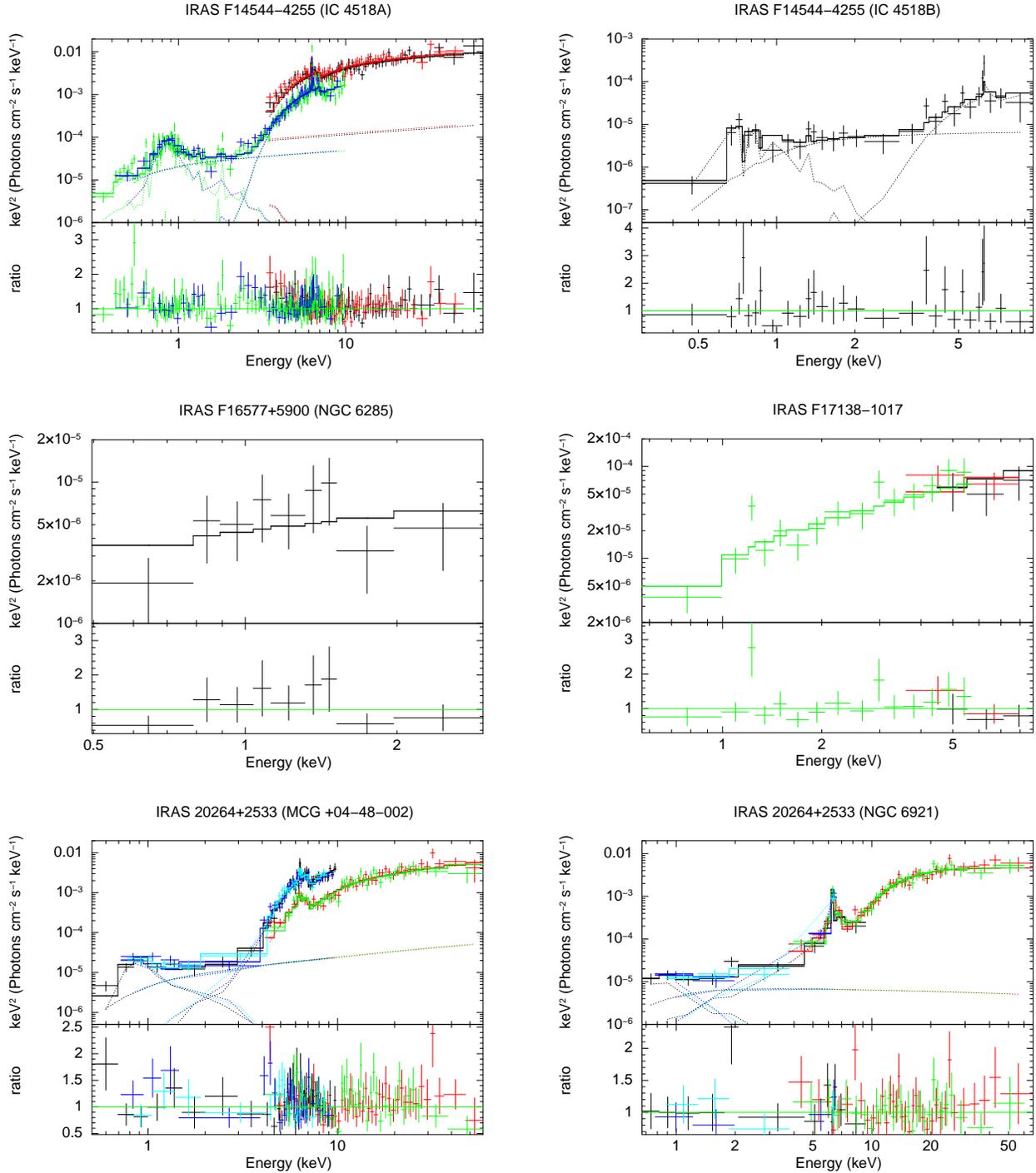

\centering
\begin{minipage}[!h]{.49\textwidth}
\centering
\includegraphics[height=8cm, angle=270]{figures/IC4518A_spec.ps}\end{minipage}
\begin{minipage}[!h]{.49\textwidth}
\centering
\includegraphics[height=8cm, angle=270]{figures/IC4518B_spec.ps}\end{minipage}
\par\bigskip
\begin{minipage}[!h]{.49\textwidth}
\centering
\includegraphics[height=8cm, angle=270]{figures/NGC6285_spec.ps}\end{minipage}
\begin{minipage}[!h]{.49\textwidth}
\centering
\includegraphics[height=8cm, angle=270]{figures/F17138-1017_spec.ps}\end{minipage}
\par\bigskip
\begin{minipage}[!h]{.49\textwidth}
\centering
\includegraphics[height=8cm, angle=270]{figures/MCG+04-48_spec.ps}\end{minipage}
\begin{minipage}[!h]{.49\textwidth}
\centering
\includegraphics[height=8cm, angle=270]{figures/NGC6921_spec.ps}\end{minipage}
% %% caption
 \begin{minipage}{1\textwidth}
  \caption{X-ray spectra of IC\,4518A [{\it NuSTAR} and {\it XMM-Newton}; $\S$\ref{sect:F14544-4255}, $N_{\rm\,H}=(2.4\pm0.2)\times10^{23}\rm\,cm^{-2}$], IC\,4518B [{\it XMM-Newton}; $\S$\ref{sect:F14544-4255}, $N_{\rm\,H}=(3.2^{+4.1}_{-1.4})\times10^{23}\rm\,cm^{-2}$], NGC\,6285 ({\it Chandra}, $\S$\ref{sect:F16577+5900}), IRAS\,F17138$-$1017 ({\it NuSTAR} and {\it Chandra}; $\S$\ref{sect:17138}), MCG +04$-$48$-$002 [{\it NuSTAR} and {\it XMM-Newton}, $\S$\ref{sect:20264+2533}, $N_{\rm\,H}=(5.8^{+0.7}_{-0.4})\times10^{23}\rm\,cm^{-2}$], and NGC\,6921 ({\it NuSTAR} and {\it XMM-Newton}, $\S$\ref{sect:20264+2533}, $N_{\rm\,H}=(1.78^{+0.30}_{-0.53})\times10^{24}\rm\,cm^{-2}$).}
\label{fig:spectra3}
 \end{minipage}
\end{figure*}

\begin{figure*}
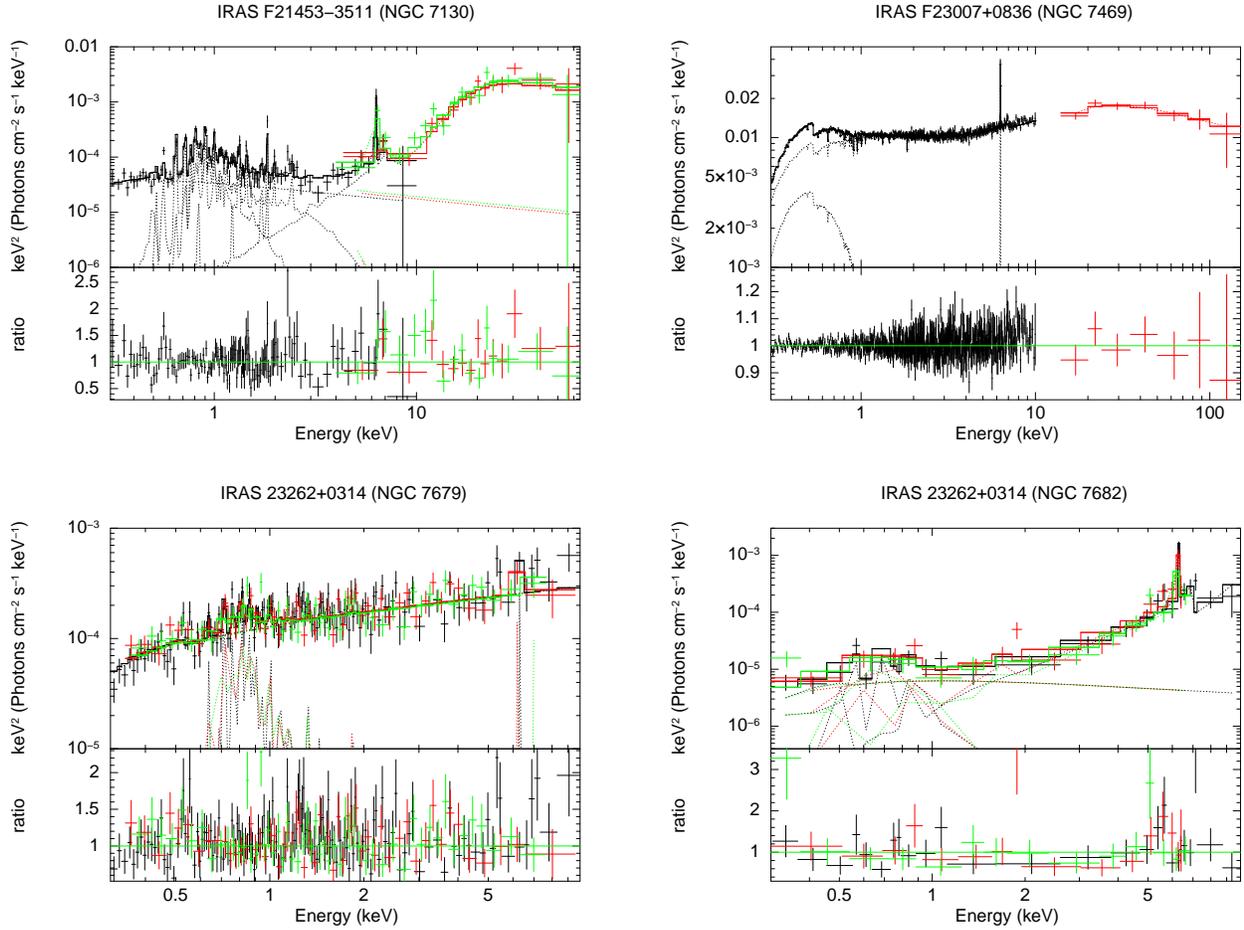

\centering
\begin{minipage}[!h]{.49\textwidth}
\centering
\includegraphics[height=8cm, angle=270]{figures/NGC7130_spec.ps}\end{minipage}
\begin{minipage}[!h]{.49\textwidth}
\centering
\includegraphics[height=8cm, angle=270]{figures/N7469_spec.ps}\end{minipage}
\par\bigskip
\begin{minipage}[!h]{.49\textwidth}
\centering
\includegraphics[height=8cm, angle=270]{figures/N7679_spec.ps}\end{minipage}
\begin{minipage}[!h]{.49\textwidth}
\centering
\includegraphics[height=8cm, angle=270]{figures/N7682_spec.ps}\end{minipage}
%
% %% caption
 \begin{minipage}{1\textwidth}
  \caption{X-ray spectra of NGC\,7130 [{\it NuSTAR} and {\it Chandra}; $\S$\ref{sect:n7130}, $N_{\rm\,H}=(4.07^{+1.52}_{-0.91})\times10^{24}\rm\,cm^{-2}$], NGC\,7469 [{\it XMM-Newton} and {\it Swift}/BAT; $\S$\ref{sect:F23007}, $N_{\rm\,H}=(6\pm2)\times10^{19}\rm\,cm^{-2}$], NGC\,7679 [{\it XMM-Newton}; $\S$\ref{sect:23262}, $N_{\rm\,H}\leq 2\times10^{20}\rm\,cm^{-2}$], and  NGC\,7682 [{\it XMM-Newton}; $\S$\ref{sect:23262}, $N_{\rm\,H}=(2.43^{+0.60}_{-0.44})\times10^{24}\rm\,cm^{-2}$]. }
\label{fig:spectra4}
 \end{minipage}
\end{figure*}

The X-ray spectra and the best-fit models are illustrated in Figures \ref{fig:spectra1}--\ref{fig:spectra4}. For visualization purposes, spectra are typically rebinned to have at least a significance of $3\sigma$ per bin.

\section*{Acknowledgements}

We thank the referee for the careful reading of the manuscript and for the prompt report that helped us improve the quality of the paper. We thank the {\it NuSTAR} Cycle 1 TAC for the {\it NuSTAR} data on which this paper is based, Chin-Shin Chang, George Lansbury, Dave Alexander and Johannes Buchner for useful comments on the manuscript.
This work made use of data from the {\it NuSTAR} mission, a project led by the California Institute of Technology, managed by the Jet Propulsion Laboratory, and funded by the National Aeronautics and Space Administration. We thank the {\it NuSTAR} Operations, Software and Calibration teams for support with the execution and analysis of these observations.  This research has made use of the {\it NuSTAR} Data Analysis Software (\textsc{NuSTARDAS}) jointly developed by the ASI Science Data Center (ASDC, Italy) and the California Institute of Technology (Caltech, USA), and of the NASA/ IPAC Infrared Science Archive and NASA/IPAC Extragalactic Database (NED), which are operated by the Jet Propulsion Laboratory, California Institute of Technology, under contract with the National Aeronautics and Space Administration. We acknowledge financial support from the CONICYT-Chile grants ``EMBIGGEN" Anillo ACT1101 (CR, FEB, ET), FONDECYT 1141218 (CR, FEB), FONDECYT 1160999 (ET), FONDECYT 3150361 (GP), Basal-CATA PFB--06/2007 (CR, FEB, ET), the NASA {\it NuSTAR} AO1 Award NNX15AV27G (FEB), the China-CONICYT fund (CR), the Swiss National Science Foundation (Grant PP00P2\_138979/1 and PP00P2\_166159, KS), the Chinese Academy of Science grant No. XDB09030102 (LH), the National Natural Science Foundation of China  grant No. 11473002 (LH), the Ministry of Science and Technology of China grant No. 2016YFA0400702 (LH), the Spanish MINECO (KI) under grant AYA2013-47447-C3-2-P and MDM-2014-0369 of ICCUB (Unidad de Excelencia ÒMar\'ia de Maeztu"), and the Ministry of Economy, Development, and Tourism's Millennium Science Initiative through grant IC120009, awarded to The Millennium Institute of Astrophysics, MAS (FEB).

\bibliographystyle{mnras}
 \bibliography{mergers}
 
 \end{document}